\documentclass[longauth]{aa}
\usepackage{xcolor}
\usepackage{chemfig}
\usepackage{graphicx}
\usepackage{longtable}
\usepackage{amsmath}
\usepackage{mwe}
\usepackage{fontawesome}

\definecolor{dgreen}{rgb}{0.0, 0.5, 0.0}
\usepackage{txfonts}
\usepackage{hyperref}
\hypersetup{
    colorlinks=true,
    linkcolor=blue,
    filecolor=blue,
    citecolor=blue,
    urlcolor=blue,
}

\begin{document} 

   \title{Ice inventory towards the protostar Ced 110 IRS4 observed with the James Webb Space Telescope}
   \subtitle{Results from the ERS Ice Age program}

   \author{W. R. M. Rocha\inst{1,2}
   \and M. K. McClure\inst{2}
   \and J. A. Sturm\inst{2}
   \and T. L. Beck\inst{3}
   \and Z. L. Smith\inst{4,2}
   \and H. Dickinson\inst{4}
   \and F. Sun\inst{5, 6}
   \and E. Egami\inst{5}
   \and A. C. A. Boogert\inst{7}
   \and H. J. Fraser\inst{4}
   \and E. Dartois\inst{8}
   \and I. Jimenez-Serra\inst{9}
   \and J. A. Noble\inst{10}   
   \and J. Bergner\inst{11}
   \and P. Caselli\inst{12}
   \and S. B. Charnley\inst{13}
   \and J. Chiar\inst{14,15}
   \and L. Chu\inst{16}
   \and I. Cooke\inst{17}
   \and N. Crouzet\inst{2}
   \and E. F. van Dishoeck\inst{2}
   \and M. N. Drozdovskaya\inst{18}
   \and R. Garrod\inst{19}
   \and D. Harsono\inst{20}
   \and S. Ioppolo\inst{21}
   \and M. Jin\inst{13,22}
   \and J. K. Jørgensen\inst{23}
   \and T. Lamberts\inst{2}
   \and D. C. Lis\inst{24}
   \and G. J. Melnick\inst{6}
   \and B. A. McGuire\inst{25}
   \and K. I. Öberg\inst{6}
   \and M. E. Palumbo\inst{26}
   \and Y. J. Pendleton\inst{27}
   \and G. Perotti\inst{28}
   \and D. Qasim\inst{29}
   \and B. Shope\inst{16}  
   \and R. G. Urso\inst{26}
   \and S. Viti\inst{2}   
   \and H. Linnartz\inst{2}
    }

   \institute{Laboratory for Astrophysics, Leiden Observatory, Leiden University, P.O. Box 9513, NL 2300 RA Leiden, The Netherlands.\\
    \email{rocha@strw.leidenuniv.nl}
          \and
             Leiden Observatory, Leiden University, PO Box 9513, NL 2300 RA Leiden, The Netherlands
          \and
             Space Telescope Science Institute, Baltimore, MD, USA
          \and
             School of Physical Sciences, The Open University,  Walton Hall, Milton Keynes, MK7 6AA
          \and
             Steward Observatory, University of Arizona, Tucson, AZ, USA
          \and
             Center for Astrophysics | Harvard \& Smithsonian, Cambridge, MA, USA
          \and
             Institute for Astronomy, University of Hawai'i at Manoa, Honolulu, HI, USA   
          \and
             Institut des Sciences Moléculaires d'Orsay, CNRS, Université Paris-Saclay, Orsay, France
          \and
             Centro de Astrobiología (CAB), CSIC-INTA, Ctra. de Ajalvir km 4, E-28850, Torrejón de Ardoz, Spain
          \and
             Physique des Interactions Ioniques et Moléculaires, CNRS, Aix Marseille Université, Marseille, France
          \and
             Department of Chemistry, University of California, Berkeley, Berkeley, CA, USA
          \and
             Max-Planck-Institut für extraterrestrische Physik, Garching bei München, Germany
          \and
             Astrochemistry Laboratory, Code 691, NASA Goddard Space Flight Center, Greenbelt, MD 20771, USA
          \and
             Physical Science Department, Diablo Valley College, 321 Golf Club Road, Pleasant Hill, CA 94523, USA
          \and
             Carl Sagan Center, SETI Institute, 189 Bernardo Avenue, Mountain View, CA 94043, USA
          \and
             Infrared Processing and Analysis Center, California Institute of Technology, Pasadena, CA, USA
          \and
             Department of Chemistry, University of British Columbia, 2036 Main Mall, Vancouver, BC, V6T 1Z1, Canada
          \and
             Physikalisch-Meteorologisches Observatorium Davos und Weltstrahlungszentrum, Davos Dorf, Switzerland
          \and
             Departments of Astronomy \& Chemistry, University of Virginia, Charlottesville, VA, USA
          \and
             Institute of Astronomy, Department of Physics, National Tsing Hua University, Hsinchu 30013, Taiwan
          \and
             Center for Interstellar Catalysis, Department of Physics and Astronomy, Aarhus University, Aarhus, Denmark
          \and
             Department of Physics, Catholic University of America, Washington, DC 20064, USA
          \and
             Niels Bohr Institute, University of Copenhagen, 1350 Copenhagen, Denmark
          \and
             Jet Propulsion Laboratory, California Institute of Technology, 4800 Oak Drove Drive, Pasadena, CA 91109, USA
          \and
             Department of Chemistry, Massachusetts Institute of Technology, Cambridge, MA 02139, USA
         \and
             INAF - Osservatorio Astrofisico di Catania, Catania, Italy
          \and
             Department of Physics, University of Central Florida, Orlando, FL, USA
          \and
             Max Planck Institute for Astronomy, Heidelberg, Germany
          \and
             Southwest Research Institute, San Antonio, TX, USA     
            }

   \date{Submitted xxx; Received xxxx; accepted yyyy}

 
  \abstract
   {Protostars contains icy ingredients necessary for the formation of potential habitable worlds, and therefore it is crucial to understand their chemical and physical environments. This work is focuses on the ice features toward the binary protostellar system Ced~110~IRS4A and IRS4B, separated by 250~au, and observed with the \textit{James Webb} Space Telescope (JWST) as part of the Early Release Science Ice Age collaboration.}
   {We aim at exploring the JWST observations of the binary protostellar system Ced~110~IRS4A and IRS4B to primarily unveil and quantify the ice inventories toward these sources. Finally, we compare the ice abundances with those found for the same molecular cloud.}
   {We use data from multiple JWST instruments (NIRSpec, NIRCam and MIRI) to identify and quantify ice species in the Ced~110~IRS4 system. The analysis is performed by fitting or comparing laboratory infrared spectra of ices to the observations. Spectral fits are carried out with the \texttt{ENIIGMA} fitting tool that searches for the best fit out of a large number of solutions. The degeneracies of the fits are also addressed, and the ice column densities are calculated. In cases where the full nature of the absorption features is not known yet, we explore different laboratory ice spectra to compare with the observations.}
   {We provide a list of securely and tentatively detected ice species towards the primary and the companion sources. For Ced~110~IRS4B, we detected the major ice species H$_2$O, CO, CO$_2$ and NH$_3$. All species are found in a mixture except for CO and CO$_2$, which have both mixed and pure ice components. In the case of Ced~110~IRS4A, we detected the same major species as in Ced~110~IRS4B, as well as the following minor species CH$_4$, SO$_2$, CH$_3$OH, OCN$^-$, NH$_4^+$ and HCOOH. Tentative detection of N$_2$O ice (7.75~$\mu$m), forsterite dust (11.2~$\mu$m) and CH$_3^+$ gas emission (7.18~$\mu$m) in the primary source are also presented. Compared with the two lines of sight toward background stars in the Chameleon I molecular cloud, the protostar has similar ice abundances, except in the case of the ions that are higher in IRS4A. The most clear differences are the absence of the 7.2 and 7.4~$\mu$m absorption features due to HCOO$^-$ and icy complex organic molecules in IRS4A and evidence of thermal processing in both IRS4A and IRS4B as probed by the CO$_2$ ice features.} 
   {We conclude that the binary protostellar system Ced~110~IRS4A and IRS4B has a large inventory of icy species. The similar ice abundances in comparison to the starless regions in the same molecular cloud suggests that the chemical conditions of the protostar were set at earlier stages in the molecular cloud. It is also possible that the source inclination and complex geometry cause a low column density along the line of sight, which hides the bands at 7.2 and 7.4~$\mu$m. Finally, we highlight that a comprehensive analysis using radiative transfer modelling is needed to disentangle the spectral energy distributions of these sources.}

\authorrunning{Rocha et al.}
\titlerunning{A\&A, XXX, YYYY (202Z)}
\maketitle
%

\section{Introduction}
The astrochemical investigation of low-mass young stellar objects (LYSOs) with the James Webb Space Telescope (JWST) has provided astronomers with a unique vantage point to explore the intricate interplay between physics and chemistry during the early stages of star formation. For example, recent JWST observations resulted in the detection of shallow absorption features due to minor solid-phase molecular species, which, however, are crucial to understand the chemical composition of those ices \citet{Yang2022}, \citet{Rocha2024}, \citet{Brunken2024, Brunken2024iso}, \citet{Nazari2024}, \citet{Chen2024}, \citet{Tyagi2024}, \citet{Slavicinska2024nh4, Slavicinska2024hdo}. 

In this context, the  Director's Discretionary-Early Release Science (DD-ERS) program ``Ice Age: Chemical evolution of ices during star formation'' (ID 1309, PIs: Melissa McClure, Adwin Boogert, Harold Linnartz - in memoriam) aims to trace the chemical evolution of ices within regions at different evolutionary stages. For example, observations of two background stars, NIR38 and J110621, by \citet{McClure2023} have shown that the molecular cloud Chameleon I hosts the major ice species (H$_2$O, CO, CO$_2$, NH$_3$, CH$_4$, CH$_3$OH), minor species (OCN$^-$, OCS, $^{13}$CO$_2$ and $^{13}$CO), and likely complex organic molecules (COMs) in ices. Furthermore, \citet{Dartois2024} and \citet{Noble2024} found evidence of grain growth and the OH dangling feature, respectively, in the Chameleon I molecular cloud. In the later stage of evolution, the protoplanetary disk phase, \citet{Sturm2023} demonstrated that, except for the COMs and $^{13}$CO, the same features were observed in the protoplanetary disk HH48. Our work aims to compare the ice inventory and abundances found towards NIR38 and J110621 in the Chameleon I molecular cloud and those in the protostar Cederblad~110~IRS4 (hereafter Ced~110~IRS4). Additionally, we search for evidence of ice thermal processing which is not present in the molecular cloud, as also found in previous works \citep{Boogert2011, Boogert2013}. 

Ced~110~IRS4 is nestled within the Chameleon I Molecular Cloud (R.A. 11$^{\rm{h}}$06$^{\rm{m}}$46.3682$^{\rm{s}}$ and DEC  -77$^{\rm{o}}$22$^{\arcmin}$32.882$^{\arcsec}$). The distances between Ced~110~IRS4 and NIR38 and J110621 are 81 and 114$^{\arcsec}$, respectively. This region is bright in the near and mid-infrared (IR), and four other names exist for the same target in the literature, namely,  NIR76, IRAS~11051-7706 and 2MASS~J11064638-7722287, [PMG2001] NIR~72. \citet{Zinnecker1999} observed this source in the near-IR with the Infrared Spectrometer And Array Camera (ISAAC) camera mounted on the Very Large Telescope (VLT). It is described as a bisected reflection nebula with a dark lane and a highly reddened central source. A later work by \citet{Persi2001} used ESO-SOFI, a sub-arcsecond resolution near-IR camera, to study the Ced~110 region, and derived an average extinction of $A_{\rm{V}}$ = 30~mag. Ced~110~IRS4 is classified as Class I source \citep{Lehtinen2001}, which implies the presence of a large envelope and a young disk. However, a Class 0 designation was attributed to this source based on the bolometric temperature derived from the Photodetector Array Camera and Spectrometer (PACS) data from Herschel \citep[$T_{\rm{bol}}$ = 56~K;][]{Kristensen2012}. The same classification (Class 0, $T_{\rm{bol}}$ = 68~K) is reported by \citet{Ohashi2023}, which is derived from a complete spectral energy distribution (SED) of the source, that includes data between 2 and 1000~$\mu$m. Another important discovery about Ced~110~IRS4 was recently made by the ``Early Planet Formation in Embedded Disks (eDisk)'' large program (PI: Nagayoshi Ohashi), using observations from the Atacama Large Millimeter/submillimeter Array (ALMA) . It revealed that Ced~110~IRS4 is actually a binary system \citep{Sai2023} with a separation of 250~au, thus being renamed as Ced~110~IRS4A (same coordinates as above) and IRS4B (R.A. 11$^{\rm{h}}$06$^{\rm{m}}$46.8194$^{\rm{s}}$ and DEC  -77$^{\rm{o}}$22$^{\arcmin}$32.505$^{\arcsec}$). This pair presents disk-like structures where the primary and secondary sources have a dust disks with estimated radii of 91.7$\pm$0.2 and 33.6$\pm$0.6~au, respectively.

Gas-phase observations of the Ced~110~IRS4A system have shown molecular \citep[H$_2$, H$_2$O, CO, HCO$^+$; e.g.,][]{Hiramatsu2007, vankempen2009, Kristensen2012, Sai2023} and atomic \citep[Fe II; e.g.,][]{Lahuis2010} emissions. Based on C$^{18}$O lines, \citet{vankempen2009} concluded that no foreground cloud is present along the line of sight towards this source, but there is warm gas, which is related to an outflow from IRS4A \citep{Belloche2006}. Non-detections towards IRS4A include CH$_3$OH and SiO, known as shock tracers, based on Atacama Submillimeter Telescope Experiment (ASTE) observations \citep{Hiramatsu2007}. \citet{vgelder2022PhD} further confirmed that no CH$_3$OH gas detection is present at Ced~110~IRS4 system. On the other hand, SO, another shock tracer, was detected by \citet{Sai2023} and shows an arc-like structure on the north side of the system. Solid-phase species, such as H$_2$O, CO$_2$ and silicates were identified toward this source based on the Infrared Spectrograph (IRS) onboard the \textit{Spitzer} Space Telescope \citep{Manoj2011}. Regarding larger species, so far no complex organic molecules (COMs) has been reported toward this source in either in the solid or gas phases. 

The solid-phase nature of the Chameleon I dense cloud where Ced~110~IRS4 is located was recently addressed in two papers from the Ice Age collaboration, \citet{McClure2023}, \citet{Dartois2024} and \citet{Noble2024}, who provided an important comparison to the protostellar environment. For example, the two background stars, NIR38 and J110621228, which were targeted in \cite{McClure2023}, reveal the presence of many ice species, including H$_2$O, CO, CO$_2$, NH$_3$ and OCS, as well as CH$_3$OH, which highlights the complex ice chemistry already occurring prior to protostellar formation. Additionally, \citet{Noble2024} shows the presence of the elusive OH dangling bond at 2.7~$\mu$m. Regarding grain sizes, \citet{Dartois2024} show that grain growth occurs in the Chameleon I cloud by accretion and aggregation given the presence of icy grains of a larger radius (0.9~$\mu$m) in comparison to the interstellar medium \citep[ISM, 0.25~$\mu$m;][]{Mathis1977}.

In this work, we analyze the spectrum of Ced~110~IRS4A and IRS4B using multiple JWST instruments and covering the critical spectral ranges for tracing thermal processing (CO$_2$; 4.27 and 15.2~$\mu$m) and COMs (7$-$9~$\mu$m). It is important to mention that for the first time, we present an ice study of the companion protostar, IRS4B, recently discovered by \citet{Sai2023}. This work primarily focuses on the ice features, and although we point out a few gas-phase lines, a detailed study of gaseous species will be deferred to a future paper. The structure of this work is as follows. Section~\ref{sec_obs} presents the JWST observations and the data reduction process, and the ice laboratory data used for spectral fitting and comparison is presented in Section~\ref{icelab_sec}. The methodology with the different approaches adopted in this manuscript is presented in Section~\ref{sec_met}. We described the analysis and the results in Section~\ref{sec_res}. The discussion is presented in Section~\ref{sec_diss}, and the conclusions are outlined in Section~\ref{sec_conc}.

\section{Observations}
\label{sec_obs}

\begin{figure*}
   \centering
   \includegraphics[width=\hsize]{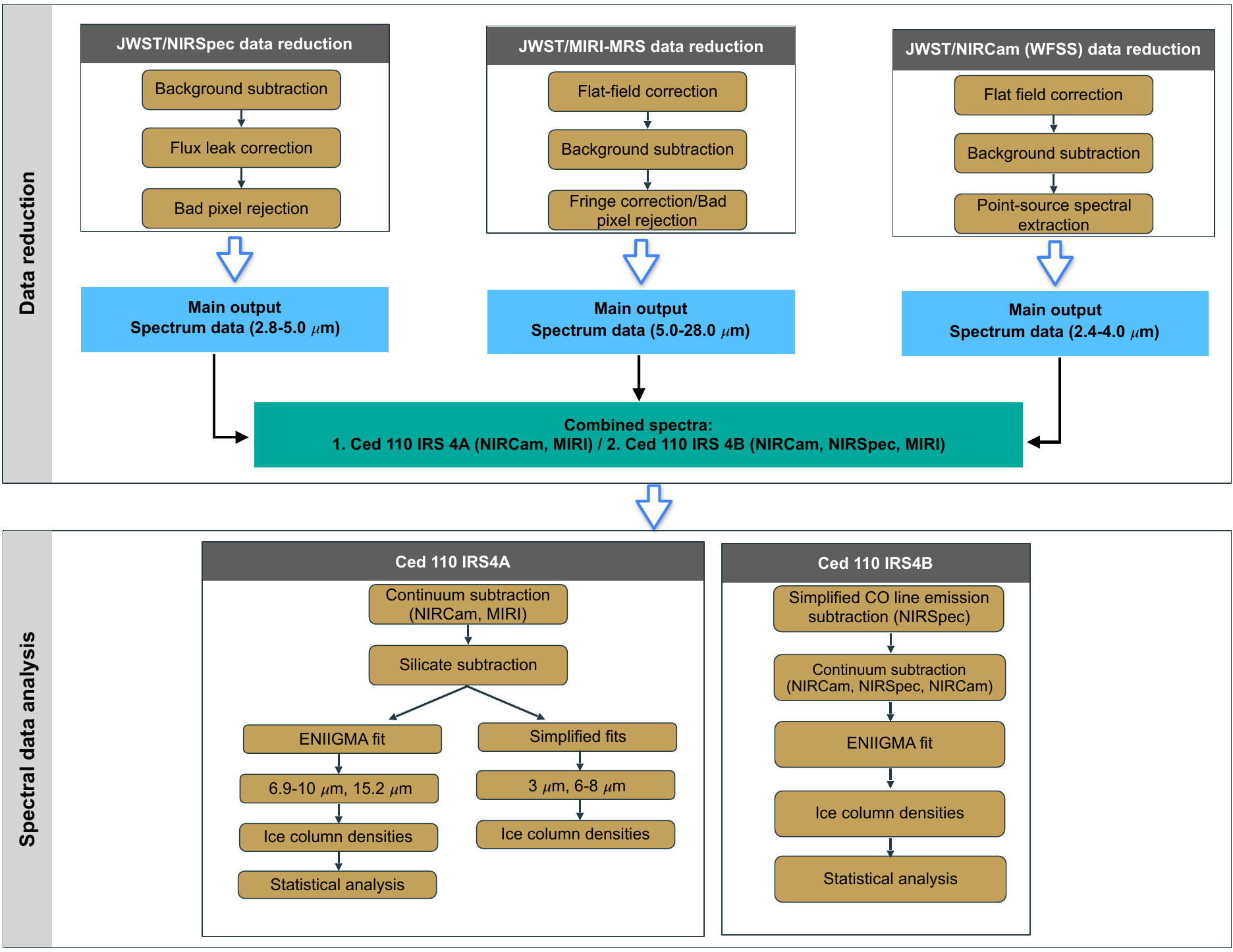}
      \caption{Summary of steps performed in the data reduction and data analysis of Ced~110~IRS4A and IRS4B.}
         \label{flowchart}
   \end{figure*}

In this section we describe the steps performed for data reduction of the JWST data from multiple instruments. In Figure~\ref{flowchart} we present a flowchart with key procedures with specific highlights of the data analysis described in Section~\ref{sec_met}.

\subsection{NIRCam data reduction}

The JWST/NIRCam Wide-Field Slitless Spectroscopy (WFSS) observations of Ced~110 IRS4 were taken over two observing periods on August 11$^{th}$ and August 12$^{th}$ 2022 as part of the Ice Age program \citep{McClure2017}. These two observation sets used the F322W2 filter with the Grism R dispersion direction for the spectral observations and F150W, F410M and F140M for the direct imaging observations, though we only discuss the F150W images in this study. These observations each used a dither setup of four sub-pixel and three primary dither positions with a mosaic pattern of 1 row $\times$ 2 columns, leading to Ced~110 IRS4A and IRS4B being observed within 24 F322W2 spectral images. Ced~110 IRS4A and IRS4B were placed within NIRCam's module B during these observations, resulting in a $\sim$25\% sensitivity decrease compared to module A. This was because Ced~110 IRS4 was not the primary target of the IceAge NIRCam WFSS observations. The full IceAge NIRCam WFSS observing programme, image processing and spectral extraction and reduction are described in \citet{McClure2023}, \citet{Noble2024}, and \citet{Smith2024sub}. The data reduction steps are described again here.

Direct image processing of the F150W images is the first stage of data reduction, where the techniques outlined in \cite{Sun2022} are used. This is performed by combining the data outputs from up to and including stage 2 of the JWST NIRCam direct imaging standard pipeline. Following this, members of the NIRCam calibration team performed additional reduction steps, as described in \cite{Sun2023}, where the F150W images were cleaned, astrometry corrected to Gaia stars in the field of view, cross-matched to each other, and mosaicked together.\\ 

\noindent To obtain the spectral data presented in this study, each individual grism spectroscopic image frame was reduced using the standard JWST calibration pipeline v1.6.2 up to and including stage 2a, using the default CRDS set up with JWST’s operational pipeline, OPS, and no modifications (CRDS context 0953). Flat-field correction was then applied using the imaging flat data obtained with the same filter (F322W2), followed by two-dimensional sky-background subtraction using sigma-clipped median images that were constructed for each individual spectral image in our dataset in which the Ced~110 IRS4 sources were found. 

\begin{figure*}
   \centering
   \includegraphics[width=\hsize]{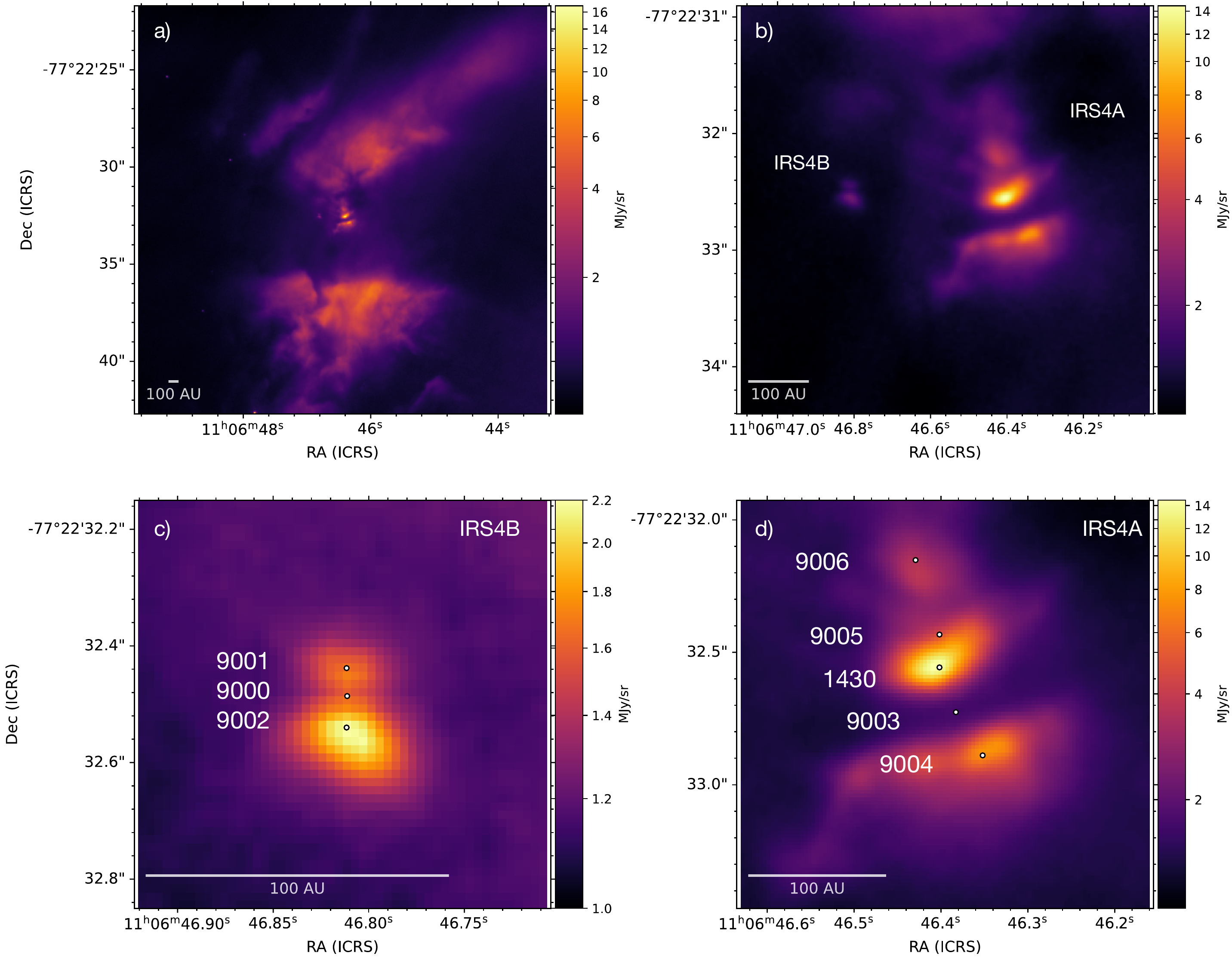}
      \caption{NIRCam image at 1.5~$\mu$m of Ced~110~IRS4A and IRS4B. Panel a displays the large scale image of the Ced~110~IRS4 system and panel b shows a zoom in panel a. Panels c and d highlight the IRS4B and IRS4A source morphologies, respectively. The small white circles show the starting positions of the PSF extractions and the numbers are identifications of the individual spectra shown in Figure~\ref{figpsfnc}.}
         \label{FigNIRCAM}
   \end{figure*}

Due to the strong cloud emission and radiative transfer effects around Ced~110 IRS4 precluded almost any data from around these sources being extracted via the original NIRCam spectral extraction in the IceAge programme. Therefore, we artificially placed sources across the two emission lobes extending vertically, and perpendicular to the disc shadow of IRS4A and 4B, as shown in Figure \ref{FigNIRCAM}, to complement MIRI data for Ced~110IRS 4A and the NIRSpec and MIRI data for Ced~110IRS 4B. We then ran the bespoke NIRCam WFSS extraction code, as described \cite{Noble2024} and \citet{Smith2024sub}, to successfully extract “point source” spectra at these positions. 

\noindent Once the source positions were determined, a custom spectral extraction methodology developed by the IceAge ERS team, described in \cite{Noble2024} and \citet{Smith2024sub}, was used to extract the presented spectra. This is because the standard JWST NIRCam WFSS spectral extraction pipeline is unsuited to extracting crowded and highly varying background fields like that surrounding IRS4. A brief outline is provided here. A spectrum was extracted at each source position from each of the 24 spectral images across wavelengths 2.4-4~$\mu$m.  These extracted spectra were then wavelength and flux calibrated using the in-flight measurements obtained with JWST Commissioning Program \#1076. We then took these individual spectral image spectra and used the \texttt{SpectRes} \citep{Carnall2017} Python package to regrid them all onto a common wavelength grid with 1.006~nm wavelength steps. This allows these spectra to be reliably recombined using the weighted median values of each spectral data point from every individual dispersed frame spectra to produce the final spectrum for each source position in Figure \ref{FigNIRCAM}. The recombination slightly decreases the spectrum length from 1670 to 1638 wavelength elements as expected from pre-flight measurements. This is a minor change that still results in an R value very close to the measured $R\sim 1,600$ at 3.95$~\mu$m and $R\sim 1,150$ at 2.5$~\mu$m from the JWST pre-flight measurements. Each source position's final spectrum was then summed together to produce the final spectrum for Ced~110 IRS4A, where we show the individual source position and summed spectra in Appendix~\ref{psfnirc}.

\subsection{NIRSpec data reduction}
The protostar Ced 110 IRS4 was targeted for observations with the Near InfraRed Spectrograph (NIRSpec) onboard JWST on 2022 July 19th. The requested pointing was defined to be 11$^\mathrm{h}$06$^\mathrm{m}$47.1$^\mathrm{s}$, $-77$\degr22\arcmin34.0\arcsec (J2000), based on the resolved K band imaging in \citet{Persi2001}. However, the Persi et al. coordinates of this source and many of the infrared sources in this field proved to be systematically offset with respect to the Spitzer IRAC coordinates. Because we did not perform a target acquisition, the NIRSpec observation was therefore mispointed by 2.4\arcsec and did not capture the image of Ced~110~IRS4. Serendipitously, the NIRSpec observations did capture a companion protostar (IRS4B) at a separation of 1.3\arcsec.

The spectroscopic data were taken with a standard 4-point dither pattern for a total integration of 1225~s.  The G395H/F290LP disperser/filter combination was used with NIRSpec, which covers wavelengths from 2.87 -- 4.08~$\mu$m and 4.19 -- 5.25~$\mu$m at a resolving power of $R\sim 2700$. Leakcal exposures were acquired which observe the 4-point dither pattern with the IFU aperture closed so that excess flux leaking through the finite contrast MSA shutters was effectively removed.

The data were processed through the November 2022 development version of the JWST NIRSpec pipeline (2022\_2a; 1.8.3.dev26+g24aa9b1d). 
Calibration reference file database version 11.16.16 was used that included the updated flight flat field and throughput calibrations for absolute flux calibration accuracy estimate on the order of $\sim$6-8\%. The standard steps in the JWST pipeline were carried out to process the data from the raw (uncal files) ramp format to the cosmic ray corrected slope image (rate files).  Further processing of the 2D slope image for WCS, flat fielding, flux calibration, etc. was also carried out using standard steps in the ``Level 2'' data pipeline calwebb\_spec2 (cal files). To build the calibrated 2D IFU slice images into the 3D datacube, the ``Stage 3" pipeline was run step-wise and intermediate products were investigated for accuracy.   The four dithers were combined using the "drizzle" algorithm with equal weighting and full pixel regions used.  The final pipeline processed product presented here was built into 3D with the outlier bad pixel rejection step turned off, as running this over-corrected and removed target flux associated with the protostar.

\subsection{MIRI data reduction}
The JWST/MIRI observations of Ced 110 IRS4 were conducted on December 7, 2023. Spectroscopic measurements were acquired using a conventional 4-point dither scheme, totaling an integration time of 1665~s, with the telescope pointing set at 11$^\mathrm{h}$06$^\mathrm{m}$46.45$^\mathrm{s}$, -77\degr22\arcmin32.93\arcsec. An additional observation was made on March 23, 2023, at 11$^\mathrm{h}$04$^\mathrm{m}$24.48$^\mathrm{s}$ -77\degr18\arcmin38.77\arcsec to eliminate the background signal from the target flux. The data were processed using JWST pipeline version 1.11.4 \citep{Bushouse2023_jwstmiri}, incorporating time-dependent corrections for channel 4 throughput. Calibration reference files from version 11.16.21 and jwst\_1119.pmap were utilized, featuring updated onboard flat-field and throughput calibrations to achieve an absolute flux accuracy estimate of approximately 10\%. Standard procedures within the JWST pipeline were executed to transform the data from the 3D ramp format to the cosmic ray corrected slope image. Subtraction of the scientific background occurred post the "Level 1" processing stage. Subsequent processing of the 2D slope image for WCS, flat fielding, and flux calibration was performed using the standard procedures in the "Level 2" data pipeline calwebb\_spec2. The creation of the calibrated 2D IFU slice images into the 3D datacube involved the execution of the "Stage~3" pipeline. The final product generated by the pipeline was a 3D datacube, where the outlier bad pixel rejection step was disabled to prevent excessive correction and removal of target flux.

\begin{figure*}
   \centering
   \includegraphics[width=\hsize]{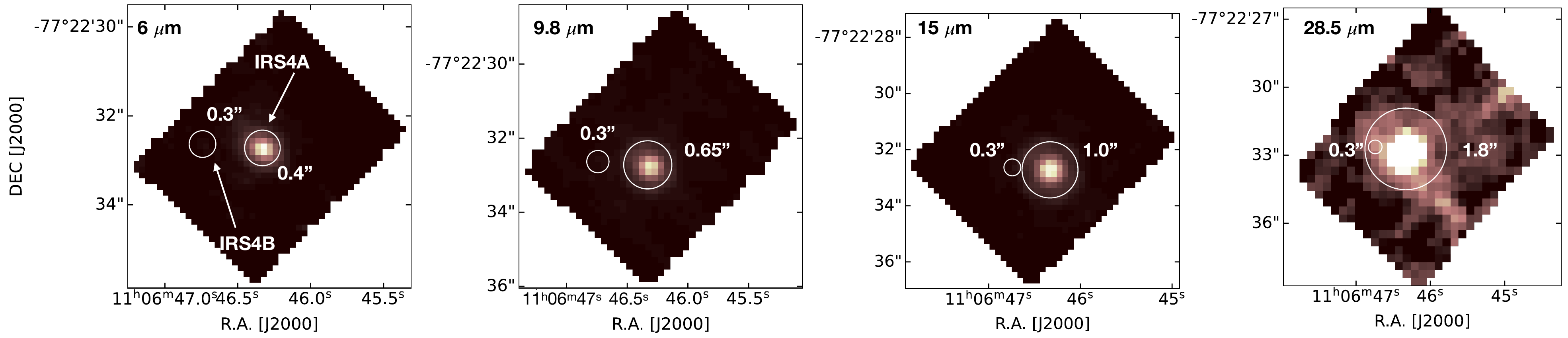}
      \caption{MIRI images from key spectral regions, namely, 6~$\mu$m: H$_2$O  ice, 9.8~$\mu$m: silicate, 15.2~$\mu$m: CO$_2$ ice and 28.5~$\mu$m: continuum. The white circles show the regions used for spectral extraction. For IRS~4A we use an aperture size proportional to the size of the PSF, and for IRS~4B a constant size of  0.3" is adopted.}
         \label{MIRIextract}
   \end{figure*}

Spectral extraction for the primary source (IRS4A) was performed using an aperture size proportional to the PSF distribution (see Figure~\ref{MIRIextract}). The spectrum of the companion source (IRS4B) was extracted by assuming a constant aperture radius of 0.3~arcsec. As shown in Figure~\ref{MIRIextract}, at 6 and 9.8~$\mu$m, the two sources are well separated, whereas the PSF of the two sources are blended longwards of 15~$\mu$m. A comparison between the spectrum from MIRI/MRS and the \textit{Spitzer/}Infrared Spectrograph (IRS) is presented in Appendix~\ref{JWST_Spitzer_sec}.  

\subsection{Combined spectra}
\begin{figure*}
   \centering
   \includegraphics[width=\hsize]{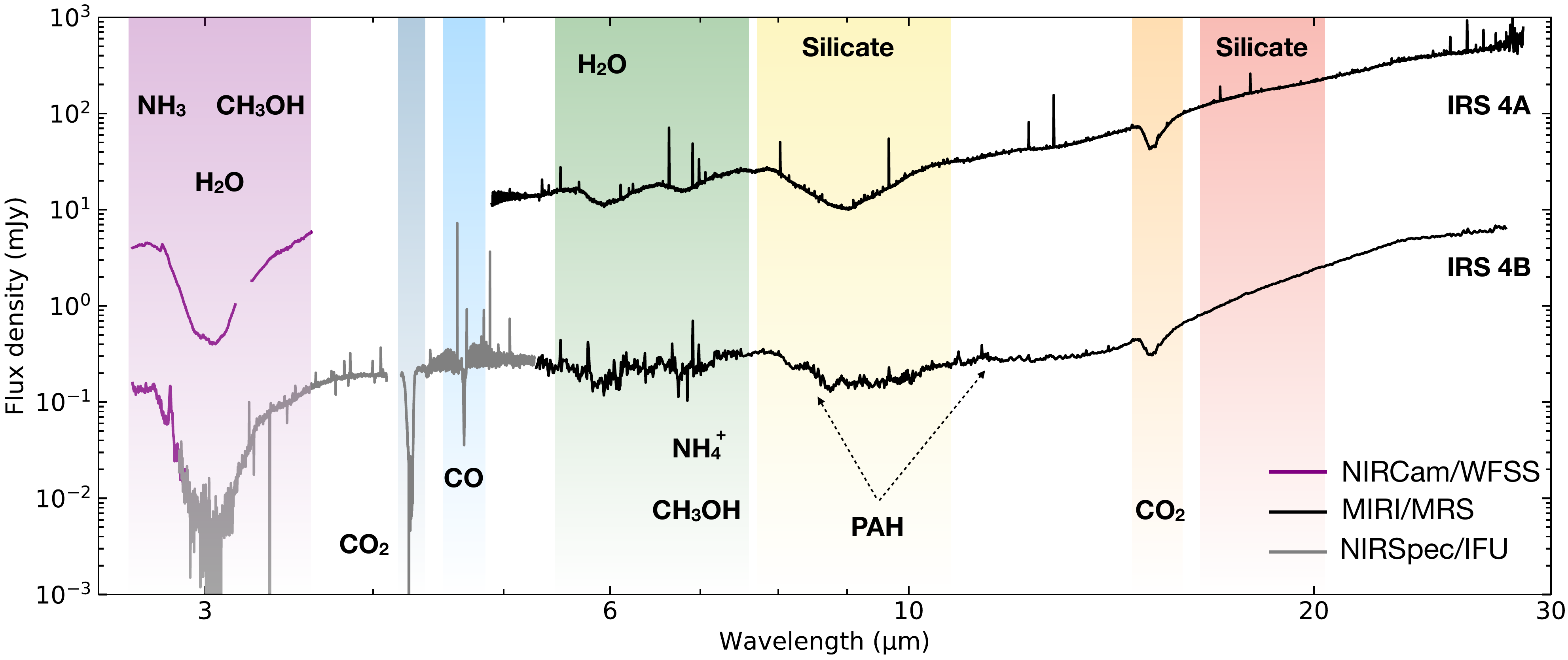}
      \caption{Spectra of Ced~110~IRS4A and IRS4B between 2.5$-$28~$\mu$m. These spectra use data from NIRSpec/IFU (grey), NIRCam/WFSS (purple) and MIRI/IFU (black). Major ice species and PAH features are indicated. The hatched areas highlight different spectral regions. Gas-phase line emissions are not masked or assigned in this figure. No offset is employed.}
         \label{rainbow_spec}
   \end{figure*}

Figure~\ref{rainbow_spec} shows the combined spectra of Ced~110~IRS4A and IRS4B extracted from NIRCam, NIRSpec and MIRI/MRS. Apart from the spectral gaps already mentioned, no data is available between 3.8 and 5~$\mu$m for the primary source. The primary source is brighter than the companion by a factor of 50$-$100 across the spectrum and both sources show clear ice and silicate features which are labelled. The silicate features in both sources also have peculiar shapes. For example, in IRS4A, the peak of the absorption feature is at 9~$\mu$m instead of around 9.6~$\mu$m as commonly seen in protostars \citep[e.g.,][]{Chiar2006}. However, other protostars showing a similar absorption band at 9~$\mu$m have been reported in the literature \citep[e.g.,][]{Furlan2008}. In IRS4B, a flatter profile is visible, which is likely caused by polycyclic aromatic hydrocarbons (PAH) emission where features are also labelled. The silicate absorption at 18~$\mu$m is absent in both sources (See Section~\ref{SEDemp}). In the next sections, we explore the full spectral range, with special focus on the ice features.   

\section{Ice laboratory data}
\label{icelab_sec}

The laboratory data considered in this paper are listed in Appendix~\ref{list_lab}. These data were taken mainly from the Leiden Ice Database for Astrochemistry\footnote{\url{https://icedb.strw.leidenuniv.nl/}} \citep[LIDA;][]{Rocha2022} and from the Goddard NASA database\footnote{\url{https://science.gsfc.nasa.gov/691/cosmicice/spectra.html}}. For this paper, this includes a list of 71 species, corresponding to pure and ice mixtures, and at different temperatures.

We highlight in this section the series of CO:CO$_2$ ice IR spectra because it plays an important role in the ice fits presented in Section~\ref{sec_nirB}. This dataset was characterized by \citet{vanBroekhuizen2005}, where several spectroscopic details are provided. In particular, warmer ices (>50~K) have a prominent blue shoulder at 4.25~$\mu$m, which is caused by ice trapping \citep[e.g.,][]{Sandford1988}. Although it is not mentioned in \citet{vanBroekhuizen2005}, this data has H$_2$O contamination that could also play a role in trapping molecules in the ice. For this reason, we relabelled this ice mixture at 70~K as H$_2$O:CO$_2$ (0.7:1) based on the H$_2$O ice column density and the non-presence of CO ice in the spectrum at that specific temperature. Further details about this spectrum are provided in Appendix~\ref{bin_mix}. We kept this data in our analysis for three reasons: 1) this dataset was characterized in detail by \citet{vanBroekhuizen2005}, 2) the presence of H$_2$O ice supports the trapping effect, 3) the blue shoulder is visible in the H$_2$O:CO$_2$ (1:1) at 70~K from \citet{Ehrenfreund1999}, and therefore it is consistent with the data from \citet{vanBroekhuizen2005}.

\section{Methodology}
\label{sec_met}

This section shows the methods used to perform global continuum subtraction covering both NIRSpec and MIRI ranges, as well as two approaches to subtract the silicate feature. The first approach is physically constrained with a SED empirical model, whereas the second method uses a polynomial function anchored at specific wavelengths. At the end of this section, we briefly describe the \texttt{ENIIGMA} spectral fitting methodology.

\subsection{Continuum determination}
To determine the baseline for the absorption features towards Ced~110~IRS4A and IRS4B, we used a low-order polynomial function as also adopted in previous works \citep[e.g.,][]{Chiar2006, Boogert2008, Rocha2024}. Since the observations do not cover all wavelengths in the near-IR, we used different anchor points to trace the continuum. In addition, for IRS4A, we also considered photometry data from the Two Micron All Sky Survey \citep[2MASS;][]{Skrutskie2006}\footnote{\url{https://irsa.ipac.caltech.edu/Missions/2mass.html}} catalogue to further constrain the continuum determination below 2.8~$\mu$m. Similarly, in the MIRI range, different anchor data points are used to determine the continuum. Specifically, longwards of 15~$\mu$m, where there is PSF overlap of IRS4A with IRS4B, a larger spectral range is considered for IRS4B when tracing the continuum. This choice is made because it is no longer possible to constrain the long wavelength wing of the silicate bending mode at this range. The procedure does not affect the subsequent analysis of the IRS4B source that was limited to the NIRSpec range. Table~\ref{phot} shows the spectral data used to determine the continuum for both Ced~110~IRS4A and IRS4B.

\begin{table*}
\caption{\label{phot} List of wavelengths and flux densities used as anchor points to determine the continuum.}
\centering 
\begin{tabular}{cc}
\hline\hline
$\lambda$ ($\mu$m) & Flux density$^a$ (mJy)\\
\hline
\multicolumn{2}{c}{Ced~110~IRS4A}\\
\hline
J (1.23) & 0.20$\pm$0.08\\
H (1.66) & 1.21$\pm$0.51\\
K (2.16) & 2.46$\pm$0.72\\
NIR (2.68, 2.71, 3.79) & 4.31, 4.49, 8.43\\
MIR (5.45$-$5.57, 27.1$-$27.7) & 15.84$-$16.46, 492.11$-$505.08\\
\hline
\multicolumn{2}{c}{Ced~110~IRS4B}\\
\hline
J (1.23) & ...\\
H (1.66) & ...\\
K (2.16) & ...\\
NIR (2.65-2.75, 4.0$-$4.1, 4.74, 4.95, 5.19) & 0.13$-$0.14, 0.19$-$0.20, 0.23, 0.24, 0.25\\
MIR (5.45$-$5.57, 17.9$-$27.7) & 0.25$-$0.26, 1.36$-$6.26\\
\hline
\end{tabular}
\tablefoot{\footnotesize
$^a$ Flux error is assumed to be 10\% in both NIR and MIR.}
\end{table*}

\begin{figure*}
   \centering
   \includegraphics[width=\hsize]{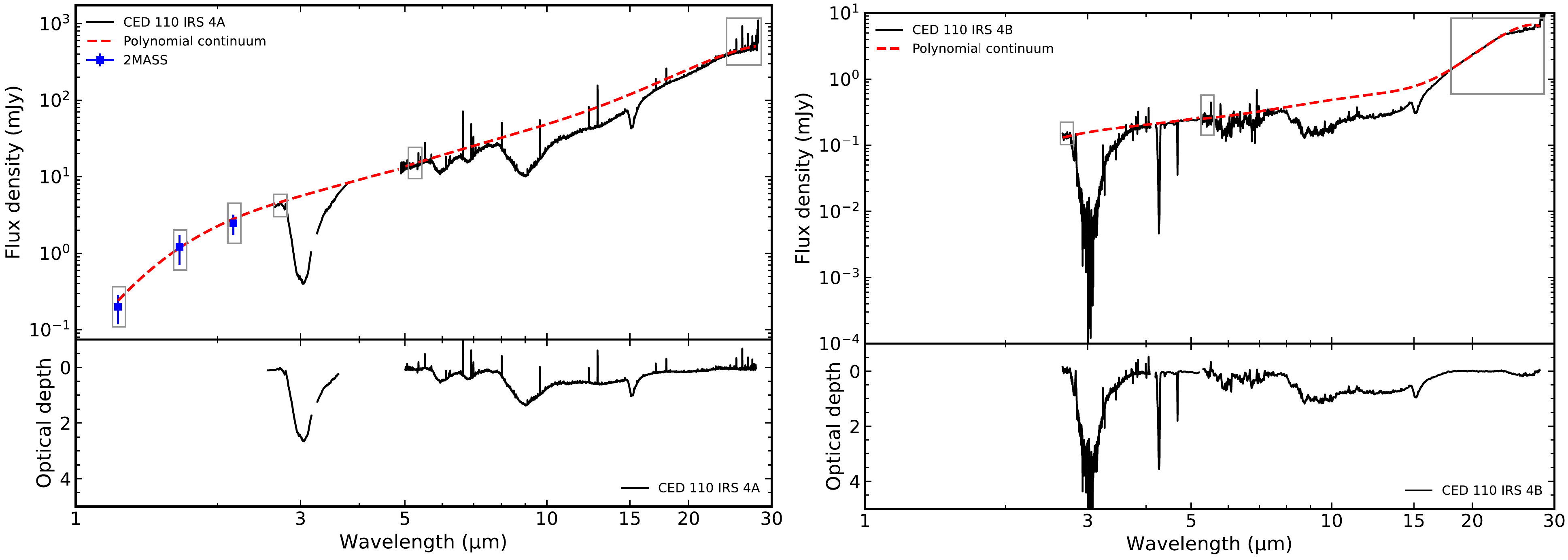}
      \caption{Dust continuum profiles (top panels) and optical depth spectra (bottom panels) of Ced~110~IRS4A (left) and IRS4B (right). For IRS4A source, 2MASS data are used to trace the continuum below 2.5~$\mu$m. The red dashed line shows the continuum profile traced by a polynomial function and the grey boxes indicate the spectral regions used as anchor to guide the fit. The polynomials for IRS4A are 3rd order for NIRSpec and MIRI respectively. Similarly, for IRS4B, they are 2nd and 5th orders, respectively. The strong upward inflexion in the IRS4B source longwards of 15~$\mu$m is due to the PSF overlap in the MIRI range with the IRS4A source (see Figure~\ref{MIRIextract}).}
         \label{polycontinuum}
   \end{figure*}

Once the global continuum is traced for each source, the spectrum is converted to an optical depth scale using the following equation:
\begin{equation}
    \tau_{\lambda} = -\mathrm{ln} \left( \frac{F_{\lambda}^{\rm{source}}}{F_{\lambda}^{\mathrm{cont}}} \right),
    \label{tau_obs}
\end{equation}
where $F_{\lambda}^{\rm{source}}$ is the observed spectrum and $F_{\lambda}^{\mathrm{cont}}$ is the global continuum of the source.

Figure~\ref{polycontinuum} displays the continuum traced for Ced~110~IRS4A and IRS4B and the derived optical depth. We also highlight, that in the case of IRS4B, we first removed the CO emission lines between 4.3 and 5.0~$\mu$m before tracing the continuum. This procedure is presented in Appendix~\ref{corovib}. It is worth mentioning that PAH emission features are not seem towards the IRS4A source, whereas they are visible at 6.2, 8.6 and 11.2~$\mu$m towards the IRS4B protostar. For that source, no subtraction of the PAH features are performed because they would require a more detailed radiative transfer modelling that simulates the excitation of those PAHs. This is beyond the scope of this work. Since we do not perform the analysis of the absoption features in the MIRI range of IRS4B because of the low S/N, the non subtraction of the PAH emission features does not compromise our analysis and results presented for that source.

\subsection{Silicate subtraction}
\label{silic_sub}
\begin{figure}
   \centering
   \includegraphics[width=8cm]{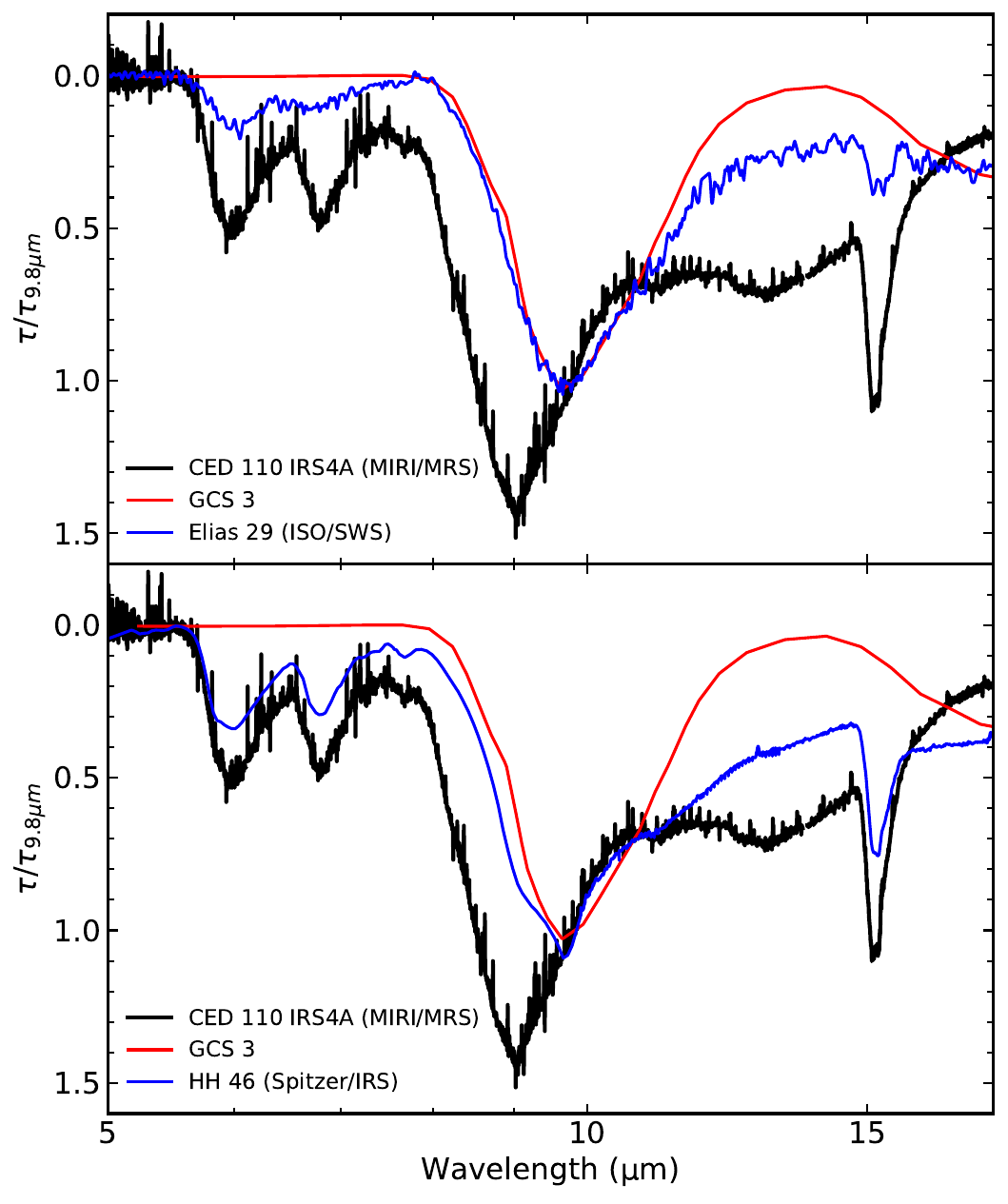}
      \caption{Silicate profiles of GCS3 \citep{Kemper2004}, Elias~29 \citep{Boogert2000} and HH~46 \citep{NoriegaCrespo2004} compared to the MIRI/MRS spectrum of IRS4A. All spectra are normalized by the peak optical depth at 9.8~$\mu$m.}
         \label{silicmatch}
\end{figure}

In order to analyze the ice bands, the strong silicate features at 10 and 20~$\mu$m must be removed. As demonstrated in the literature \citep[e.g.,][]{Boogert2008, Bottinelli2010, Poteet2013}, the standard approach is to subtract a scaled ice-free silicate profile from the Wolf-Rayet source GCS3 \citep{Kemper2004}. We first tried this methodology. However, Figure~\ref{silicmatch} shows an atypical silicate profile towards IRS4A compared with other protostars. While the 9.8~$\mu$m band of the low-mass protostars Elias~29 \citep{Boogert2000} and HH~46 \citep{NoriegaCrespo2004} have a profile similar to GCS3, the spectrum of IRS4A is rather different given the strong absorption excess at the short-wavelength side of the observational feature around 9.8~$\mu$m. The small excess seen on the short-wavelength side of HH46 can be attributed to the composition of the silicate and ices, and in fact, was already discussed in the literature \citep[e.g.,][]{Boogert2013, McClure2023, Rocha2024}. The silicate profile towards IRS4A seems rather strong to be attributed to only the composition of the dust and might also be affected by radiative transfer effects. Such an atypical profile was also observed previously in the literature \citep[e.g.,][]{Furlan2008}, but no explanation has been provided. 

The complex environment of the Ced~110~IRS4 system, suggests the presence of a disk surrounded by an envelope and may be the cause of the unusual silicate profile \citep{Pontoppidan2005circ}. Addressing this source geometry in detail or the dust mineralogy for the Ced~110~IRS4 system is beyond the scope of this paper. Therefore we used two alternative methodologies to perform the silicate subtraction for IRS4A, as described in Sections~\ref{SEDemp} and \ref{sec_Polysilic}. No silicate analysis or subtraction is performed for the companion source, IRS4B. First, this data has a low S/N which hinders a profile analysis of the band. Second, this source has clear PAH emission features, which causes the flattened shape around 9.8~$\mu$m. A more refined model is needed for the silicate analysis of IRS4B.

\subsubsection{SED empirical model}
\label{SEDemp}
Given the nature of the unusual shape of the silicate profile towards IRS4A, we used a simplified model to perform continuum and silicate subtractions simultaneously as presented below:
\begin{equation}
    F_{\lambda}^{\rm{source}} = F_{\lambda}^{\rm{BB}}\rm{exp} \left(-\sum_i\tau_{i,\lambda}\right) + \sum_j \left\{ \frac{B_{\lambda}(T)}{\pi}[1 - exp(-\tau_{j,\lambda})]\right\},
    \label{RT_eq}
\end{equation}
where $F_{\lambda}^{\rm{source}}$ is the received flux, $F_{\lambda}^{\rm{BB}}$ is the flux emitted from the source, which is simplified by a blackbody function, $\tau$ is the dust grain optical depth and $B_{\nu}(T)$ is the optically thick emission from the dust. The optical depth is calculated from the opacity data ($\kappa$), and is given by $\tau_{\lambda} = N \cdot \kappa_{\lambda}$, where N is the mass column density in units of g cm$^{-2}$. The first part of Equation~\ref{RT_eq} accounts for grain (dust and ice) absorption, whereas the second part accounts for dust emission (warm, cold). The indexes $i$ and $j$ indicate the kind of solid particle in the model. If the same material is in emission and absorption, then $i = j$.

To model the silicates in this approach, we calculated opacity models ($\kappa$) by assuming a power-law size distribution ($n(a) \propto a^{-3.5}$) of hollow spheres (DHS) with sizes ranging from 0.1 to 1.0~$\mu$m. The effects of the grain-ice porosity and scattering are not considered in this model, which would be better reproduced with a full radiative transfer modelling. The literature focused on disk mineralogy generally reports that silicates with olivine stoichiometry dominate the 9.8~$\mu$m emission band \citep[e.g.,][]{Olofsson2009, Sargent2009, Juhasz2010, Fogerty2016, Grimble2024}, whereas, silicates with pyroxene stoichiometry dominates the 9.4~$\mu$m absorption band towards protostars \citep{Chiar2006, Rocha2015, DoDuy2020}. Although one type of silicate dominates the band profile, it is important to mention that often a combination of both kinds is needed in the fits (See \citet{Henning2013} for a review). The two dust components used in this paper are pyroxene (Mg$_{0.7}$Fe$_{0.3}$SiO$_{3}$) and olivine (MgFeSiO$_{4}$), and their optical constants were taken from \citet{Dorschner1995}. In addition, optical constants taken from \citet{Zubko1996} are used to include amorphous carbon into the dust composition, following the standard mass ratio of 0.87/0.13 for the silicate and carbon, respectively \citep[e.g.,][]{Woitke2016}. Similarly, for the H$_2$O ice population, we use the same size limits as for the dust. We adopt continuous distribution of ellipsoids (CDE) for the grain-shape correction, and the optical constants were taken from \citet{Rocha2024_ice}. 

\begin{figure}
   \centering
   \includegraphics[width=\hsize]{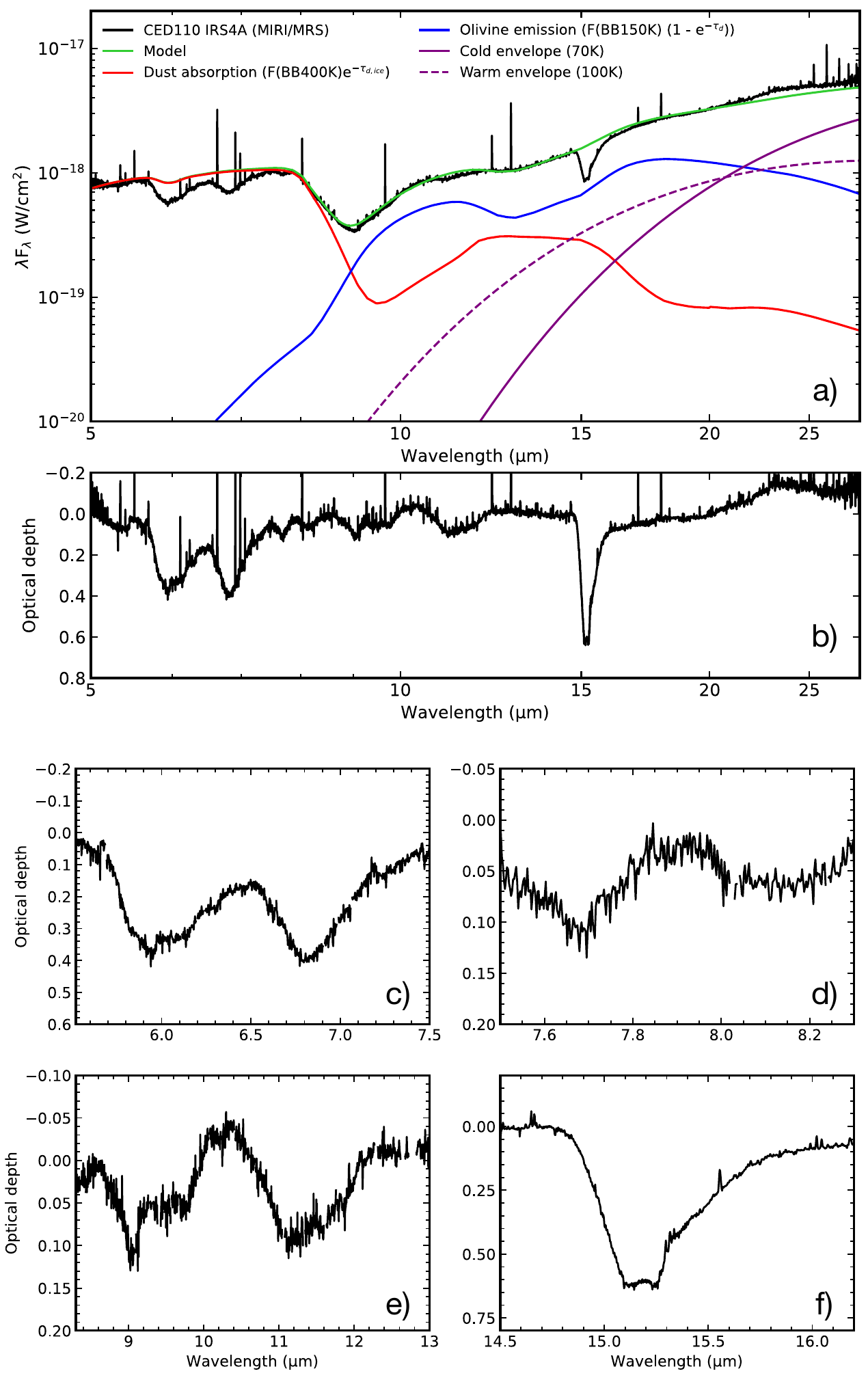}
      \caption{Panel a shows the SED fit using Method 2 shown by the green line. The absorption component (red line) considers pyroxene and H$_2$O ice absorption. The emission component (blue line) considers olivine thermal emission, as well as colder dust emission (purple lines) at longer wavelengths. Panel b shows the optical depth spectrum after subtracting the green profile. Panels c$-$f show different absorption features that are investigated in this work.}
         \label{sedmod}
   \end{figure}

In our simplified model, we consider two populations of dust. Given the inclination of IRS4A \citep[70$^{\circ}$;][]{Manoj2011}, we assume that pyroxene dust dominates the absorption and comes mostly from the envelope. In contrast, olivine originates in the disk and dominates the emission profile. The model using Equation~\ref{RT_eq} is presented in Figure~\ref{sedmod}a. The values of the mass column densities of pyroxene and olivine are 5.5$\times$10$^{-4}$~g~cm$^{-2}$ and 4.3$\times$10$^{-4}$~g~cm$^{-2}$, respectively, which are of the same order of magnitude as found for protostar HOPS$-$68 \citep{Poteet2011}. A blackbody profile with a temperature of 400~K is used to model the pyroxene and H$_2$O ice absorption in the MIRI range, and the olivine emission is modelled with a black-body with a temperature of 150~K. The blackbody at 400~K is mostly caused by the extinct blackbody from the central source \citep[see][for individual radiative transfer components]{Boogert2002}. The radiation field associated with the extinct blackbody interacts with the dust and icy material located in regions with temperatures above and below 150~K, respectively, where 150~K is the sublimation temperature of H$_2$O ice \citep{Collings2004}. In the case of the dust thermal emission, radiative transfer models show that disks embedded in protostellar envelopes have temperatures around 100$-$200K between 100$-$400~au \citep[e.g.,][]{Nazari2023}. In our simplified model, it is assumed that such olivine emission comes from those regions. Two other blackbody emission representing optically thick dust at temperatures of 70 and 100~K are needed to fit the spectrum longwards of 20~$\mu$m. This level of theory is enough to subtract the silicate profile and the dust continuum towards Ced~110~IRS4A. Naturally, more sophisticated models aiming at reproducing both spectrum and image, as well as the silicate emission and absorption interplay are required to fully unveil the dust nature of in that source.

After subtracting the continuum profile, clear absorption features due to ices are revealed, which are presented in Figure~\ref{sedmod}b, and zoom-ins are shown in panels ~\ref{sedmod}c$-$f. We note that a larger fraction of H$_2$O ice at 6~$\mu$m was already subtracted with the continuum based on the 3~$\mu$m, where more details are presented in Section~\ref{fullrange}. In that case, other contributions must be assumed in order to explain this absorption feature.

\subsubsection{Polynomial subtraction}
\label{sec_Polysilic}
\begin{figure}[h!]
   \centering
   \includegraphics[width=\hsize]{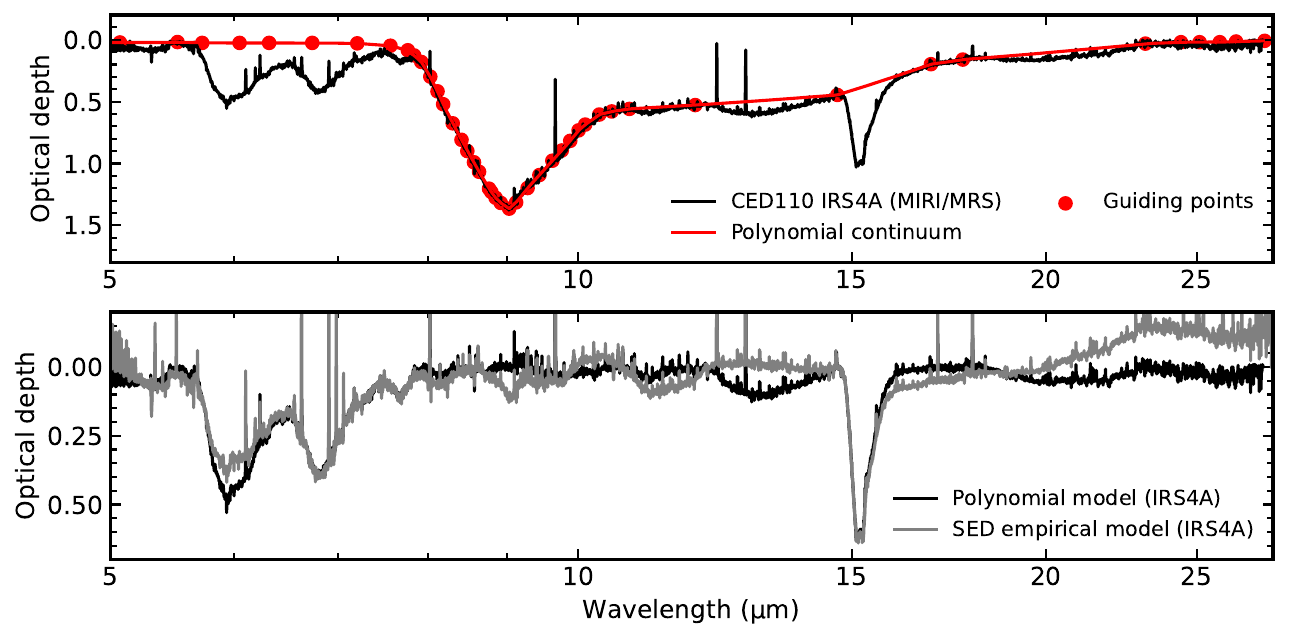}
      \caption{Silicate profile removed using polynomial functions anchored at specific guiding points. The top panel shows the fitted profile and the bottom panel compares the optical depth spectrum obtained from the polynomial and SED empirical models.}
         \label{polysilic}
   \end{figure}

Because of the uncertainty in the nature of the silicate profile towards IRS4A, we also perform a silicate subtraction using a guided continuum (top panel of Figure~\ref{polysilic}). The criteria for selecting the position of these points was that they should not subtract clear absorption features other than the silicate itself. For instance, in the region between 5$-$7.2~$\mu$m the points are positioned at $\tau = 0$. From 7.5~$\mu$m onwards, the points follow the shape of the broadband at 9~$\mu$m. Longwards of 11~$\mu$m, the points trace the baseline for some strong absorption features, specifically, at 11.2, 13.0, 15.2 and 20~$\mu$m. The downside of this method is that the total subtraction of the H$_2$O ice libration band at 12~$\mu$m, whereas the contribution of the H$_2$O bending mode around 6~$\mu$m is not removed. Another limitation is that the 9~$\mu$m band is fully removed since the physical motivation of this approach is that this entire band profile is caused by radiative transfer effects instead of chemistry.
We highlight that this is not a problem for the analysis of this paper since we only perform ice analysis for specific wavelength ranges.  

In the bottom panel of Figure~\ref{polysilic} we compare the optical depth spectra of IRS4A after subtracting the silicate using the polynomial and the SED empirical method. Both spectra show rather similar features between 6.4 and 7.9~$\mu$m, and for most of the CO$_2$ bending mode at 15.2~$\mu$m. On the other hand, some differences are seen at 6, 8.7$-$9.3 and 14~$\mu$m, which are discussed in this paper.

\subsection{Ice fitting approach}
We perform a global fit of the primary and companion source spectra covering specific spectral ranges using the \texttt{ENIIGMA} fitting tool \citep{Rocha2021}. This method has the advantage of accounting for multiple molecular absorption features. \texttt{ENIIGMA} combines laboratory spectra using a linear combination and adopts genetic modelling algorithms to search for the best fit. Genetic algorithms are robust optimization techniques based on the processes of biological natural selection that aim at finding the global minimum solution for complex problems \citep{Holland1975, Koza1992}. The code searches for the global minimum solution that fits the observations using the reduced chi-square function, calculated with the equation below:
\begin{equation}
\chi^2 = \frac{1}{dof}\sum_{i=0}^{n-1} \left(\frac{\tau_{\nu,i}^{\rm{obs}}  - \sum_{j=0}^{m-1} w_j \tau_{\nu,j}^{\rm{lab}}}{\gamma_{\nu,i}^{\rm{obs}}} \right)^2
\label{deltachi}
\end{equation}
where $dof$ is the number of degrees of freedom, $\gamma$ is the error in the observational optical depth spectrum propagated from the flux error ($\sim$10\%), $n$ is the number of data points, $\nu$ is the wavenumber, $\mathrm{\tau_{\nu,i}^{obs}}$ is the optical depth of the observational spectrum defined in Equation~\ref{tau_obs}, $\mathrm{\tau_{\nu,j}^{lab}}$ is the optical depth of the laboratory data, and $w_j$ is the scaling factor. The absorbance laboratory data ($Abs$) is converted to an optical depth scale by equation $\tau_{\nu}^{lab} = 2.3 Abs$.

The ice column density from the best fit is calculated with the following equation:
\begin{equation}
    N_{ice} = \frac{1}{A} \int_{\nu_1}^{\nu_2} w \cdot \tau_{\nu}^{lab} d\nu,
    \label{eq_cd}
\end{equation}
where $A$ is the vibrational mode band strength of the molecule, which is listed in Table~\ref{ice_bs}.

\begin{table*}
\caption{\label{ice_bs} List of vibrational transitions and band strengths of molecules presented in this work.}
\centering 
\begin{tabular}{ccclccccc}
\hline\hline
Structure & Chemical formula & Name & $\lambda \; [\mu \mathrm{m}]$ & $\nu \; \mathrm{[cm^{-1}]}$ & Identification & $\mathcal{A} \; \mathrm{[cm \; molec^{-1}]}$ & References$^a$\\
\hline
\raisebox{-.5\height}{\includegraphics[height=0.25in]{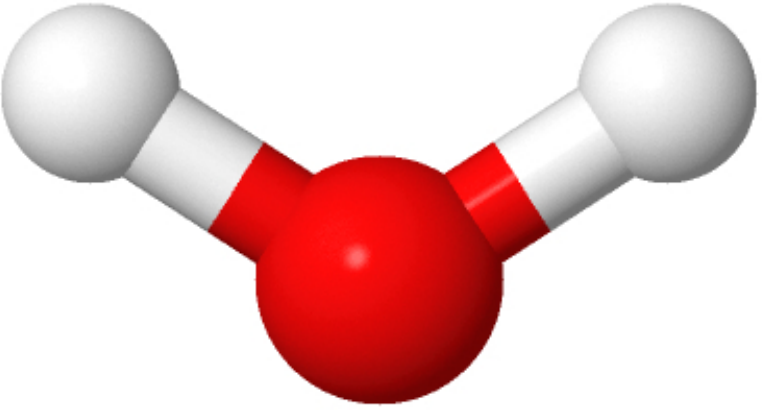}} & H$_2$O & Water & 3.01    & 3322 & OH str. & $\mathrm{2.2 \times 10^{-16}}$ & [1]\\
\raisebox{-.5\height}{\includegraphics[height=0.25in]{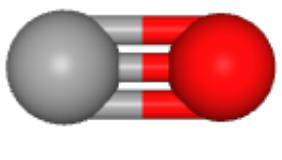}} & $^{12}$CO & Carbon monoxide & 4.67  & 2141 & CO str. & $\mathrm{1.4 \times 10^{-17}}$ & [1]\\
\raisebox{-.5\height}{\includegraphics[height=0.25in]{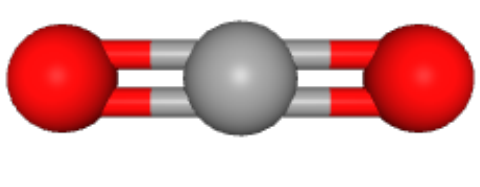}} & $^{12}$CO$_2$ & Carbon dioxide & 4.27    & 2141 & CO str. & $\mathrm{1.1 \times 10^{-16}}$ & [1]\\
\raisebox{-.5\height}{\includegraphics[height=0.25in]{Figures/CO2.pdf}} & $^{13}$CO$_2$ & Carbon dioxide & 4.38    & 2283 & CO str. & $\mathrm{1.1 \times 10^{-16}}$ & [1]\\
\raisebox{-.5\height}{\includegraphics[height=0.32in]{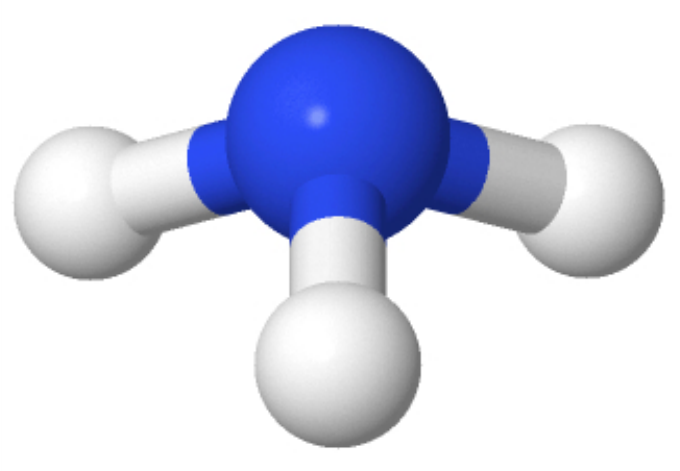}} & NH$_3$ & Ammonia & 9.0  & 1111 & NH str. & $\mathrm{2.1 \times 10^{-17}}$ & [1]\\
\raisebox{-.5\height}{\includegraphics[height=0.32in]{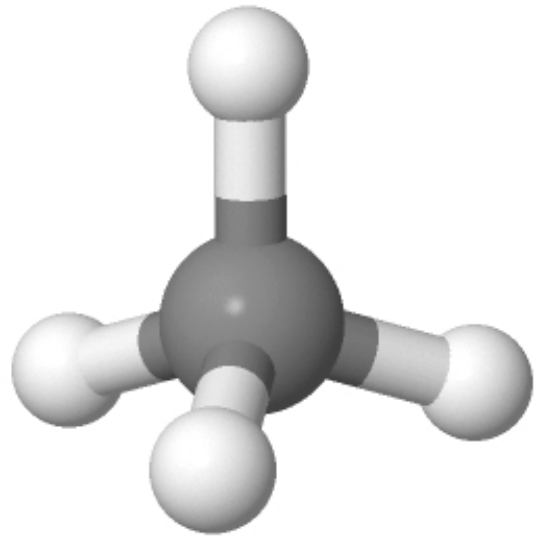}} & CH$_4$ & Methane & 3.32  & 3007 & CH str. & $\mathrm{1.27 \times 10^{-17}}$ & [1]\\
\raisebox{-.5\height}{\includegraphics[height=0.32in]{Figures/CH4v0.pdf}} & CH$_4$ & Methane & 7.67  & 1303 & CH$_4$ def. & $\mathrm{8.4 \times 10^{-18}}$ & [1]\\
\raisebox{-.5\height}{\includegraphics[height=0.3in]{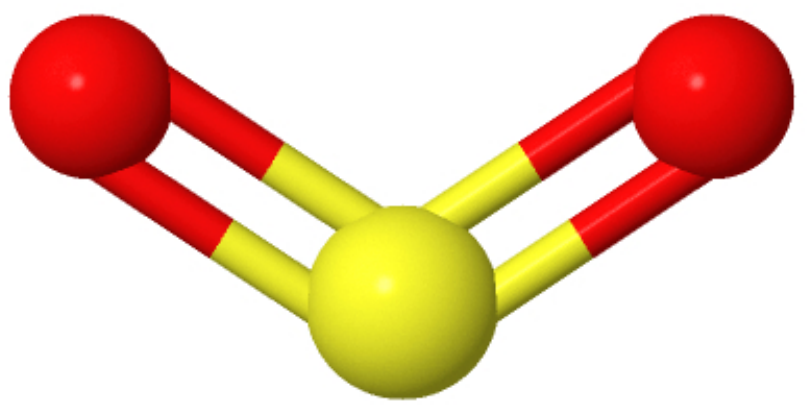}} & SO$_2$   & Sulfur dioxide    & 7.60    & 1320 & SO$_2$ str.     & $\mathrm{3.4 \times 10^{-17}}$ & [2]\\
\raisebox{-.5\height}{\includegraphics[height=0.18in]{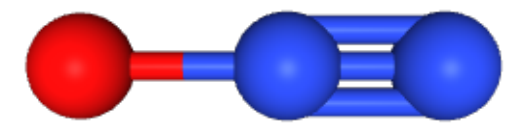}} & N$_2$O   & Nitrous oxide  & 4.47    & 2337 & NO str.    & $\mathrm{1.22 \times 10^{-17}}$ & [3]\\
\raisebox{-.5\height}{\includegraphics[height=0.23in]{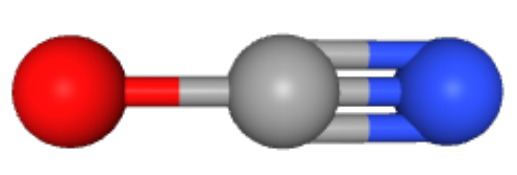}} & OCN$^-$ & Cyanate ion   & 4.59    & 2166 & CN str.  & $\mathrm{1.3 \times 10^{-16}}$ & [4]\\
\raisebox{-.5\height}{\includegraphics[height=0.23in]{Figures/OCN.pdf}} & OCN$^-$ & Cyanate ion   & 7.62  & 1312 & Comb. (2$\nu_2$)  & $\mathrm{7.45 \times 10^{-18}}$ & [5]\\
\raisebox{-.5\height}{\includegraphics[height=0.5in]{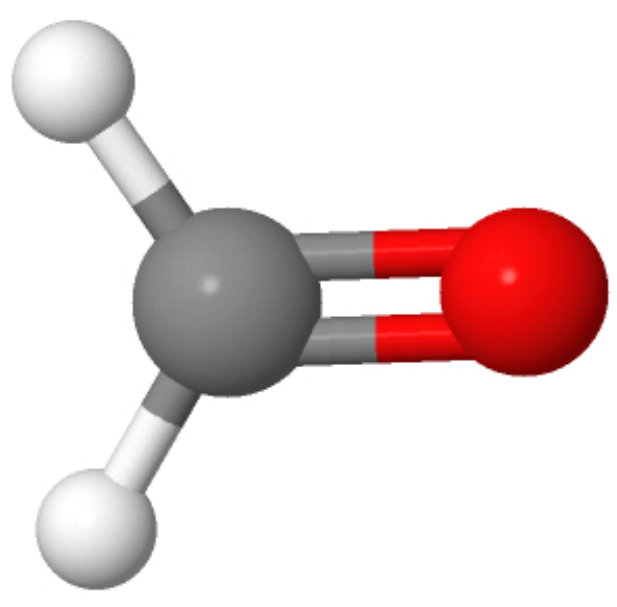}} & H$_2$CO   & Formaldehyde    & 8.04    & 1244 & CH$_2$ rock     & $\mathrm{1.0 \times 10^{-18}}$ & [6]\\
\raisebox{-.5\height}{\includegraphics[height=0.6in]{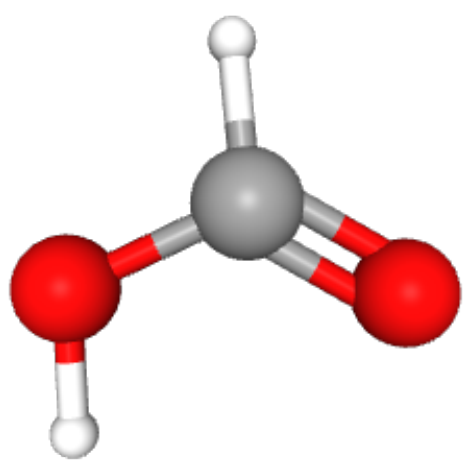}} & HCOOH   & Formic acid     & 8.22    & 1216 & C$-$O stretch     & $\mathrm{2.9 \times 10^{-17}}$ & [1]\\
\raisebox{-.5\height}{\includegraphics[height=0.32in]{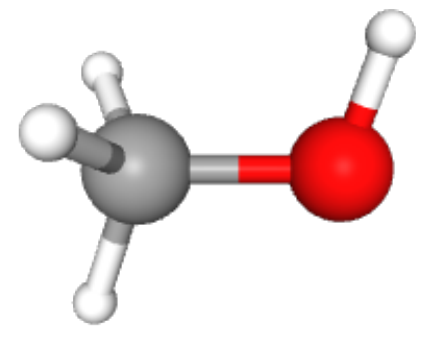}} & CH$_3$OH & Methanol & 3.2$-$3.8  & 2631$-$3125 & CH str. & $\mathrm{1.3 \times 10^{-16}}$ & [1]\\
\raisebox{-.5\height}{\includegraphics[height=0.32in]{Figures/CH3OH.pdf}} & CH$_3$OH & Methanol & 9.74  & 1026 & CO str. & $\mathrm{1.8 \times 10^{-17}}$ & [1]\\
\raisebox{-.5\height}{\includegraphics[height=0.5in]{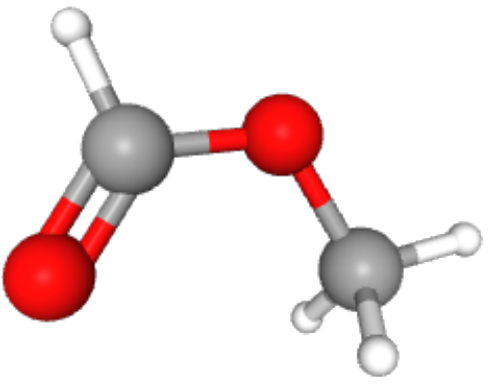}} & CH$_3$OCHO & Methyl formate  & 8.25    & 1211 & C$-$O str.   & $\mathrm{2.52 \times 10^{-17}}$ & [7]\\
\hline
\end{tabular}
\tablefoot{\footnotesize
[1] \citet{Bouilloud2015}, [2] \citet{Boogert1997}, [3] \citet{Fulvio2009}, [4] \citet{vanBroekhuizen2005}, [5] \citet{Rocha2024}, [6] \citet{Schutte1993}, [7] \citet{Scheltinga2021}. $^a$References to \citet{Bouilloud2015} refer to the corrected band strength value considering a different ice density mentioned by the authors.}
\end{table*}

\texttt{ENIIGMA} searches for the best fit in a vast solution space. Because of the degeneracy involved in the fit, \texttt{ENIIGMA} also performs a confidence interval analysis of the linear combination coefficients using a reduced $\chi^2$ map. Based on the confidence intervals, \texttt{ENIIGMA} calculates the lower and upper ice column densities inside a 3$\sigma$ confidence interval. In total the \texttt{ENIIGMA} code evaluates around 4000 solutions based on this dataset. This represents a huge number of solutions, but it is important to mention that some mixtures may not be included in the current dataset, such as specific variations in the mixture fraction.

\section{Results}
\label{sec_res}

In this section, we show the ice fits and results for Ced~110~IRS4A and IRS4B. Finally, we derive the ice column densities of the species securely or tentatively identified.

\subsection{Ice analysis: Ced~110~IRS4A}
We performed a more detailed analysis of the MIRI spectrum of Ced~110~IRS4A, which has a high S/N. Due to the lack of data between 3.6 and 5.0~$\mu$m we cannot measure either CO$_2$ (4.27~$\mu$m) or CO ice (4.67~$\mu$m). Therefore the NIRCam spectrum (2.6$-$3.6~$\mu$m) is solely used to estimate the H$_2$O ice column density, which is later used to calculate the ice abundances. In this section, we analyzed the optical depth spectra between 7 and 10~$\mu$m calculated from the SED and polynomial silicate subtraction methods, as presented in Section~\ref{sec7_10}. Similarities and differences from the two data are compared. In Section~\ref{sec_6}, we also analyzed the 6$-$8~$\mu$m optical depth data from both subtraction methods. In the case of SED subtraction, we performed analysis using the C1$-$C5 empirical components \citep{Boogert2008} because this method requires subtraction of H$_2$O ice first. For the polynomial subtraction method, where H$_2$O ice is not removed, this range is used to cross check the intensity of the features derived from the fit between 7$-$10~$\mu$m. The chemical nature causing the 6$-$8~$\mu$m range is highly unknown, and a dedicated analysis of this part of the spectrum will be deferred to future work. Other absorption features observed in the MIRI range, such as 11.2~$\mu$m and 15.2~$\mu$m are also analyzed. Since this paper does not target gas-phase species, we only mention briefly the 7.18~$\mu$m feature in Appendix~\ref{CH3ion}.

\subsubsection{H$_2$O ice: 2.5$-$28~$\mu$m}
\label{fullrange}

The NIRCam and MIRI spectra of Ced~110~IRS4A (Fig.~\ref{rainbow_spec}) show clear H$_2$O ice absorption features at 3 (O$-$H stretch), 6 (bending) and 12~$\mu$m (libration). Among these features, we use the 3~$\mu$m band to estimate the H$_2$O ice column density. The 6~$\mu$m band has strong overlap with other species, and the 12~$\mu$m is affected by the atypical silicate profile coming from the complex circumstellar environment of Ced~110~IRS4A \citep[e.g.,][]{Pontoppidan2005circ}. However, the 3~$\mu$m band is the sum of PSF spectra at selected regions from NIRCam, and for this reason, we derive the H$_2$O ice column density by scaling laboratory data to the observations. The analysis of the individual PSF spectra will be performed in a future work.

\begin{figure}
   \centering
   \includegraphics[width=\hsize]{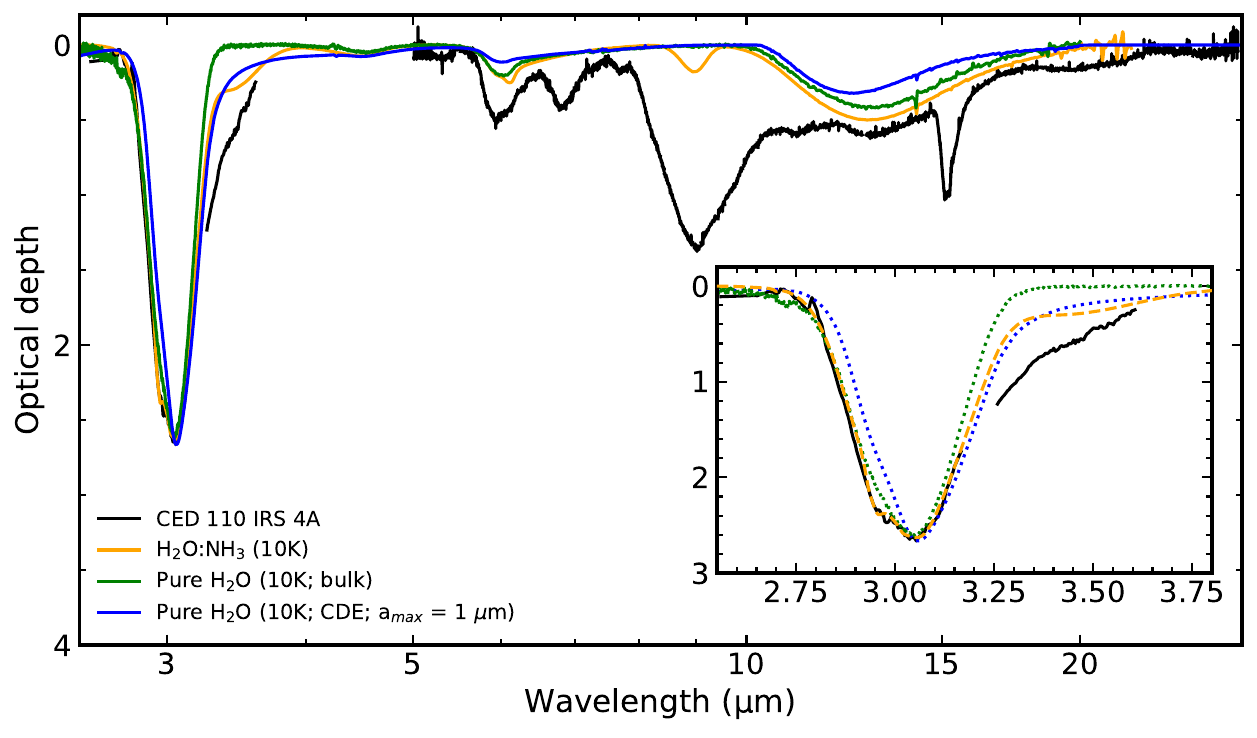}
      \caption{H$_2$O ice IR data compared with the optical depth spectrum of Ced~110~IRS4A prior silicate subtraction. The green and orange lines represent the pure H$_2$O bulk ice and H$_2$O:NH$_3$ spectra. Blue shows the CDE-corrected ice spectra of pure H$_2$O ice. The inset panel shows a zoom-in of the 3~$\mu$m band.}
         \label{H2Oice_scale}
\end{figure}

The IR spectra of bulk and grain-shape corrected ice spectrum of pure H$_2$O ice are scaled to the JWST data as shown in Figure~\ref{H2Oice_scale}. Both data at 10~K account for most of the 3~$\mu$m band, but some differences can be mentioned. For example, the grain-shape corrected spectrum using the CDE approach with $a_{max}$ = 1~$\mu$m, does not explain the blue wing excess of the 3~$\mu$m feature, whereas it accounts for some of the red wing excess compared to the bulk ice spectrum. More differences between pure ice bulk and geometric-corrected spectrum are notable at 6 and 12~$\mu$m, where the large grain spectrum has flatter absorption features, and the 12~$\mu$m band is blue-shifted. A third laboratory spectrum, H$_2$O:NH$_3$ (10:1.6) is used in this analysis motivated by the best fits towards the two background stars in \citet{McClure2023}. This spectrum shows an excellent match with the 3~$\mu$m band, and the small inflection around 2.9~$\mu$m, related to NH$_3$ ice. Using this ice mixture shows other NH$_3$ related ice features are seen at 6.2 and 9.0~$\mu$m. An absorption excess at 3.4~$\mu$m due to the formation of ammonia hydrates \citep{Dartois2001} is seen in this laboratory data, which is not present in the pure H$_2$O ice bulk spectra. Interestingly, the intensity of this excess is comparable to the profile from large-grain spectra and highlights that both physical and chemical phenomena play a role in creating this absorption profile.

Finally, we note that the OH dangling feature for these two protostars are not addressed in this paper since it will require first deriving accurate flux error from the NIRCam observations. The OH dangling band was observed towards several lines of sight in the Chameleon I molecular cloud as demonstrated by \citet{Noble2024}.

\subsubsection{7$-$10~$\mu$m}
\label{sec7_10}

\begin{figure*}
   \centering
   \includegraphics[width=15cm]{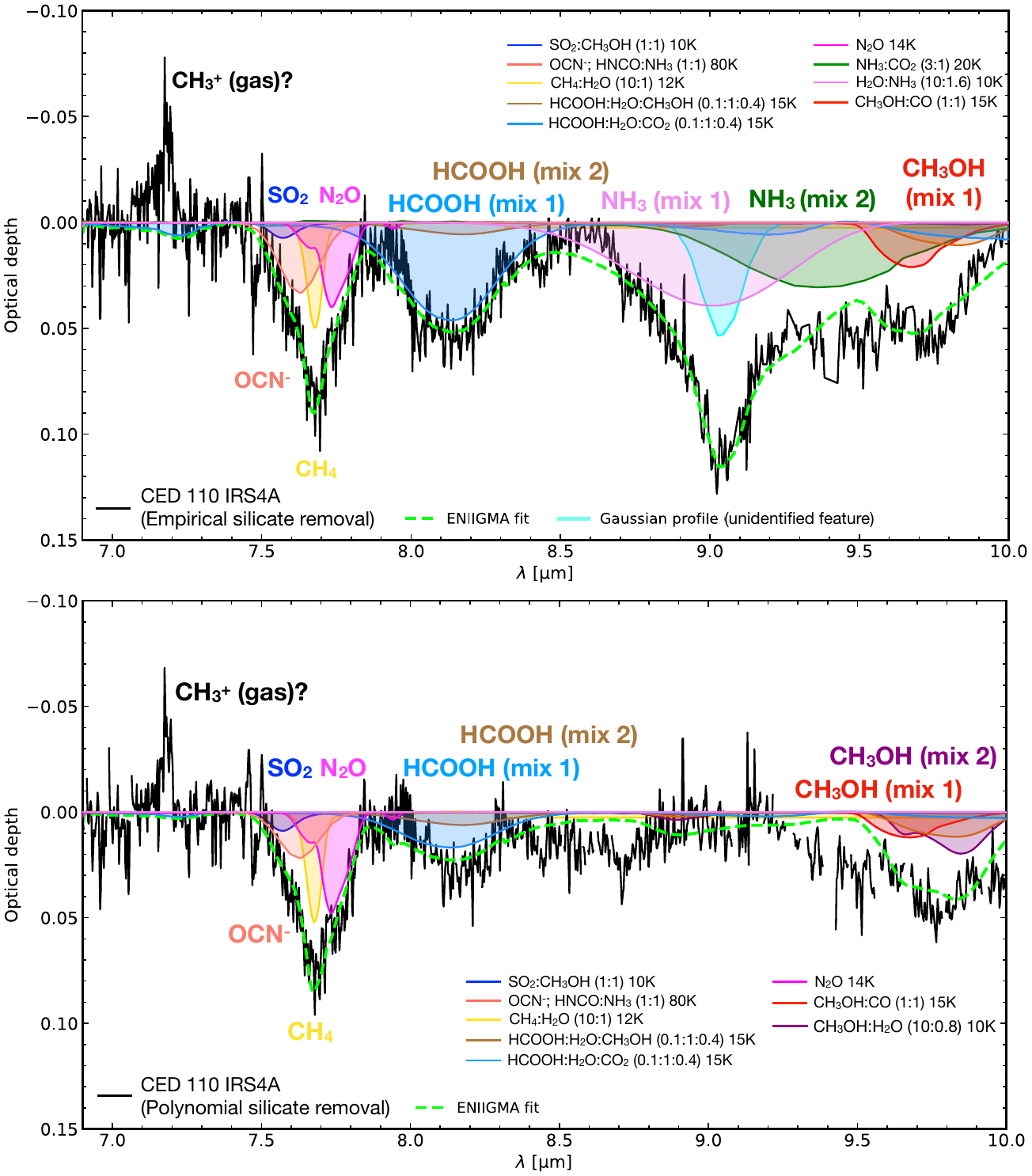}
      \caption{\texttt{ENIIGMA} fit (dashed green) in the range between 6.9 and 10~$\mu$m of Ced~110~IRS4A spectrum (black) after the empirical (top) and polynomial (bottom) silicate subtraction. The IR spectra of many ice species contribute to this fit and are indicated by labels and coloured areas. The terms ``mix 1'' and ``mix 2'' are used when two mixtures containing the same species are needed. The information of these mixtures is presented in the figure labels. We highlight that these best fits are based on the large laboratory ice database listed in Appendix~\ref{list_lab} used to explore a variety of solutions. Other solutions can be further explored by expanding the database of species tested with new data. We refer the Appendix~\ref{CH3ion} for a brief discussion about the CH$_3^+$ emission feature.}
         \label{eniigmafitSA}
   \end{figure*}

The analysis of the MIRI spectrum between 7 and 10~$\mu$m of Ced~110~IRS4A is performed independently of the rest of the observational data and follows the methodology described in \citet{Rocha2024}. This approach allows for a better constraining of the species contributing to this range without addressing the nature of the 6 and 6.8~$\mu$m bands. Despite the independent analysis of this spectral range, we used the best fit to assist the analysis of the 6 and 6.8~$\mu$m bands in Section~\ref{sec_6}. 

Here, we use the \texttt{ENIIGMA} fitting tool to search for the best fit out of a large dataset of laboratory data listed in Table~\ref{tab_list}. The local continuum subtraction around this spectral range and for the empirical silicate subtraction is shown in Appendix~\ref{localfit78}. In the case of the polynomial method, this subtraction follows the profile reported in Figure~\ref{polysilic}. 

Figure~\ref{eniigmafitSA} presents the \texttt{ENIIGMA} fit for the 7$-$10~$\mu$m range of Ced~110~IRS4A for both silicate subtraction methods. In the two cases, there are clear absorption features at 7.67, 8.2~$\mu$m and around 9.8~$\mu$m. No absorption features at 7.2 and 7.4~$\mu$m are seen towards Ced~110~IRS4A. Additionally, in the case of the polynomial silicate subtraction, no absorption is present at 9~$\mu$m, where we would expect to see NH$_3$ ice. The best fits indicate that the feature at 7.67~$\mu$m is composed of CH$_4$:H$_2$O, SO$_2$:CH$_3$OH, OCN$^-$ (HNCO:NH$_3$) and N$_2$O. Previous analyses of this band \citep{Gibb2004, Oberg2008, Zasowski2009} using {\it ISO} and {\it Spitzer} data have concluded that two species, namely, CH$_4$ and SO$_2$, contribute to this absorption feature. With the most recent JWST observations, a detailed analysis performed of NGC~1333~IRAS~2A and IRAS~23385+6053 by \citet{Rocha2024} confirmed a good spectral match with CH$_4$, SO$_2$, and also OCN$^-$.

The next strong feature in this spectral range is seen at 8.2~$\mu$m. The fits demonstrate that HCOOH mixed with H$_2$O:CO$_2$ and with H$_2$O:CH$_3$OH contribute to this band, where the former is stronger in the data after empirical silicate subtraction. This result is aligned with \citet{Bisschop2007} and \citet{Rocha2024}, which found a good spectral match with HCOOH:H$_2$O:CH$_3$OH. In addition, this result indicates that a mixture where CO$_2$ ice is presented is also possible, which brings important insights into the origin of HCOOH \citep{Qasim2019}.  

The subsequent spectral range between 8.6 and 10~$\mu$m is dominated by NH$_3$ and CH$_3$OH absorption features and the \texttt{ENIIGMA} fits are different in the two spectra given their distinct profiles. In the case of empirical silicate subtraction, a large excess is present in this entire spectral range, which is fitted with NH$_3$:H$_2$O, NH$_3$:CO$_2$, CH$_3$OH:CO, CH$_3$OH:H$_2$O and a narrow Gaussian profile at 9.04~$\mu$m representing an unidentified feature. On the other hand, no absorption remains at 9~$\mu$m after using the polynomial method with the specified guided points. Other guiding point avoiding the 9~$\mu$m band would leave some space to fit NH$_3$ ice, but exactly where to place the continuum is rather challenging. The band around 9.8~$\mu$m is fitted with CH$_3$OH:CO, and two CH$_3$OH:H$_2$O components, namely CH$_3$OH:H$_2$O and HCOOH:CH$_3$OH:H$_2$O. Most of these features were previously found in the literature, as are the cases of NH$_3$:H$_2$O, CH$_3$OH:CO and CH$_3$OH:H$_2$O \citep{Lacy1998, Bottinelli2010, Cuppen2011, McClure2023}. The reliability of the other two features, NH$_3$:CO$_2$ and the Gaussian profile, are discussed in Section~\ref{otherfeatures}.

Finally, we performed two complementary analyses that are shown in Appendix~\ref{stats_analysis}. First, we performed a consistency check of the components used in this fit with the entire spectral range available for the primary source, which is shown in Figure~\ref{con_check}. This assessment demonstrates that the intensities of the features found in this local analysis are consistent with features of those components in other wavelengths. It also indicates that additional components are needed to fit those spectral ranges. The second analysis consists of deriving Akaike Information Criterion (AIC) values to estimate the minimum number of components to the fit (Figure~\ref{aic_model}). This analysis indicates that eight of the 10 components are crucial to the fit. The least needed components are SO$_2$ and HCOOH:H$_2$O:CH$_3$OH. However, among these two, SO$_2$ keeps the model within the interval of a robust fit, whereas HCOOH:H$_2$O:CH$_3$OH does not.

In summary, the analysis of the spectral range between 7 and 10~$\mu$m, allows us to report IR features from CH$_4$, SO$_2$, OCN$^-$, NH$_3$, HCOOH, and CH$_3$OH and the tentative assignment of N$_2$O, as well as an unidentified feature at 9~$\mu$m.

\subsubsection{6 and 6.8~$\mu$m bands}
\label{sec_6}

\begin{figure}
    \centering
    \includegraphics[width=1.0\linewidth]{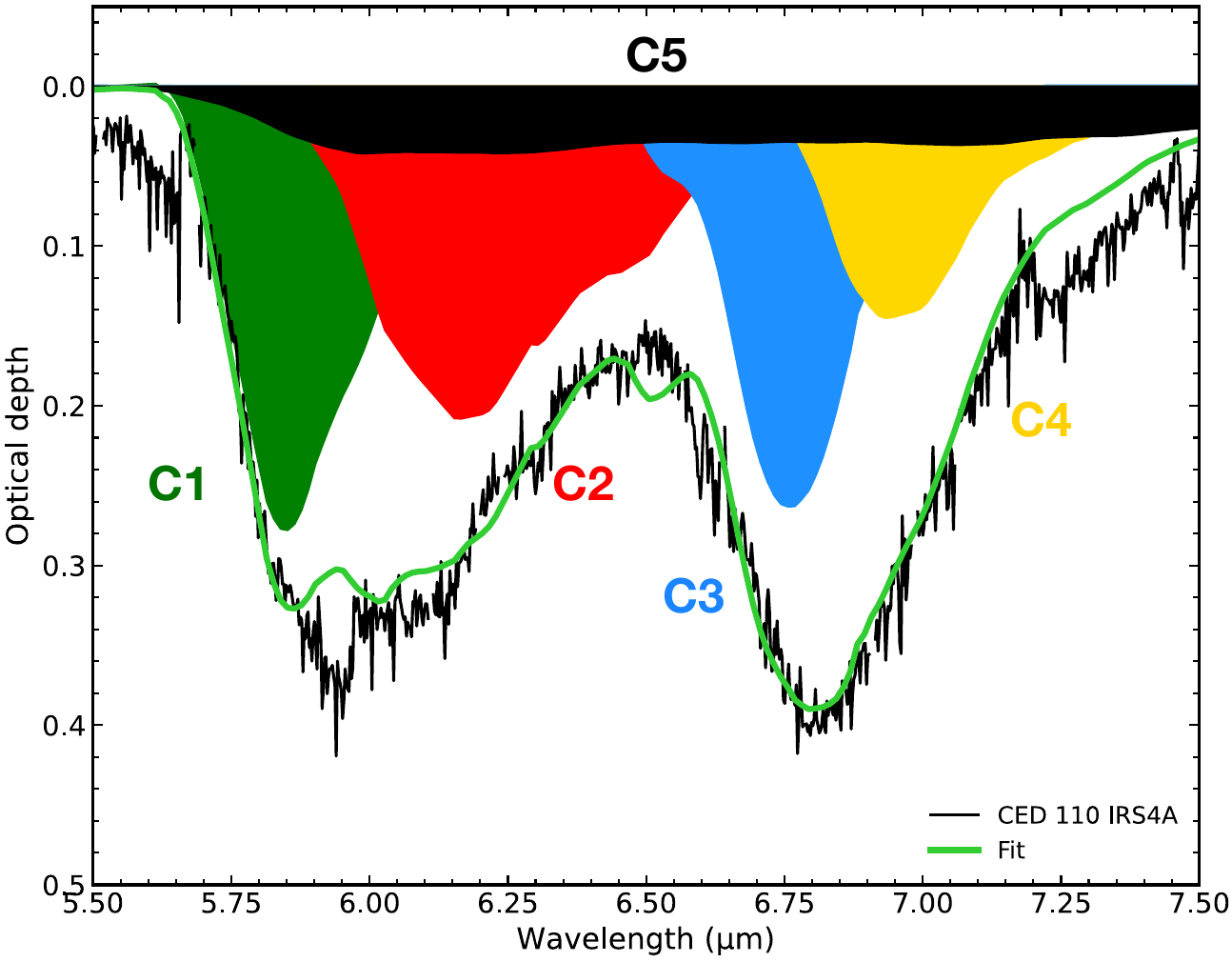}
    \caption{Spectral decomposition of the H$_2$O-subtracted spectrum of Ced 110 IRS4A using five empirical components from \citet{Boogert2008}. See accompanying text for the meaning of components C1-C5.}
    \label{fig:c1c5}
\end{figure}

The 6$-$7~$\mu$m interval is analysed in a similar way as in \citet{Boogert2008}, that proposed the use of five empirical components to model this spectral range. As required in this method, the H$_2$O pure ice spectrum is subtracted, which for simplicity, is done using the bulk ice spectrum at 10~K without grain-shape correction (Fig.~\ref{H2Oice_scale}). The residual spectrum fitted using the five components is shown in Figure~\ref{fig:c1c5}, and is likely related to different chemical species. Components C1 and C2 account for the 6~$\mu$m absorption band, where C1 can be related with a carbonyl group (C=O) and C2 with NH$_3$ and formate ion (e.g., HCOO$^-$) deformation modes. C3 and C4 fit together the 6.8~$\mu$m, and they can be tracers of NH$_4^+$ at different temperatures and chemical environments \citep[e.g.,][]{Novozamsky2001, Galvez2010}. Finally, C5 is a broad feature covering the full range. This component can be linked with crystalline H$_2$O ice and to organic refractory material \citep[e.g.,][]{Gibb2002}.

\begin{figure}
    \centering
    \includegraphics[width=1.0\linewidth]{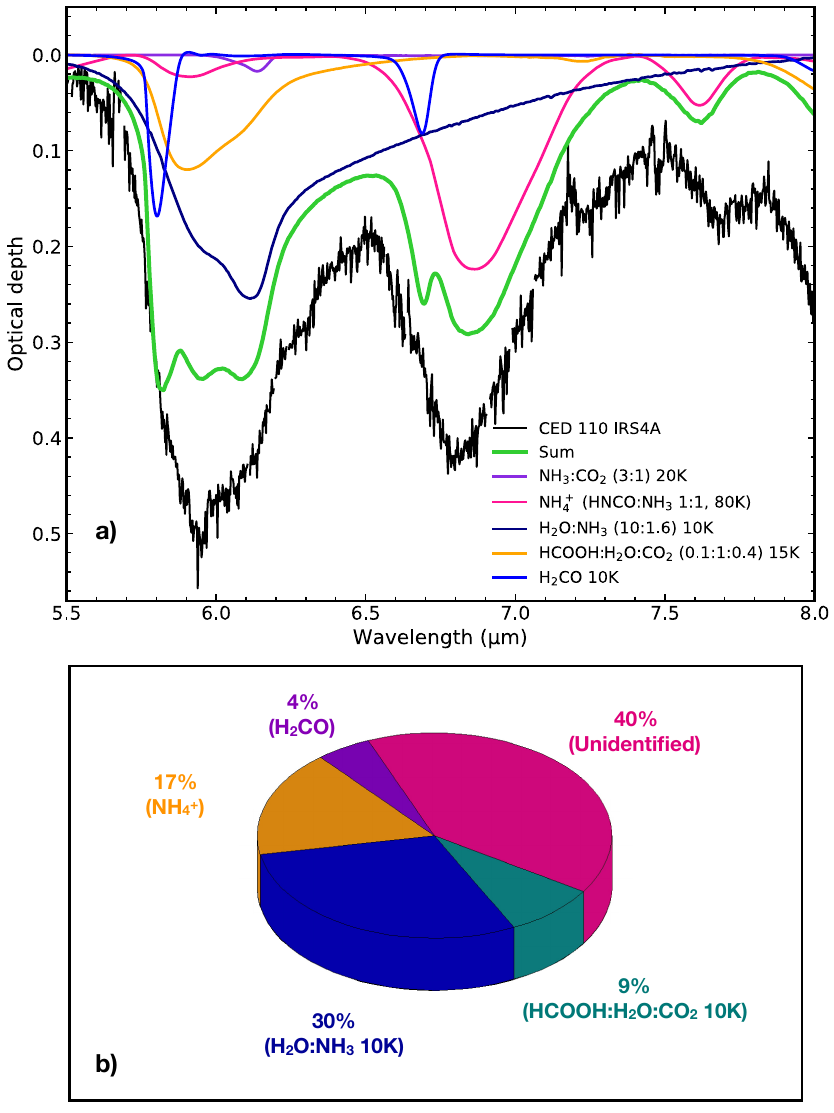}
    \caption{Simplified analysis of the 6 and 6.8~$\mu$m bands of Ced~110~IRS4A using laboratory data. H$_2$O ice is not subtracted in this case. Panel a shows the sum (green) of four laboratory spectra of known species. Panel b shows the fractional area of each component relative to the total area in the MIRI spectra between 5.5 and 7.5~$\mu$m. Unidentified refers to the difference between the MIRI data and the sum of the four components in the specified range. The contribution of NH$_3$:CO$_2$ is not added to the chart because it represents less than 1\% of the total area.}
    \label{pizza}
\end{figure}

We performed an additional analysis of the 6$-$7~$\mu$m range by accounting for areas of the bands fitted in the range between 7 and 10~$\mu$m. For instance, we consider H$_2$O:NH$_3$ ice at 10~K constrained from Figure~\ref{H2Oice_scale}. The contributions of other species is constrained from the local fits performed in Section~\ref{sec7_10}. For example, the intensity of NH$_4^+$ at 6.8~$\mu$m, is derived from the same spectral data fitting the OCN- feature at 7.6~$\mu$m in Figure~\ref{eniigmafitSA}, which is HNCO:NH$_3$ at 80~K \citep{Novozamsky2001}. The same is repeated for HCOOH (mixed with H$_2$O and CO$_2$) and NH$_3$ (mixed with CO$_2$). The H$_2$CO contribution comes from the upper limit based on the 8~$\mu$m band. Figure~\ref{pizza}a displays the individual contributions of the five laboratory spectra as well as their cumulative sum. Notably, the analysis reveals that both 6 and 6.8~$\mu$m bands are not entirely adequately explained by the species used in this analysis. In particular, the NH$_4^+$ feature constrained from the local fits presented in Section~\ref{sec7_10}, does not account for all absorption at 6.8~$\mu$m.
   
The percentage contribution of each species is illustrated in Figure~\ref{pizza}b. The component H$_2$O:NH$_3$, that can fit entirely the 3~$\mu$m, contributes to 30\% of the 6$-$8~$\mu$m band. The second and third most prominent features are the NH$_4^+$ (17\%), and HCOOH:H$_2$O:CO$_2$ accounts for 9\%. Finally, H$_2$CO ice contributes with 4\% as estimated from the upper limits based on the 8~$\mu$m band. Subtracting these four species, an area of 40\% remains to be fitted. This simple analysis illustrates that an important portion of the 6$-$7~$\mu$m remains unknown. Candidates can be some of the carriers of the C1 to C5 components shown in Figure~\ref{fig:c1c5}.

\subsubsection{11.2~$\mu$m}

\begin{figure}
   \centering
   \includegraphics[width=\hsize]{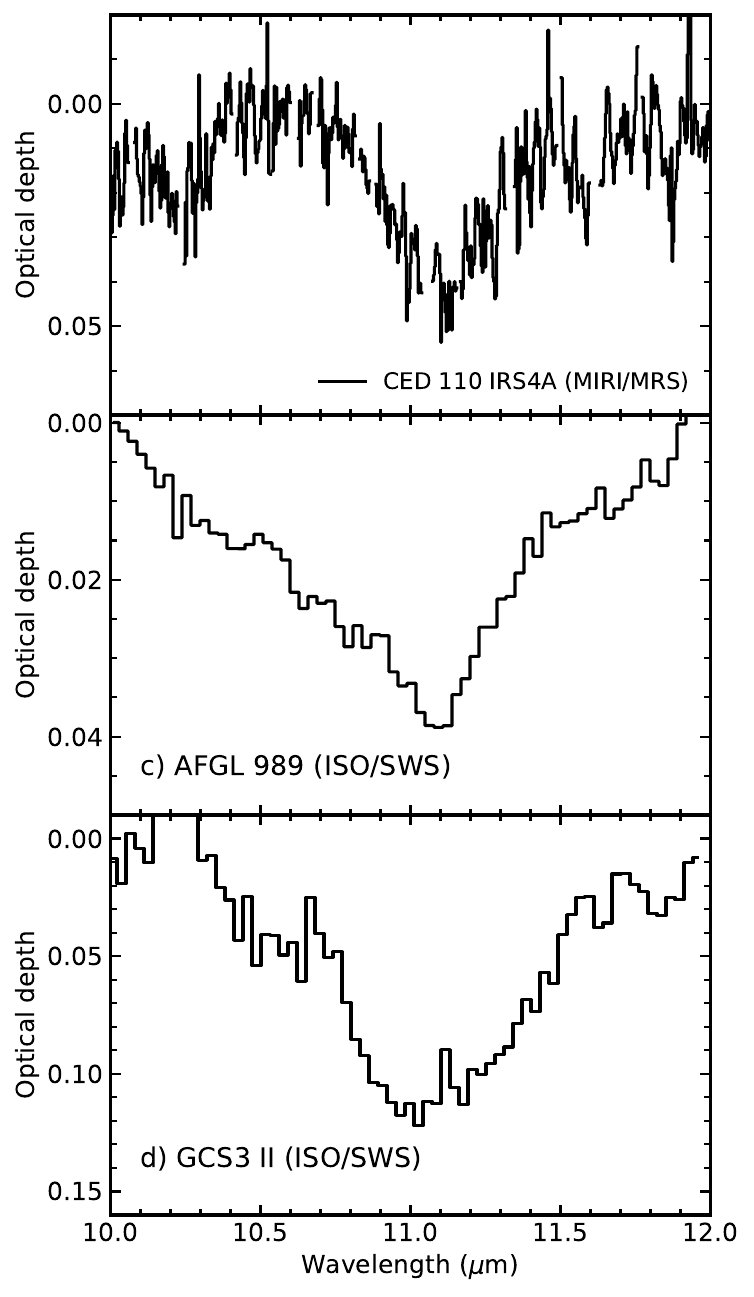}
      \caption{Comparison between the 11~$\mu$m feature towards Ced~110~IRS4A (panel a) with ISO/SWS towards a high-mass protostar AFGL~989 and the galactic center source GCS II (panels b and c).}
         \label{Forst}
   \end{figure}

The analysis of the 11.2~$\mu$m band is performed by using the optical depth spectrum after polynomial silicate subtraction as presented in Figure~\ref{Forst}a. This procedure was needed because the silicate subtraction performed with the SED modelling underpredicts the absorption profile at this range, which is noticed by the negative optical depth values (see Fig.~\ref{sedmod}). For comparison, data from the \textit{Infrared Space Observatory} using the Short Wavelength Spectrometer (SWS) data of AFGL~989 and GCS3 II from \citet{DoDuy2020} are presented, both with an absorption feature at 11.2~$\mu$m.

We calculated the fraction of crystalline silicate using the following equation from \citet{Poteet2013}:

\begin{equation}
    \frac{N_{\rm{cr}}}{N_{\rm{cr}} + N_{\rm{am}}} = \frac{\tau_{\rm{cr}}(\lambda_{\rm{p}}) \kappa_{\rm{am}}}{\tau_{\rm{cr}}(\lambda_{\rm{p}}) \kappa_{\rm{am}} + \tau_{\rm{am}}(\lambda_{\rm{p}}) \kappa_{\rm{cr}}}
\end{equation}
where $N_{\rm{cr}}$ and $N_{\rm{am}}$ are the dust column densities of the crystalline and amorphous dust, respectively. Similarly, $\tau_{\rm{cr}}$ and $\tau_{\rm{am}}$ are the optical depths in the two ice morphologies, $\kappa_{\rm{cr}}$ and $\kappa_{\rm{am}}$, the mass-opacity values and $\lambda_{\rm{p}}$ the wavelength at the peak of the feature.

The optical depth values for the crystalline and amorphous dust are respectively 0.06 (Fig.~\ref{Forst}) and 2.8 (estimated from the empirical silicate subtraction). The mass-opacity values at 11.2~$\mu$m are $\kappa_{\rm{am}} = 3.5 \times 10^3$ cm$^2$ g$^{-1}$ and $\kappa_{\rm{cr}} = 6.0 \times 10^3$ cm$^2$ g$^{-1}$. The crystalline to amorphous dust ratio is 1.2\%, which is consistent with values in the majority of YSOs \citep[e.g.,][]{Demyk1999, Wright2016, DoDuy2020}.

\subsubsection{15.2~$\mu$m}
We analysed the 15.2~$\mu$m feature using the continuum subtraction method from the SED model. In this particular example, we have adopted the same ice components used previously by \citet{Pontoppidan2008} to fit the spectrum, namely, CDE corrected spectrum of pure CO$_2$ ice, CO$_2$:H$_2$O (0.14:1), CO$_2$:CO (1:1), CO$_2$:CO (0.21:1). In addition, to these data, we added CO$_2$:CH$_3$OH mixtures with five different proportions, 1:1, 1:3, 1:10, 10:1, 3:1. This spectral decomposition reveals that the double peak in this feature is mostly caused by the presence of pure CO$_2$ ice, which indicates ice segregation or distillation due to thermal processing \citep{Dartois1999, Ehrenfreund1999}. Pure CO$_2$ also has a prominent blue wing that contributes to the short-wavelength shoulder of the 15~$\mu$m band. The red wing of this band is dominated by CO$_2$ in a H$_2$O rich environment and CO$_2$:CH$_3$OH (1:1). Other two components due to CO$_2$ in a CO-rich ice and CO$_2$:CH$_3$OH (3:1) contribute to an increase in the optical depth at 15.2~$\mu$m. Overall, this fit illustrates that the majority of the 15.2~$\mu$m band is dominated by CO$_2$ in H$_2$O ice-rich environment, which is aligned with \citet{Pontoppidan2008}. In addition, the mixture with CH$_3$OH ice agrees with recent results \citet{Brunken2024} for $^{13}$CO$_2$ ice at 4.38~$\mu$m.     

\begin{figure}
   \centering
   \includegraphics[width=\hsize]{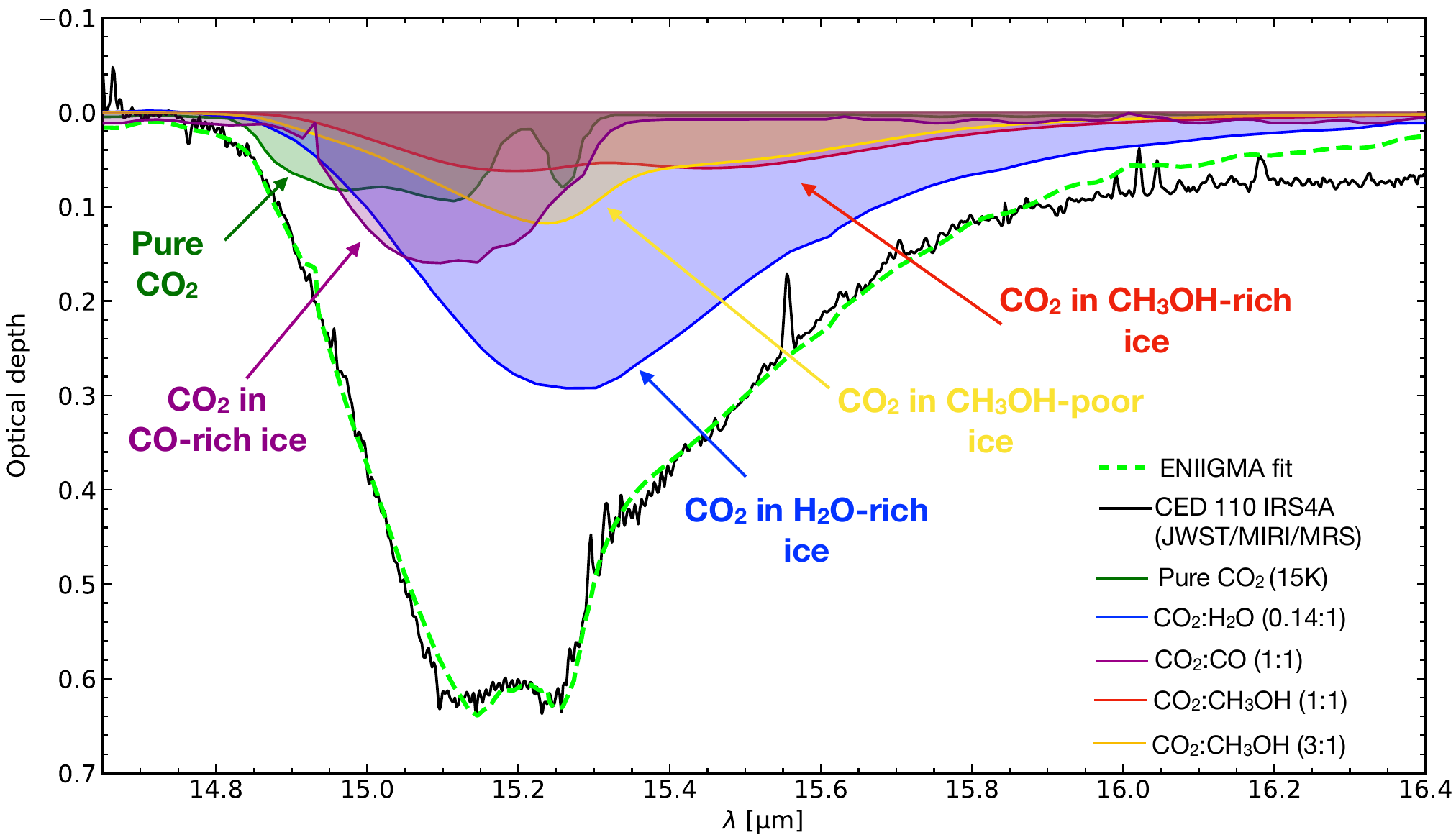}
      \caption{ENIIGMA fit of the CO$_2$ ice bending mode towards Ced~110~IRS4A. The fit performed with five components indicate contributions of CO$_2$ pure and mixed with H$_2$O, CO and CH$_3$OH.}
         \label{CO2fit}
   \end{figure}

A statistical analysis was performed to decide on the minimum number of components providing the best fit, and what is the effect of adding more components to the fit. For this analysis, both $\chi^2$ and AIC are calculated. This result is presented in Appedix~\ref{stats_analysis}. It indicates that a number of five components are needed to have both low $\chi^2$ and AIC, which are the components displayed in Figure~\ref{CO2fit}.

\subsection{Ice fits: Ced~110~IRS4B}
\label{sec_nirB}

\begin{figure*}
   \centering
   \includegraphics[width=\hsize]{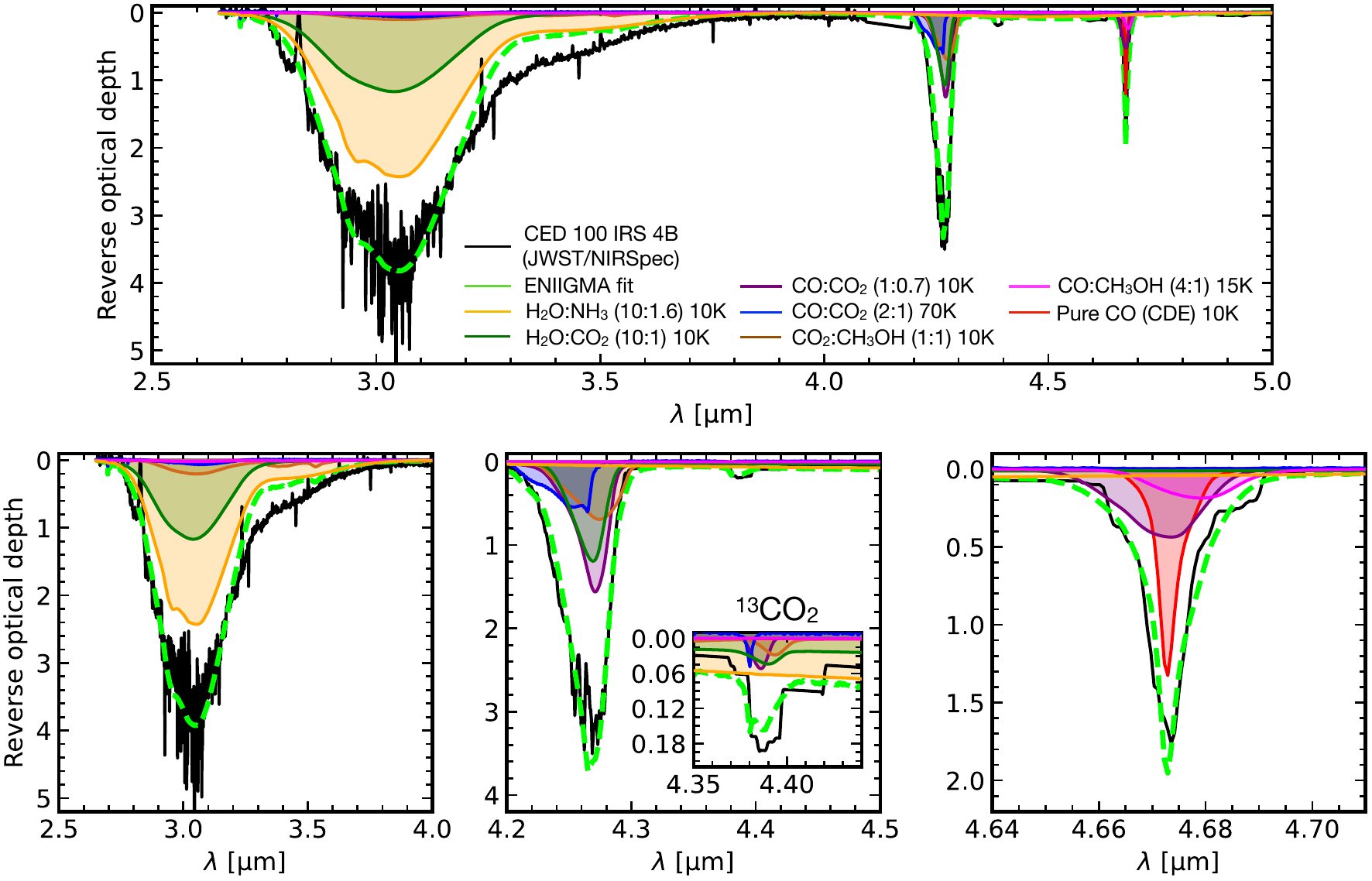}
      \caption{\texttt{ENIIGMA} fit (green) between 2.5 and 5~$\mu$m of the Ced~110~IRS4B spectrum (black). Top panel shows the global fit, and bottom panels displays details of specific regions, namely, H$_2$O, CO$_2$ and CO.}
         \label{eniigmafitSB}
   \end{figure*}

A global fit covering the range between 2.5 and 5.0~$\mu$m of the NIRSpec data of Ced~110~IRS4B was performed using the \texttt{ENIIGMA} fitting tool. The MIRI/MRS data is not included in this fit because of the low SNR and flux overlap with the primary source at longer wavelengths. Figure~\ref{eniigmafitSB} shows the best fit with 7 components, namely, H$_2$O:NH$_3$, H$_2$O:CO$_2$, CO:CO$_2$ (10 and 70~K), CO$_2$:CH$_3$OH, CO:CH$_3$OH, and pure CO. The top panel of Figure~\ref{eniigmafitSB} shows the overall fit highlighting the major absorption features at 3~$\mu$m, 4.27~$\mu$m and at 4.67~$\mu$m. The bottom panels show zoom-ins of the three strong absorption features, where there is an inset displaying the 4.38~$\mu$m band in 
the middle panel.

The 3~$\mu$m feature is dominated by H$_2$O ice (O$-$H stretch), which is mixed with CO$_2$ and NH$_3$. Ammonia has a natural vibrational mode at 2.9~$\mu$m (N$-$H stretch), and mixed with H$_2$O ice, a broad profile peaking at 3.5~$\mu$m, called ammonia
hydrate is activated \citep{Dartois2001, Moore2007}. A small contribution to the broad 3~$\mu$m feature also comes from CH$_3$OH mixed with CO. Another important aspect to highlight is the absorption excess in the red wing of the 3~$\mu$m band that is not fitted with the ice components adopted in this work. The nature of this band is mostly associated with grain growth effects and larger chemical species.

The second major absorption feature in the Ced~110~IRS4 spectrum is the 4.27~$\mu$m band, which is associated with $^{12}$CO$_2$ ice. The fits indicate that this band is composed of CO$_2$:H$_2$O, and CO:CO$_2$ at two temperatures, 10~K and 70~K. For the warm component, we found a range between 60 and 80~K that fit well the observations, as presented in Figure~\ref{coco2oT}. CO$_2$ ice is known to be a reliable temperature tracer of protostellar ices from analysis of the 4.38~$\mu$m \citep{Boogert200013co2, Brunken2024} and 15~$\mu$m bands \citep{Gerakines1999, Keane2001, Pontoppidan2008, Kim2012}.

The $^{13}$CO$_2$ band at 4.38~$\mu$m is another important feature to highlight. All four CO$_2$ components used to fit the 4.27~$\mu$m band are also present here and agree with recent $^{13}$CO$_2$ ice results from \citet{Brunken2024}. Nevertheless, the absorption profile is not fully reproduced in the global fit. Mostly the peak at 4.38~$\mu$m and the long-wavelength side are not well fitted. In principle, increasing the contributions from CO$_2$:CH$_3$OH, H$_2$O:CO$_2$ and CO:CO$_2$ (low T) would improve the fit. However, this approach is not reasonable because the concurrent fit between 2.5 and 5~$\mu$m already takes into account contributions in other wavelengths. This is still an open question that requires further analysis with a larger number of sources, and ideally considering both $^{12}$CO$_2$ and $^{13}$CO$_2$ bands.      

The feature at 4.67~$\mu$m, associated with the CO ice, is the third major ice band in the NIRSpec data of Ced~110~IRS4. The fits indicate that three CO components are needed to interpret the shape of the CO band, i.e., pure CO, CO:CO$_2$ and CO:CH$_3$OH, which is aligned with previous studies \citep{Pontoppidan2003, Cuppen2011, McClure2023}. Only the cold CO:CO$_2$ component contributes to the CO ice absorption since 60$-$80~K is beyond the CO ice desorption. The pure CO ice in this work is fitted using grain-shape correction assuming the CDE approach and grain size ranging from 0.1 to 1~$\mu$m. The lack of accurate optical constants for ice mixtures precludes the use of grain-shape correction for all species in the fit. Additionally, weaker bands are less affected by grain-shape correction \citep[e.g.,][]{Tielens1991}.



A consistency check of this fit was performed in the mid-IR spectral range as presented in Figure~\ref{nir_check} of Appendix~\ref{stats_analysis}. The lower quality of the MIRI/MRS data precludes further analyses of this range. However, the components fitting the NIR range are consistent with the Mid-IR, and no overshooting of the absorption features is found. In fact, the H$_2$O ice libration band and the CO$_2$ bending mode at 12 and 15.2~$\mu$m, respectively, are relatively well fitted.

Other minor features, such as OCN$^-$ (4.6~$\mu$m), $^{13}$CO (4.78~$\mu$m) and OCS (4.9~$\mu$m) were not addressed in this paper. If present, these features are strongly blended with CO gas emission lines, which is required to be accurately removed prior to the ice analysis. 

\subsection{Ice column densities}
The ice column densities of the chemical species identified towards Ced~110~IRS4A and IRS4B were calculated using Equation~\ref{eq_cd} and are presented in Table~\ref{ice_cd}. The ice abundances with respect to H$_2$O ice are also calculated and compared with the literature values in LYSOs and comet 67P/Churyumov–Gerasimenko (67P/C-G).

The ice column densities of H$_2$O in Ced~110~IRS4A and $^{13}$CO$_2$ in Ced~110~IRS4B were calculated using simplified methods. In the former, the H$_2$O ice was calculated from the H$_2$O:NH$_3$ ice mixture scaled to the NIRCam spectrum as shown in Figure~\ref{H2Oice_scale}. This approach was used because the libration band is affected by the silicate emission and therefore are excluded from the global fits. In the latter, the band area at 4.38~$\mu$m is used to derive the $^{13}$CO$_2$ ice column density because the \texttt{ENIIGMA} fits do not account for all the absorption in the NIRSpec data. Upper limits for NH$_4^+$ and H$_2$CO towards Ced~110~IRS4A were derived. In the former case, we used the sum of components C3 and C4 (Fig.~\ref{fig:c1c5}), whereas in the latter, we adopted the band of the 5.8~$\mu$m band shown in Figure~\ref{pizza}. Excluding these cases, all ice column densities are calculated directly from the \texttt{ENIIGMA} fitting tool, and the error bars are derived from the confidence intervals presented in Appendix~\ref{stats_analysis}. In the case of SO$_2$, Ced~110~IRS4A has the same abundance of sulfur dioxide ice compared to NGC1333~IRAS~2A \citep{Rocha2024}, despite the lower ice column density. The abundance of CO ice in Ced~110~IRS4B being close to the lower limit of what is found in LYSOs could point to a heating event where part of CO ice sublimates. 

An important result from these values is the $^{12}$CO$_2$/$^{13}$CO$_2$ ratio, which is 71$_{68}^{88}$. This interval is similar to the values found for the two background stars presented in \citet{McClure2023}, which range between 69 and 87. This ratio cannot be derived for IRS4A since the observations do not cover the 4.38~$\mu$m band. A recent work using JWST data by \citet{Brunken2024ratio} derived $^{12}$CO$_2$/$^{13}$CO$_2$ ratios of 85$\pm$23, 76$\pm$12 and 97$\pm$17 when derived from the $^{12}$CO$_2$ ice bands at 2.7, 4.27 and 15.2~$\mu$m, which are consistent with the isotope ratio towards Ced~110~IRS4B.   

Comparisons of the ice column densities and abundances between two protostars are not conclusive since different wavelengths were analysed for both sources. Nevertheless, it seems that the companion source hosts more ice material than the primary source. The only exception is $^{12}$CO$_2$. Despite the saturation of the 4.27~$\mu$m band in the NIRSpec data of Ced~110~IRS4B, the error bars still indicate lower amount of CO$_2$ ice. In comparison with LYSO abundances, all values are within the ranges previously found in the literature. It is also worth mentioning that the ice column density values estimated from the data after empirical and polynomial continuum subtraction are different in the cases of NH$_3$, HCOOH and OCN$^-$, which are lower by a 100\%, 55\% and 30\%, respectively.

\begin{table*}[h]
\caption{\label{ice_cd} Ice column densities and abundances with respect to H$_2$O ice towards Ced~110~IRS4A and IRS4B. These values are compared to literature values for other objects, including LYSOs, the two field stars probing lines of sight towards Chamaleon I, NIR38 and J110621, as well as the comet 67P/C-G.}
\renewcommand{\arraystretch}{1.5}
\centering 
\begin{tabular}{lcccccccc}
\hline\hline
Species &  \multicolumn{2}{c}{$N_{\rm{ice}}$ ($10^{17}$ cm$^{-2}$)} & \multicolumn{2}{c}{$X_{\rm{H_2O}}$ (\%)} & \multicolumn{4}{c}{Literature (\% H$_2$O)}\\
\cline{2-9}
 & IRS4A$^a$ & IRS4B & IRS4A & IRS4B & LYSOs$^b$ & NIR38$^e$ & J110621$^f$ &Comet 67P/C-G$^f$\\
\hline
H$_2$O & 45.2$\pm$2.3 & 63.5$_{48.1}^{69.3}$ & 100 & 100 & 100 & 100 & 100 &  100\\
$^{12}$CO$_2$ & 11.1$_{9.7}^{15.0}$ & 7.1$_{6.8}^{7.9}$ & 24.5 & 11.2 & 28$_{23}^{37}$ & 20.1${_{7.3}^{38.7}}$ & 13.0${_{6.1}^{19.0}}$ &  4.7$\pm$1.4\\
$^{13}$CO$_2$ & ... & 0.1$\pm$0.01 & ... & 0.2 & ... & 0.3  & 0.3 & ... \\
$^{12}$CO & ... & 6.6$_{5.9}^{7.2}$ & ... & 10.4 & 21$_{12}^{35}$ & 43.1${_{20.0}^{65.4}}$ & 27.5${_{13.0}^{47.2}}$ & 3.1$\pm$0.9\\
CH$_4$ & 0.78$_{0.52}^{1.0}$ (0.80$_{0.55}^{1.12}$) & ... & 1.7 (1.8) & ... & 4.5$_{3}^{6}$ & 2.6${_{1.4}^{4.8}}$ & 1.9${_{0.9}^{2.4}}$ & 0.340$\pm$0.07\\
SO$_2$ & 0.11$_{0.06}^{0.17}$ (0.09$_{0.07}^{0.15}$) & ... & 0.2 (0.2) & ... & 0.2$^{c}$ & 0.05 & 0.04 & 0.127$\pm$0.100\\
NH$_3$ & 2.88$_{1.90}^{3.86}$ (0.0) & 5.7$_{2.1}^{6.3}$ & 6.4 (0.0) & 8.9 & 6$_{4}^{8}$ & 4.4${_{2.1}^{10.6}}$ & 5.0${_{2.5}^{8.6}}$ & 0.67$\pm$0.20\\
HCOOH & 0.92$_{0.56}^{1.28}$ (0.41$_{0.22}^{0.61}$) & ... & 2.0 (0.9) & ... &$<0.5-$4$^d$, 1.0$^c$ & ... & ... & 0.013$\pm$0.008\\
CH$_3$OH & 0.94$_{0.70}^{1.3}$ (0.89$_{0.67}^{1.05}$) & 3.85$_{2.7}^{5.2}$ & 2.1 (2.0) & 6.1 &6$_{5}^{12}$ & 8.9${_{2.6}^{17.7}}$ & 3.9${_{2.5}^{8.2}}$ & 0.21$\pm$0.06\\
OCN$^-$ & 0.92$_{0.46}^{1.4}$ (0.63$_{0.35}^{0.82}$) & ... & 2.0 (1.4) & ... & 0.6$_{0.4}^{0.8}$, 1.2$^c$ & 0.3 & 0.3 & ...\\
N$_2$O & 0.81$_{0.53}^{1.10}$ (0.85$_{0.55}^{1.21}$) & ... & 1.8 (1.9) & ... & ... & ... & ... & ...\\
\hline
\multicolumn{9}{c}{Upper limits}\\
\hline
NH$_4^+$ & $<$6.5 & ... & $<$14.3 & ... & 11$_{7}^{15}$ & $<$6 & $<$8 & ...\\
H$_2$CO & $<$3.4 & ... & $<$7.5 & ... & $\sim$6 & ... & ... & 0.32$\pm$0.1\\
\hline
\end{tabular}
\tablefoot{$^a$ Values inside parentheses are calculated based on the data after polynomial silicate subtraction. $^b$ When not indicated otherwise, the values correspond to the median and lower and upper quartile values from summarized by \citet{Boogert2015},
$^c$\citet{Rocha2024}: IRAS~2A, $^d$\citet{Oberg2011},
$^e$\citet{McClure2023},
$^f$\citet{Rubin2019}.}
\end{table*}

\begin{figure*}
   \centering
   \includegraphics[width=\hsize]{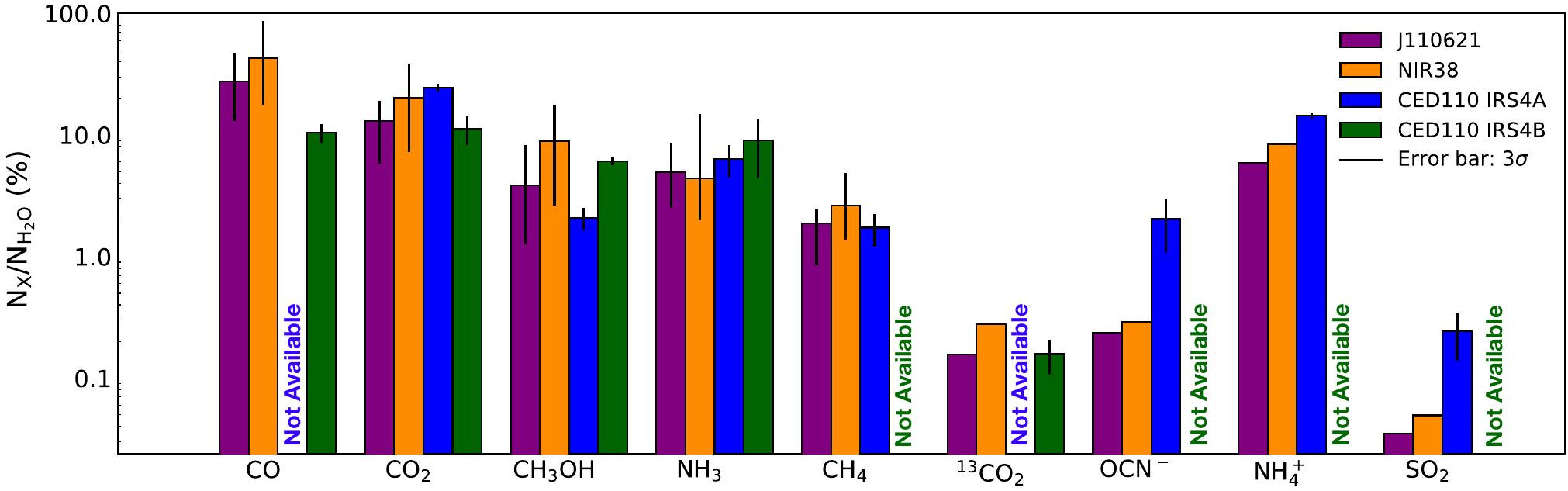}
      \caption{Ice abundances relative to H$_2$O ice for the protostars Ced~110~IRS4A and Ced~110~IRS4B compared to the two lines of sight towards field stars behind the Chameleon I molecular cloud, NIR38 and J110621.}
         \label{abundplot}
   \end{figure*}

Finally, Figure~\ref{abundplot} displays a bar plot with the ice abundances relative to H$_2$O ice compared to the molecular cloud values towards the background stars NIR38 and J110621. In the case of the minor species in IRS4A, we used abundances derived after SED empirical subtraction of the continuum and silicate. No significant difference in the ice abundance is observed for the molecules addressed in this work. Only OCN$^-$, NH$_4^+$ and SO$_2$ seem to be enhanced in Ced~110~IRS4A compared to the molecular cloud. In fact, in the cases of these two ions, \citet{Boogert2015} suggest an enhancement in the median values from background stars to low-mass protostars.

\section{Discussion}
\label{sec_diss}

\subsection{Thermal processing}

Both Ced~110 sources show evidence of thermal processing characterized by the CO$_2$ ice band. In the case of IRS~4A, the bending mode of CO$_2$ ice at 15.2~$\mu$m shows that the double peak can be due to pure CO$_2$ ice that has stem from ice segregation in the case of H$_2$O:CO$_2$ or ice distillation for CO$_2$:CO, as well as CO$_2$ interacting with a complex molecule \citep[e.g.,][]{Dartois1999}. Distinguishing between the two scenarios is beyond the scope of this work. The fraction of pure CO$_2$ in the fits is 8.2\%, which is consistent with the values found by \citet{Brunken2024} for other protostars. The two CO$_2$:CH$_3$OH data can also be used as thermal ice tracers. For example, the mixture with 3:1 proportion can indicate a colder region where a larger fraction of CO$_2$ is mixed with CH$_3$OH. On the other hand, the 1:1 mixture would be associated with a higher temperature where most CO$_2$ ice has sublimated.


In the case of IRS~4B, the evidence of thermal processing also comes from the CO$_2$ stretch mode at 4.27~$\mu$m. In past works, a short-wavelength shoulder (4.23~$\mu$m) was also used as a tracer of ice thermal processing as can be noted in \citet{deGraauw1996CO2} and \citet{Gerakines1999}. This feature is present in IRS~4B, which is associated with H$_2$O:CO$_2$ (0.7:1) after CO distillation (see Section~\ref{icelab_sec} and Appendix~\ref{bin_mix}). Performing a simultaneous fit of the two strong $^{12}$CO$_2$ bands 4.27~$\mu$m and 15.2~$\mu$m would be important to rule out degeneracies in the analysis and obtain an accurate conclusion on the CO$_2$ ice composition. Unfortunately, this is not possible for IRS4A and IRS4B since there is no data coverage for both sources in these ranges. This type of analysis was performed by \citet{Keane2001} who found that annealed CO$_2$ ice mixtures \citep[e.g., H$_2$O:CO$_2$:CH$_3$OH][]{} are needed to reproduce both the 4.27~$\mu$m and 15.2~$\mu$m in the spectrum of the source S~140:IRS1. The CO$_2$ bending mode in S~140:IRS1 is more pronounced than in Ced~110~IRS4A, which likely indicate a warmer environment \citep[see][]{Dartois1999}. Our results are similar to \citet{Keane2001} regarding the need for a polar ice mixture to hold CO$_2$ ice when warming up, and also that the short-wavelength shoulder of the CO$_2$ stretching mode in the observations can be attributed to ice thermal processing.   

In a recent paper, \citet{Brunken2024} used JWST data to show that warmer sources present a prominent short-wavelength profile in the $^{13}$CO$_2$ ice band at 4.38~$\mu$m, where the intensity is proportional to the level of thermal processing. A similar trend was previously found toward high-mass protostars \citep{Boogert2000}. The 4.38~$\mu$m absorption profile is not covered in IRS~4A, while in IRS~4B, the presence of a blue profile is not fully clear. However, we note that this feature is convoluted with a forest of CO gas-phase lines, which can affect the shape of the $^{13}$CO$_2$ band in the secondary source of the Ced~110 system.

\subsection{Chemical environment}
This section is focused on discussing the chemical environment of the Ced~110~IRS4 system and how it compares with the Chameleon I molecular cloud.

\subsubsection{Absence of the 7.2 and 7.4~$\mu$m bands in IRS4A}
\label{sec_nocoms}

Unlike NIR38 and J110621, which display clear absorption features around 7.2 and 7.4~$\mu$m \citep[see Fig. 3 in][]{McClure2023} indicative of COMs, Ced 110 IRS4A exhibits no absorption in this range. The absence of these two features can have either a physical or a chemical origin, or both.

\begin{figure}
   \centering
   \includegraphics[width=\hsize]{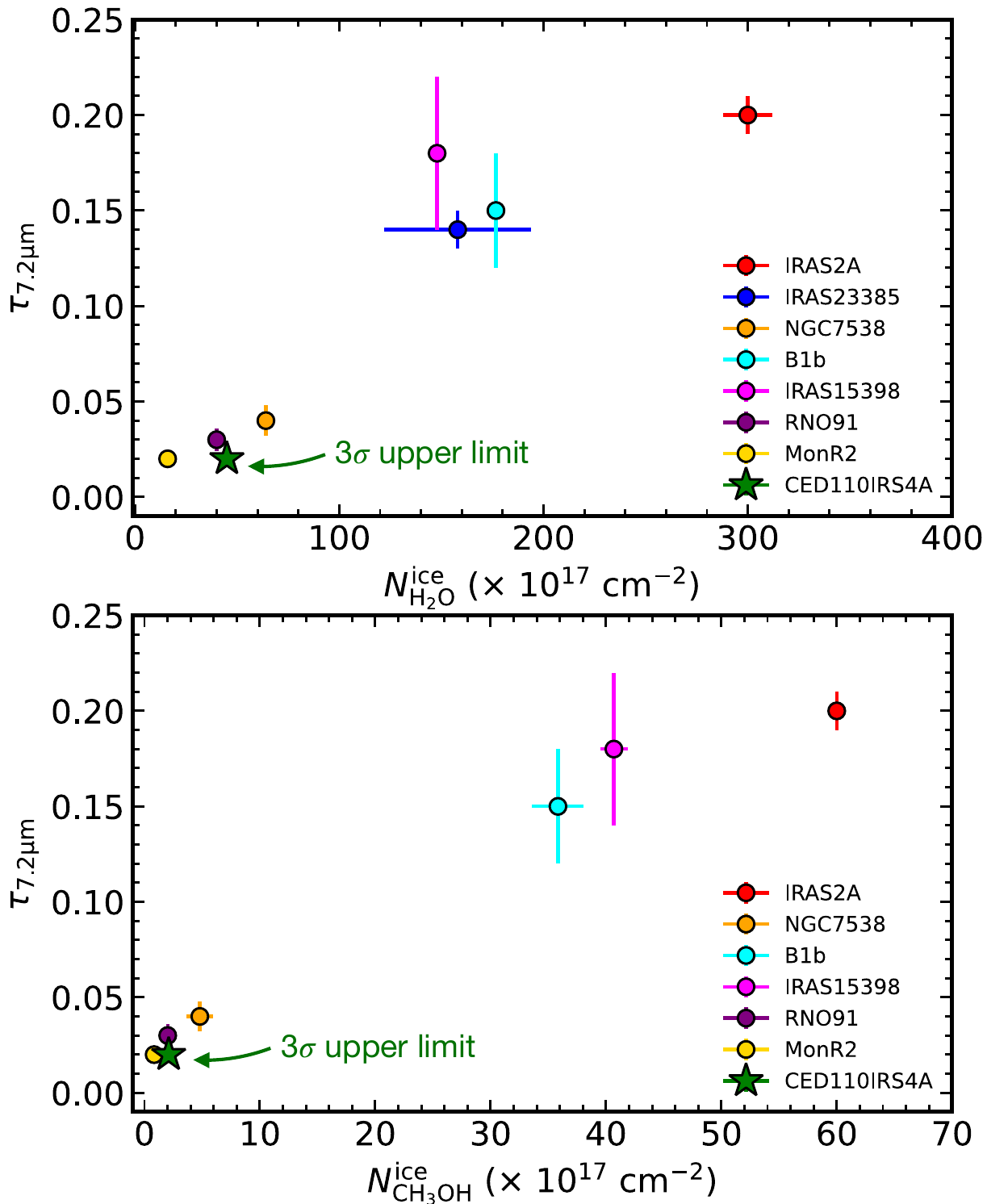}
      \caption{Correlation between the H$_2$O and CH$_3$OH ice column densities and the intensity of the 7.2~$\mu$m band for different protostars. Ced~110~IRS4A is indicated by the green star and corresponds to a 3$\sigma$ upper limit.}
         \label{corrplot}
   \end{figure}

In Figure~\ref{corrplot}, we show a potential correlation between the H$_2$O ice column density toward different YSOs and the intensity of the 7.2~$\mu$m band. H$_2$O ice is simply used here as a tracer of material along the line of sight. Extinction parameters are not used to avoid complications arising from the total-to-selective extinction and the size distribution of grains that change the extinction law \citep[e.g.,][]{cardelli1989, pontoppidan2024}. It should be noted that higher peak intensities at 7.2~$\mu$m may be related to a high ice material along the line of sight, whereas shallow features are likely due to less material between the source and the observer. This correlation still holds for Ced~110~IRS4A if we consider a 3$\sigma$ upper limit for the optical depth. This value is consistent with other sources observed with ISO and Spitzer \citep[e.g., RNO91 and Mon~R2~IRS3;][]{Boogert2008} with similar H$_2$O ice column densities.

If the absence of the 7.2~$\mu$m band in Ced~110~IRS4A is entirely due to the low column density material along the line of sight, this implies that the source inclination plays an important role in tracing these bands. Consequently, the non-detection of these bands does not imply the absence of icy COMs and HCOO$^-$. On the other hand, a chemical argument can be invoked. For example, the ice mixtures NH$_3$:HCOOH and H$_2$O:NH$_3$:HCOOH are known to form HCOO$^-$, peaking at 7.2 and 7.4~$\mu$m whereas other mixtures do not lead to this same chemical product \citep[e.g., HNCO:NH$_3$;][]{Novozamsky2001}. This would imply that no reaction between NH$_3$ and HCOOH occurred in this source. A third possibility is that strong gas-phase emission lines hide the ice's absorption profiles. For instance, \citet{vangelder2024} show that SO$_2$ gas-phase emission can also affect the shape of the 7.4~$\mu$m feature due to the P-branch emission profile. However, no SO$_2$ gas is seen towards Ced~110~IRS4A \citep[see][]{vangelder2024}.

Finally, it is important to highlight the non-detection of gas-phase COMs in this source, including CH$_3$OH gas \citep{vgelder2022PhD}. According to \citet{Nazari2022}, the absence of CH$_3$OH gas emission does not imply the lack of these molecules in the source. In fact, the presence of a disk with relatively large grains reduces the CH$_3$OH emission by more than an order of magnitude \citep[see also][]{Nazari2024gas}. From this perspective, the absence of 7.2 and 7.4~$\mu$m bands is likely more a result of the physical evolution and geometry of the source than a chemical evolution of the ices.

\subsubsection{CH$_4$, SO$_2$, OCN$^-$, NH$_4^+$}
CH$_4$ in a H$_2$O ice mixture is found to fit well the 7.67~$\mu$m towards Ced~110~IRS4A. This is aligned with other recent analyses of spectrally resolved JWST data \citep{Rocha2024, Chen2024}, and with the formation mechanism of these two species via hydrogenation \citep{Tielens1991, Oberg2011}.

SO$_2$ ice has been proposed to contribute to the short-wavelength side of the 7.67~$\mu$m band \citep{Oberg2008}, and it was confirmed statistically in IRAS2A \citep{Rocha2024} as well as for Ced~110~IRS4A (see Figure~\ref{7micanalysis}). Despite the fact that SO$_2$ mixed with CH$_3$OH produces the best fit, it is important to note that no other SO$_2$ ice data is currently available. \citet{Boogert2022} found a a good correlation between OCS and CH$_3$OH ices, which is likely related to a common formation pathway from the CO ice. Concluding whether SO$_2$ can be mixed with CH$_3$OH and if this mixture is the best carrier of the short-wavelength wing of the 7.67~$\mu$m requires more study with a comprehensive list of SO$_2$-containing ice mixtures.

Finally, OCN- was also proposed by \citet{Rocha2024} as one of the carriers of the short-wavelength wing of the 7.67~$\mu$m. In the case of \citet{Rocha2024}, the presence of OCN$^-$ in IRAS2A was confirmed by the band at 4.6~$\mu$m (CN stretch), also due to this negative ion. Unfortunately, there is no data in this range for Ced~110~IRS4A. However, this band has been ubiquitously identified in YSOs, and is likely to be present in this source. OCN$^-$ can be formed via acid-base reactions or ice irradiation \citep[e.g.,][]{Schutte2003}. As discussed by \citet{vanBroekhuizen2005}, an OCN$^-$ abundance of $\sim$1\% is observed for a large number of YSOs. This is consistent OCN$^-$ abundances formed in irradiated ices ($\sim$1.5\% ) with a NH$_3$ ice abundance of around 5\% with respect to H$_2$O ice. These values are aligned with what is found in this work. A better estimation of OCN$^-$ ice in IRS4A will highly benefit of observations at 4.6~$\mu$m. In the case of NH$_4^+$, it is known to be the carrier of the band at 6.85~$\mu$m since CH$_3$OH cannot account for the entire absorption \citet{Bottinelli2010}. Similarly to OCN$^-$, this ion can be formed via acid-base reactions or ice irradiation \citep[e.g.,][]{Schutte2003, Pilling2010}.

A direct comparison between the MIRI data of Ced~110~IRS4A and NIR38 and J110621 is hampered because of the spectral resolution in both cases. However, it is possible to conclude that CH$_4$ share similar chemical conditions in both cases, whereas SO$_2$ and OCN$^-$ would benefit from a higher spectral resolution towards NIR38 and J110621 data to obtain secure conclusions.

\subsubsection{HCOOH}
The best fits of Ced~110~IRS4A indicate that HCOOH may exist in two different chemical environments with different polarity levels, namely, HCOOH:CO$_2$:H$_2$O and HCOOH:CH$_3$OH:H$_2$O, where the former is more prominent. Laboratory experiments from \citet{Qasim2019} showed that HCOOH:CO$_2$ can be formed in both H$_2$O-rich or CO-rich environments via H + HOCO. In the CO-rich case, H$_2$CO may also be formed, which is a precursor of CH$_3$OH. In the H$_2$O-rich case, HCOOH and CO$_2$ are the main products (adsorption of atomic carbon as source of C), with the latter being the most abundant. In that case, it is possible that the Ced~110~IRS4A environment where HCOOH is found is dominated by H$_2$O instead of CO ice. It is worth noting that the amount of HCOOH used to fit the 8.2~$\mu$m feature do not overpredict the C=O stretch at 5.8~$\mu$m (see Fig.~\ref{con_check}).

No detailed spectral fits were made for NIR38 and J110621 in the range of 8~$\mu$m because the MIRI/LRS data suffered from wavelength calibration issues. For this reason, no HCOOH assignment was made for this dataset. Future analysis of the data with correct and improved wavelength calibration will help to address the presence of HCOOH in the molecular cloud. 

\subsubsection{CH$_3$OH and NH$_3$}
CH$_3$OH ice is detected in both sources. Based on the MIRI data of Ced~110~IRS4A, CH$_3$OH probes environments with CO, CO$_2$ and likely H$_2$O ices, where the latter is continuum dependent. All these three conditions are supported via laboratory experiments \citep[e.g.,][]{Fuchs2009, Qasim2018}. In the case of IRS4B, CH$_3$OH is found mixed with CO$_2$ and CO, which is consistent with IRS4A. The non-detection of CH$_3$OH mixed with H$_2$O is due to the lack of quality data at 9.8~$\mu$m. In the case of NH$_3$, the best fits suggest a mixture between NH$_3$ and H$_2$O, which is aligned with the formation pathways of these two species. The other component, NH$_3$:CO$_2$ is discussed in Section~\ref{otherfeatures}.

The chemical environments where CH$_3$OH and NH$_3$ were found are consistent with what is found for NIR38 and J110621. This strongly suggests that the chemical environment of CH$_3$OH ice is set prior to the protostellar stage, in the molecular cloud.

\subsubsection{Other small features}
\label{otherfeatures}
Here, we briefly discuss other species that are needed to achieve the best fit. However, confirming the presence of these particular components in ices will require further systematic studies in a large sample of sources.

\begin{figure}
   \centering
   \includegraphics[width=\hsize]{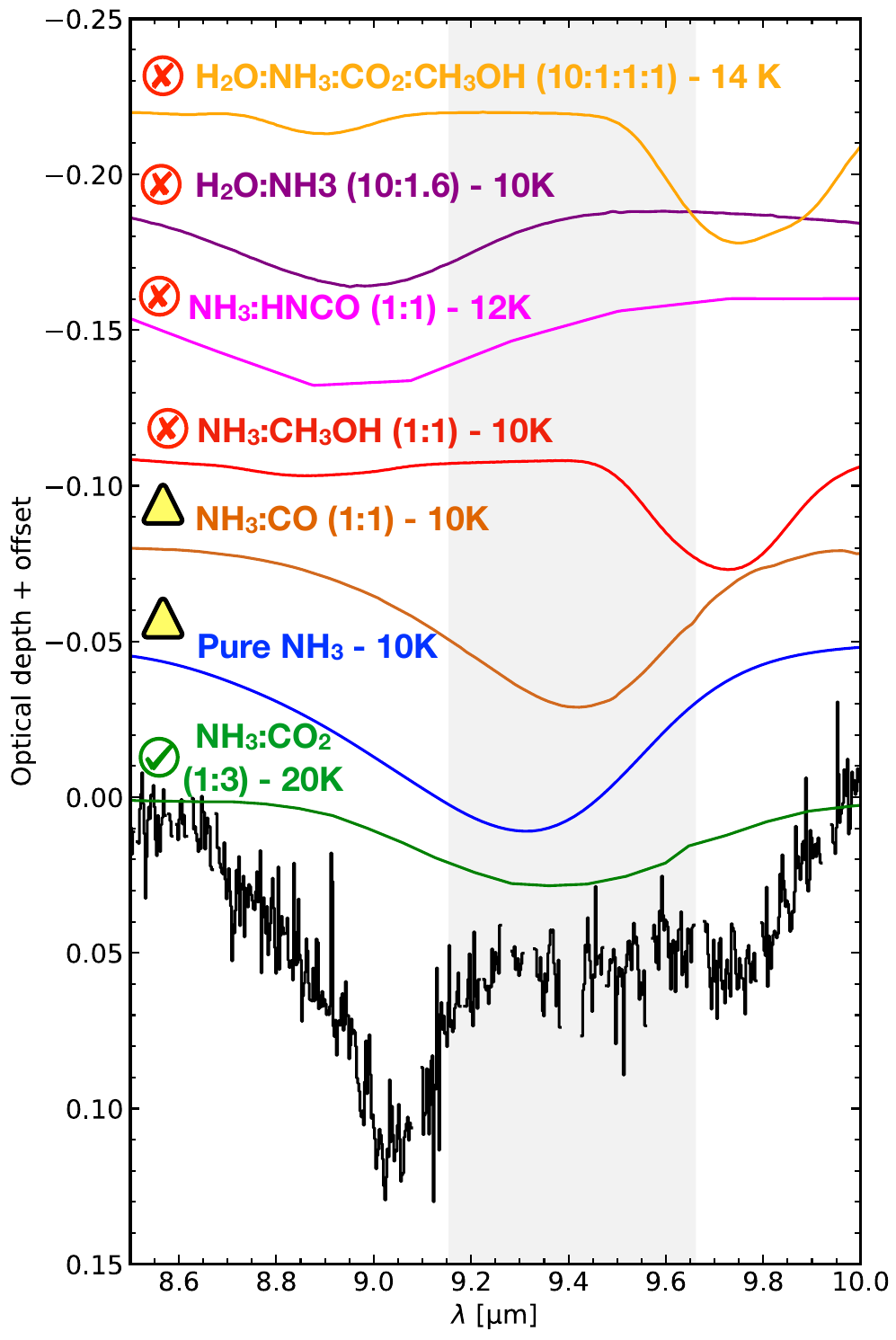}
      \caption{NH$_3$ umbrella mode in different ice mixtures compared with the MIRI spectrum of Ced~110~IRS4A. The best fit of the grey shaded area is indicated with the green check mark and other possible solutions with the yellow triangle. Excluded solutions are given by the red cross.}
         \label{NH3analysis}
   \end{figure}

 NH$_3$:CO$_2$ ice - This ice mixture fits the absorption excess at 9.4~$\mu$m. Such excess was also observed towards NIR38 and J110621 after subtracting the silicate and is not fitted with NH$_3$:H$_2$O, CH$_3$OH:CO or CH$_3$OH:H$_2$O. The ice mixture NH$_3$:CO$_2$ is plausible to be formed in ices as demonstrated in the laboratory \citep[e.g.,][]{Bossa2008, James2021}. However, the presence of this component is highly dependent of the silicate subtraction method. In fact, \citet{Bottinelli2010} noted that the choice of the continuum can affect the shape of the NH$_3$ ice band. To provide a sense of variability of the NH$_3$ band at 9~$\mu$m, Figure~\ref{NH3analysis} compares the observed spectrum with other NH$_3$-containing ices. From the fitting perspective, pure NH$_3$ and NH$_3$:CO also have the ammonia umbrella mode around 9.4~$\mu$m, but their profiles are sharper than in the case of NH$_3$:CO$_2$. A final conclusion on the presence of NH$_3$:CO$_2$ in the ice requires an accurate silicate subtraction, which is not trivial and depends on the dust composition of the source.

N$_2$O ice - Assignments for nitrogen oxides (N$_x$O$_y$) around 7.7~$\mu$m, including N$_2$O, NO$_2$ and NO were proposed in \citet{Grim1989}, and many laboratory experiments have confirmed their formation in interstellar ice analogues \citep[e.g.,][]{Congiu2012, Fedoseev2012, Fedoseev2016M}. Nevertheless, the secure detection of these features in ices remains elusive despite recent tentative detection by \citet{Nazari2024} based on JWST/NIRSpec data of protostars. N$_2$O is often not included in major chemical models. For example, \citet{Agundez2013}, addresses gas-phase formation of NO and NO$_2$, but not N$_2$O. However, N$_2$O ice is suggested as a candidate to explain the gas-phase abundance of NO likely originated from ice sublimation in the dust trap of the planet-forming disk Oph-IRS 48 \citep{Leemker2023}. 

\begin{figure}
   \centering
   \includegraphics[width=\hsize]{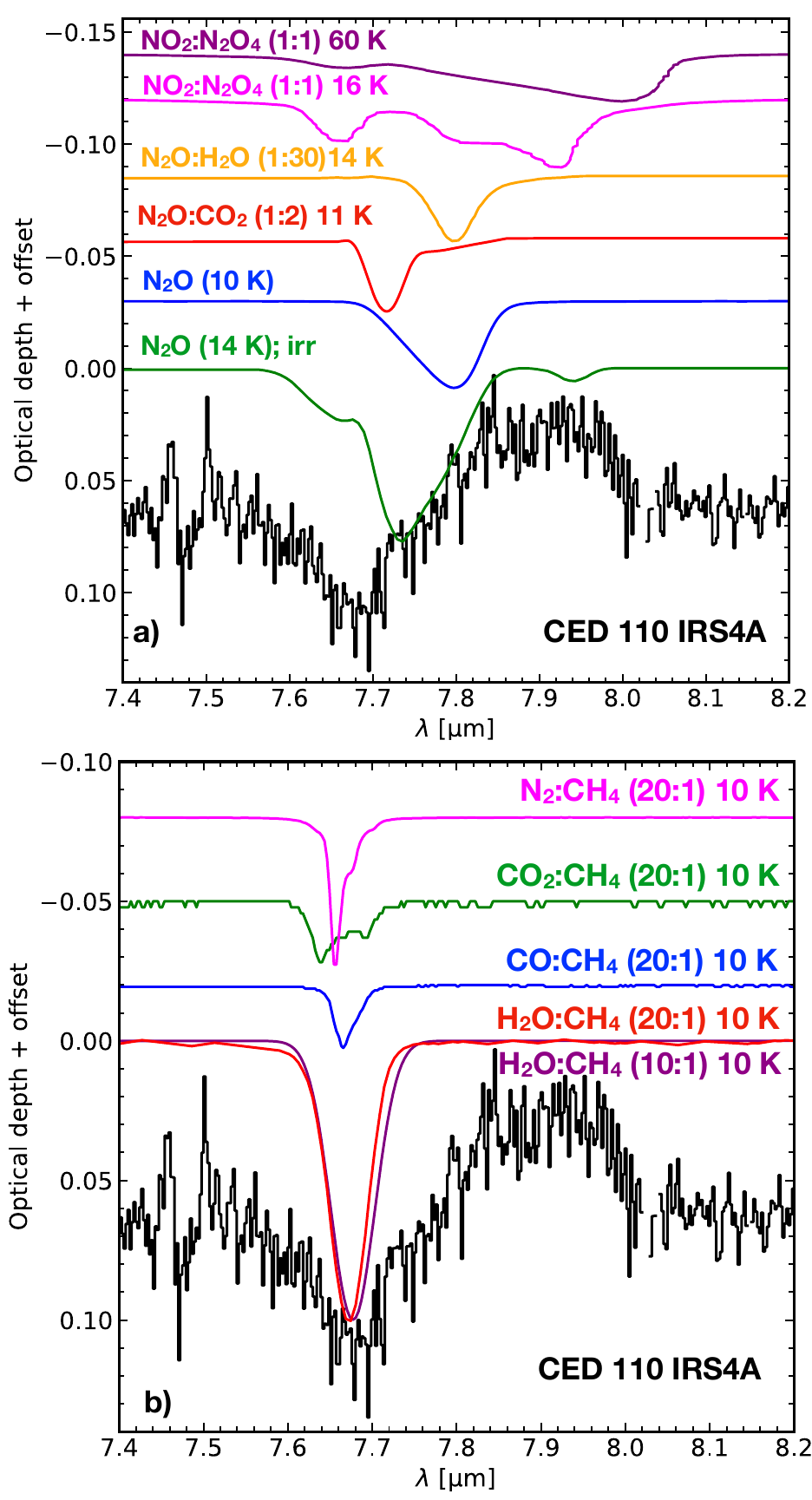}
      \caption{N$_2$O IR ice spectra from different experiments compared with the optical depth spectrum of Ced~110~IRS4A between 7.4 and 8.2~$\mu$m. The panel a displays IR features of nitrogen oxides ice data, whereas the panel b shows the CH$_4$ ice band mixed with other molecules.}
         \label{N2Oanalysis}
   \end{figure}

In Figure~\ref{N2Oanalysis}a we compare different laboratory spectra with the MIRI/MRS data of Ced~110~IRS4A between 7.4 and 8.2~$\mu$m. The N$_2$O IR features are affected by energetic processing and the chemical environment. For example, \citet{Bergantini2022} showed that the spectral profile of the 7.7~$\mu$m band changes upon bombardment with heavy ions ($^{136}$Xe$^{23+}$) of 90 MeV. While the peak position does not change with the irradiation, a prominent red wing arises associated with other nitrogen oxides, which is compatible with the spectral profile seen in Ced~110~IRS4A. A small feature at 7.67~$\mu$m from N$_2$O$_4$ is also formed. Further chemical studies are needed to clarify whether N$_2$O ice can exist in pure form or not. For example, segregation studies could elucidate whether N$_2$O behaves as other molecules that segregate and form small clusters at higher temperatures as is the case of H$_2$O:CO$_2$ \citep{Ehrenfreund1997}. Another example of how the N$_2$O band changes with the chemical environment is seen in the IR spectra of pure N$_2$O ice \citep{Hudson2017bs}, and the mixtures with H$_2$O \citep{Bergantini2022} and CO$_2$ \citep{Pereira2018}. Both pure N$_2$O and N$_2$O:H$_2$O ices have a peak at 7.8~$\mu$m, but the band of the pure ice is broader. In the case of N$_2$O:CO$_2$, the band has a peak at 7.72~$\mu$m. Other two laboratory spectra are compared to the observational data, NO$_2$:N$_2$O$_4$ at 16 and 60~K from \citet{Fulvio2009} and \citet{Fulvio2019}, but no good match with the observation is found. In Figure~\ref{N2Oanalysis}b, we show that only CH$_4$ ice in different mixtures cannot explain the blue and red wings of the 7.67~$\mu$m.

Unidentified feature at 9~$\mu$m - A narrow band is seen at 9.04~$\mu$m in the spectrum of Ced~110~IRS4A. It is unclear if this feature is a consequence of the atypical silicate profile or it is indeed associated with an additional species. In fact, no laboratory data in our list of IR spectra matched this band. Other solid-phase species can be an alternative, such as SiO$_2$ and nanosized silicate dust that peaks at the exact same position and has a narrow profile \citep{Marinoso2021}. However, the corresponding long wavelength bands counterparts are not seen in the MIRI/MRS data. Another candidate for this feature is ammonium sulfate (NH$_4$)$_2$SO$_4$ \citep{Garland2005} that shows narrow strong band at 8.96~$\mu$m. The caveat about this species is that condensation models by \citet{Janssen2023} suggest that although a fraction of sulfur condense to solids, a higher amount remains in the gas-phase. 

\subsection{Ced~110~IRS4 sources and other protostars observed with JWST}

The overall ice inventory with secure detections observed towards Ced~IRS4A and IRS4B is similar to other protostars observed with JWST presented in \citet{Brunken2024, Brunken2024ratio}, \citet{Tyagi2024}, and \citet{Slavicinska2024nh4}. The first major difference compared to other protostellar environments is the absence of both 7.2 and 7.4~$\mu$m bands. These two features are identified towards the hot core and hot corinos, IRAS~2A and IRAS~23385 \citep{Rocha2024} observed with JWST, and many other sources with other telescopes. However, a better comparison should be done with other protostars with a similar age of Ced~110~IRS4 sources. Second, Ced~110~IRS4A shows an atypical silicate profile that has not yet been seen in the sources observed with JWST. Finally, Ced~110~IRS4A show a likely CH$_3^+$ emission around 7.18~$\mu$m. So far, CH$_3^+$ was only detected towards mature protoplanetary disks, such as d203-506 \citep{Berne2023} and TW Hya \citep{Henning2024}. Other differences are motivated by the lack of data or difficulties in the spectral analysis. For example, HDO ice that is detected towards the protostars HOPS~370 and IRAS~20126+4104 \citep{Slavicinska2024hdo} cannot be investigated in the case of IRS4B due to the NIRSpec spectral gap at 4~$\mu$m, whereas for IRS4A no observational data is available. Regarding the presence of complex cyanides \citep{Nazari2024}, the spectral region around 4.4~$\mu$m is strongly blended with CO gas emission lines, and a more detailed line removal is required before the ice analysis of that band.

\section{Conclusions}
\label{sec_conc}

We have presented in this paper the first analysis of the JWST data targeting the binary system Ced~110~IRS4A and IRS4B. \texttt{ENIIGMA} fits were performed on the NIRSpec data of IRS4B and MIRI of IRS4A because of the high quality of the data. Other spectral ranges were mostly used for consistency assessments. Below we summarize the main conclusions found in this work:

\begin{itemize}
    \item Overall, the Ced~110~IRS4A and IRS4B shows the same ice mixtures regarding major species and CH$_3$OH as observed in the Chameleon I molecular cloud towards the background stars NIR38 and J110621, which includes similar ice abundances. This suggests that the chemical composition of protostars is set at earlier stages.

    \item The first major difference between the protostars and the Chameleon I molecular cloud is the absence of the 7.2 and 7.4~$\mu$m features in the protostar. This is not conclusive evidence for the lack of COMs or HCOO$^-$. Instead, it can be related to the low ice column density due to the source inclination. This is supported by the relation between the intensity of the 7.2~$\mu$m band and the H$_2$O ice column density.

    \item The second major difference between the protostars and the Chameleon I molecular cloud is the evidence of thermal processing in the protostar. In the primary source, evidence is traced by the 15.2~$\mu$m band profile of CO$_2$ ice, which indicates distillation or segregation. In the companion source, the CO$_2$ ice stretch mode at 4.27~$\mu$m shows that a warm ($\sim$60$-$80~K) ice component (H$_2$O:CO$_2$ after CO distillation) is needed to provide a good fit to the data. In both cases, CO$_2$ ices are tracing the superimposition of cold ($\sim$10~K) and warm ($\sim$60$-$80~K) regions which is consistent with a temperature gradient along the line of sight to the protostars.

    \item The $^{12}$CO$_2$/$^{13}$CO$_2$ ratio for the companion source ranges between 68 and 88. This is similar to what was found for the two background stars in the same host cloud presented in \citet{McClure2023}.

    \item Tentative ice detections of N$_2$O and NH$_3$:CO$_2$ toward Ced~110~IRS4A were presented. Both cases would contribute to enhancing the N budget observed in protostars. Finally, a narrow unidentified feature at 9~$\mu$m is shown, which does not match with the IR ice spectrum considered in this work.
    
\end{itemize}

The binary system Ced~110~IRS4A and IRS4B has proven to be a challenging protostellar system, and further studies require a sophisticated radiative transfer model to help in the interpretation of this object. In particular, a precise fit of the atypical silicate profile will be highly beneficial to understand the dust and ice evolution in these sources. Additionally, further studies are required in a large sample of sources to investigate ice mixtures that are common towards different objects aiming at unveiling the intricate chemical nature of the ices.

\begin{acknowledgements}
We thank the anonymous referee for the constructive feedback that helped to improve the clarification of this paper. We also acknowledge the following National and International Funding Agencies funded and supported the MIRI development: NASA; ESA; Belgian Science Policy Office (BELSPO); Centre Nationale d’Études Spatiales (CNES); Danish National Space Centre; Deutsches Zentrum fur Luftund Raumfahrt (DLR); Enterprise Ireland; Ministerio De Economiá y Competividad; The Netherlands Research School for Astronomy (NOVA); The Netherlands Organisation for Scientific Research (NWO); Science and Technology Facilities Council; Swiss Space Office; Swedish National Space Agency; and UK Space Agency. NIRSpec was built for the European Space Agency by Airbus Industries; the micro-shutter assembly and detector subsystems were provided by NASA. Dr. Peter Jakobsen guided NIRSpec's development until his retirement in 2011. Dr. Pierre Ferruit is the current NIRSpec PI and ESA JWST project scientist. NIRCam was built by a team at the University of Arizona (UofA) and Lockheed Martin's Advanced Technology Center, led by Prof. Marcia Rieke at UoA. This publication makes use of data products from the Two Micron All Sky Survey, which is a joint project of the University of Massachusetts and the Infrared Processing and Analysis Center/California Institute of Technology, funded by the National Aeronautics and Space Administration and the National Science Foundation. W. R. M. R. thanks the funding from the European Research Council (ERC) under the European Union’s Horizon 2020 research and innovation programme (grant agreement No. 101019751 MOLDISK. W. R. M. R. thanks Christiane Helling and Cornelia Jäger for insightful discussions about the candidates for the narrow band at 9~$\mu$m, and also Alexandre Bergantini de Souza for sharing infrared spectral data of N$_2$O ice. S. B. C. was supported by the NASA Planetary Science Division Internal Scientist Funding Program through the Fundamental Laboratory Research work package (FLaRe). I.J-.S acknowledges partial support from grants No. PID2019-105552RB-C41 and PID2022-136814NB-I00 from the Spanish Ministry of Science and Innovation/State Agency of Research MCIN/AEI/10.13039/501100011033 and by ``ERDF A way of making Europe''. M.N.D. acknowledges the Holcim Foundation Stipend. I.J-.S acknowledges funding from grant PID2022-136814NB-I00 funded by MICIU/AEI/10.13039/501100011033 and by "ERDF/EU". J.A.N. and E.D. acknowledge support from French Programme National `Physique et Chimie du Milieu Interstellaire' (PCMI) of the CNRS/INSU with the INC/INP, co-funded by the CEA and the CNES. H. J. F. acknowledges support for astrochemistry at the Open University from STFC under grant agreements ST/T005424/1, ST/X001164/1, ST/Z510087/1 . Z. L. S. acknowledges support from the Open university for his research PhD studentship and E.A. Milne Travelling Fellowship from the RAS. Part of this research was carried out at the Jet Propulsion Laboratory, California Institute of Technology, under a contract with the National Aeronautics and Space Administration (80NM0018D0004).   
\end{acknowledgements}

\bibliographystyle{aa}
\bibliography{References}

\begin{thebibliography}{148}
\expandafter\ifx\csname natexlab\endcsname\relax\def\natexlab#1{#1}\fi

\bibitem[{{Ag{\'u}ndez} \& {Wakelam}(2013)}]{Agundez2013}
{Ag{\'u}ndez}, M. \& {Wakelam}, V. 2013, Chemical Reviews, 113, 8710

\bibitem[{{Avni} \& {Bahcall}(1980)}]{Avni1980}
{Avni}, Y. \& {Bahcall}, J.~N. 1980, \apj, 235, 694

\bibitem[{{Belloche} {et~al.}(2006){Belloche}, {Parise}, {van der Tak}, {Schilke}, {Leurini}, {G{\"u}sten}, \& {Nyman}}]{Belloche2006}
{Belloche}, A., {Parise}, B., {van der Tak}, F.~F.~S., {et~al.} 2006, \aap, 454, L51

\bibitem[{Bergantini {et~al.}(2022)Bergantini, de~Barros, Toribio, Rothard, Boduch, \& da~Silveira}]{Bergantini2022}
Bergantini, A., de~Barros, A. L.~F., Toribio, N.~N., {et~al.} 2022, The Journal of Physical Chemistry A, 126, 2007, pMID: 35302766

\bibitem[{{Bern{\'e}} {et~al.}(2023){Bern{\'e}}, {Martin-Drumel}, {Schroetter}, {Goicoechea}, {Jacovella}, {Gans}, {Dartois}, {Coudert}, {Bergin}, {Alarcon}, {Cami}, {Roueff}, {Black}, {Asvany}, {Habart}, {Peeters}, {Canin}, {Trahin}, {Joblin}, {Schlemmer}, {Thorwirth}, {Cernicharo}, {Gerin}, {Tielens}, {Zannese}, {Abergel}, {Bernard-Salas}, {Boersma}, {Bron}, {Chown}, {Cuadrado}, {Dicken}, {Elyajouri}, {Fuente}, {Gordon}, {Issa}, {Kannavou}, {Khan}, {Lacinbala}, {Languignon}, {Le Gal}, {Maragkoudakis}, {Meshaka}, {Okada}, {Onaka}, {Pasquini}, {Pound}, {Robberto}, {R{\"o}llig}, {Schefter}, {Schirmer}, {Sidhu}, {Tabone}, {Van De Putte}, {Vicente}, \& {Wolfire}}]{Berne2023}
{Bern{\'e}}, O., {Martin-Drumel}, M.-A., {Schroetter}, I., {et~al.} 2023, \nat, 621, 56

\bibitem[{{Bisschop} {et~al.}(2007){Bisschop}, {Fuchs}, {Boogert}, {van Dishoeck}, \& {Linnartz}}]{Bisschop2007}
{Bisschop}, S.~E., {Fuchs}, G.~W., {Boogert}, A.~C.~A., {van Dishoeck}, E.~F., \& {Linnartz}, H. 2007, \aap, 470, 749

\bibitem[{{Boogert} {et~al.}(2022){Boogert}, {Brewer}, {Brittain}, \& {Emerson}}]{Boogert2022}
{Boogert}, A.~C.~A., {Brewer}, K., {Brittain}, A., \& {Emerson}, K.~S. 2022, \apj, 941, 32

\bibitem[{{Boogert} {et~al.}(2013){Boogert}, {Chiar}, {Knez}, {{\"O}berg}, {Mundy}, {Pendleton}, {Tielens}, \& {van Dishoeck}}]{Boogert2013}
{Boogert}, A.~C.~A., {Chiar}, J.~E., {Knez}, C., {et~al.} 2013, \apj, 777, 73

\bibitem[{{Boogert} {et~al.}(2000{\natexlab{a}}){Boogert}, {Ehrenfreund}, {Gerakines}, {Tielens}, {Whittet}, {Schutte}, {van Dishoeck}, {de Graauw}, {Decin}, \& {Prusti}}]{Boogert200013co2}
{Boogert}, A.~C.~A., {Ehrenfreund}, P., {Gerakines}, P.~A., {et~al.} 2000{\natexlab{a}}, \aap, 353, 349

\bibitem[{{Boogert} {et~al.}(2015){Boogert}, {Gerakines}, \& {Whittet}}]{Boogert2015}
{Boogert}, A.~C.~A., {Gerakines}, P.~A., \& {Whittet}, D. C.~B. 2015, \araa, 53, 541

\bibitem[{{Boogert} {et~al.}(2002){Boogert}, {Hogerheijde}, {Ceccarelli}, {Tielens}, {van Dishoeck}, {Blake}, {Latter}, \& {Motte}}]{Boogert2002}
{Boogert}, A.~C.~A., {Hogerheijde}, M.~R., {Ceccarelli}, C., {et~al.} 2002, \apj, 570, 708

\bibitem[{{Boogert} {et~al.}(2011){Boogert}, {Huard}, {Cook}, {Chiar}, {Knez}, {Decin}, {Blake}, {Tielens}, \& {van Dishoeck}}]{Boogert2011}
{Boogert}, A.~C.~A., {Huard}, T.~L., {Cook}, A.~M., {et~al.} 2011, \apj, 729, 92

\bibitem[{{Boogert} {et~al.}(2008){Boogert}, {Pontoppidan}, {Knez}, {Lahuis}, {Kessler-Silacci}, {van Dishoeck}, {Blake}, {Augereau}, {Bisschop}, {Bottinelli}, {Brooke}, {Brown}, {Crapsi}, {Evans}, {Fraser}, {Geers}, {Huard}, {J{\o}rgensen}, {{\"O}berg}, {Allen}, {Harvey}, {Koerner}, {Mundy}, {Padgett}, {Sargent}, \& {Stapelfeldt}}]{Boogert2008}
{Boogert}, A.~C.~A., {Pontoppidan}, K.~M., {Knez}, C., {et~al.} 2008, \apj, 678, 985

\bibitem[{{Boogert} {et~al.}(1997){Boogert}, {Schutte}, {Helmich}, {Tielens}, \& {Wooden}}]{Boogert1997}
{Boogert}, A.~C.~A., {Schutte}, W.~A., {Helmich}, F.~P., {Tielens}, A.~G.~G.~M., \& {Wooden}, D.~H. 1997, \aap, 317, 929

\bibitem[{{Boogert} {et~al.}(2000{\natexlab{b}}){Boogert}, {Tielens}, {Ceccarelli}, {Boonman}, {van Dishoeck}, {Keane}, {Whittet}, \& {de Graauw}}]{Boogert2000}
{Boogert}, A.~C.~A., {Tielens}, A.~G.~G.~M., {Ceccarelli}, C., {et~al.} 2000{\natexlab{b}}, \aap, 360, 683

\bibitem[{{Bossa} {et~al.}(2008){Bossa}, {Theul{\'e}}, {Duvernay}, {Borget}, \& {Chiavassa}}]{Bossa2008}
{Bossa}, J.~B., {Theul{\'e}}, P., {Duvernay}, F., {Borget}, F., \& {Chiavassa}, T. 2008, \aap, 492, 719

\bibitem[{{Bottinelli} {et~al.}(2010){Bottinelli}, {Boogert}, {Bouwman}, {Beckwith}, {van Dishoeck}, {{\"O}berg}, {Pontoppidan}, {Linnartz}, {Blake}, {Evans}, \& {Lahuis}}]{Bottinelli2010}
{Bottinelli}, S., {Boogert}, A.~C.~A., {Bouwman}, J., {et~al.} 2010, \apj, 718, 1100

\bibitem[{{Bouilloud} {et~al.}(2015){Bouilloud}, {Fray}, {B{\'e}nilan}, {Cottin}, {Gazeau}, \& {Jolly}}]{Bouilloud2015}
{Bouilloud}, M., {Fray}, N., {B{\'e}nilan}, Y., {et~al.} 2015, \mnras, 451, 2145

\bibitem[{{Brunken} {et~al.}(2024{\natexlab{a}}){Brunken}, {Rocha}, {van Dishoeck}, {Gutermuth}, {Tyagi}, {Slavicinska}, {Nazari}, {Megeath}, {Evans II}, {Narang}, {Manoj}, {Rubinstein}, {Watson}, {Looney}, {Linnartz}, {Garatti}, {Beuther}, {Linz}, {Klaassen}, {Poteet}, {Federman}, {Anglada}, {Atnagulov}, {Bourke}, {Fischer}, {Furlan}, {Green}, {Habel}, {Hartmann}, {Karnath}, {Osorio}, {Muzerolle Page}, {Pokhrel}, {Rahatgaonkar}, {Sheehan}, {Stanke}, {Stutz}, {Tobin}, {Tychoniec}, {Wolk}, \& {Yang}}]{Brunken2024}
{Brunken}, N. G.~C., {Rocha}, W. R.~M., {van Dishoeck}, E.~F., {et~al.} 2024{\natexlab{a}}, arXiv e-prints, arXiv:2402.04314

\bibitem[{{Brunken} {et~al.}(2024{\natexlab{b}}){Brunken}, {van Dishoeck}, {Slavicinska}, {le Gouellec}, {Rocha}, {Francis}, {Tychoniec}, {van Gelder}, {Navarro}, {Boogert}, {Kavanagh}, {Nazari}, {Greene}, {Ressler}, \& {Majumdar}}]{Brunken2024ratio}
{Brunken}, N.~G.~C., {van Dishoeck}, E.~F., {Slavicinska}, K., {et~al.} 2024{\natexlab{b}}, arXiv e-prints, arXiv:2409.17237

\bibitem[{Burnham \& Anderson(2002)}]{burnham2002model}
Burnham, K. \& Anderson, D. 2002, Model selection and multimodel inference: a practical information-theoretic approach (Springer Verlag)

\bibitem[{{Bushouse} {et~al.}(2023){Bushouse}, {Eisenhamer}, {Dencheva}, {Davies}, {Greenfield}, {Morrison}, {Hodge}, {Simon}, {Grumm}, {Droettboom}, {Slavich}, {Sosey}, {Pauly}, {Miller}, {Jedrzejewski}, {Hack}, {Davis}, {Crawford}, {Law}, {Gordon}, {Regan}, {Cara}, {MacDonald}, {Bradley}, {Shanahan}, {Jamieson}, {Teodoro}, \& {Williams}}]{Bushouse2023_jwstmiri}
{Bushouse}, H., {Eisenhamer}, J., {Dencheva}, N., {et~al.} 2023, {JWST Calibration Pipeline}

\bibitem[{{Cardelli} {et~al.}(1989){Cardelli}, {Clayton}, \& {Mathis}}]{cardelli1989}
{Cardelli}, J.~A., {Clayton}, G.~C., \& {Mathis}, J.~S. 1989, \apj, 345, 245

\bibitem[{{Carnall}(2017)}]{Carnall2017}
{Carnall}, A.~C. 2017, arXiv e-prints, arXiv:1705.05165

\bibitem[{{Changala} {et~al.}(2023){Changala}, {Chen}, {Le}, {Gans}, {Steenbakkers}, {Salomon}, {Bonah}, {Schroetter}, {Canin}, {Martin-Drumel}, {Jacovella}, {Dartois}, {Boy{\'e}-P{\'e}ronne}, {Alcaraz}, {Asvany}, {Br{\"u}nken}, {Thorwirth}, {Schlemmer}, {Goicoechea}, {Rouill{\'e}}, {Sidhu}, {Chown}, {Van De Putte}, {Trahin}, {Alarc{\'o}n}, {Bern{\'e}}, {Habart}, \& {Peeters}}]{Changala2023}
{Changala}, P.~B., {Chen}, N.~L., {Le}, H.~L., {et~al.} 2023, \aap, 680, A19

\bibitem[{{Chen} {et~al.}(2024){Chen}, {Rocha}, {van Dishoeck}, {van Gelder}, {Nazari}, {Slavicinska}, {Francis}, {Tabone}, {Ressler}, {Klaassen}, {Beuther}, {Boogert}, {Gieser}, {Kavanagh}, {Perotti}, {Le Gouellec}, {Majumdar}, {G{\"u}del}, \& {Henning}}]{Chen2024}
{Chen}, Y., {Rocha}, W.~R.~M., {van Dishoeck}, E.~F., {et~al.} 2024, \aap, 690, A205

\bibitem[{{Chiar} \& {Tielens}(2006)}]{Chiar2006}
{Chiar}, J.~E. \& {Tielens}, A.~G.~G.~M. 2006, \apj, 637, 774

\bibitem[{{Collings} {et~al.}(2004){Collings}, {Anderson}, {Chen}, {Dever}, {Viti}, {Williams}, \& {McCoustra}}]{Collings2004}
{Collings}, M.~P., {Anderson}, M.~A., {Chen}, R., {et~al.} 2004, \mnras, 354, 1133

\bibitem[{{Congiu} {et~al.}(2012){Congiu}, {Fedoseev}, {Ioppolo}, {Dulieu}, {Chaabouni}, {Baouche}, {Lemaire}, {Laffon}, {Parent}, {Lamberts}, {Cuppen}, \& {Linnartz}}]{Congiu2012}
{Congiu}, E., {Fedoseev}, G., {Ioppolo}, S., {et~al.} 2012, \apjl, 750, L12

\bibitem[{{Cuppen} {et~al.}(2011){Cuppen}, {Penteado}, {Isokoski}, {van der Marel}, \& {Linnartz}}]{Cuppen2011}
{Cuppen}, H.~M., {Penteado}, E.~M., {Isokoski}, K., {van der Marel}, N., \& {Linnartz}, H. 2011, \mnras, 417, 2809

\bibitem[{{Dartois} {et~al.}(1999){Dartois}, {Demyk}, {d'Hendecourt}, \& {Ehrenfreund}}]{Dartois1999}
{Dartois}, E., {Demyk}, K., {d'Hendecourt}, L., \& {Ehrenfreund}, P. 1999, \aap, 351, 1066

\bibitem[{{Dartois} \& {d'Hendecourt}(2001)}]{Dartois2001}
{Dartois}, E. \& {d'Hendecourt}, L. 2001, \aap, 365, 144

\bibitem[{{Dartois} {et~al.}(2024){Dartois}, {Noble}, {Caselli}, {Fraser}, {Jim{\'e}nez-Serra}, {Mat{\'e}}, {McClure}, {Melnick}, {Pendleton}, {Shimonishi}, {Smith}, {Sturm}, {Taillard}, {Wakelam}, {Boogert}, {Drozdovskaya}, {Erkal}, {Harsono}, {Herrero}, {Ioppolo}, {Linnartz}, {McGuire}, {Perotti}, {Qasim}, \& {Rocha}}]{Dartois2024}
{Dartois}, E., {Noble}, J.~A., {Caselli}, P., {et~al.} 2024, Nature Astronomy

\bibitem[{Dawes {et~al.}(2016)Dawes, Mason, \& Fraser}]{Dawes2016}
Dawes, A., Mason, N.~J., \& Fraser, H.~J. 2016, Phys. Chem. Chem. Phys., 18, 1245

\bibitem[{{de Graauw} {et~al.}(1996){de Graauw}, {Whittet}, {Gerakines}, {Bauer}, {Beintema}, {Boogert}, {Boxhoorn}, {Chiar}, {Ehrenfreund}, {Feuchtgruber}, {Helmich}, {Heras}, {Huygen}, {Kester}, {Kunze}, {Lahuis}, {Leech}, {Lutz}, {Morris}, {Prusti}, {Roelfsema}, {Salama}, {Schaeidt}, {Schutte}, {Spoon}, {Tielens}, {Valentijn}, {Vandenbusshe}, {van Dishoeck}, {Wesselius}, {Wieprecht}, \& {Wright}}]{deGraauw1996CO2}
{de Graauw}, T., {Whittet}, D.~C.~B., {Gerakines}, P.~A., {et~al.} 1996, \aap, 315, L345

\bibitem[{{Demyk} {et~al.}(1999){Demyk}, {Jones}, {Dartois}, {Cox}, \& {D'Hendecourt}}]{Demyk1999}
{Demyk}, K., {Jones}, A.~P., {Dartois}, E., {Cox}, P., \& {D'Hendecourt}, L. 1999, \aap, 349, 267

\bibitem[{{Do-Duy} {et~al.}(2020){Do-Duy}, {Wright}, {Fujiyoshi}, {Glasse}, {Siebenmorgen}, {Smith}, {Stecklum}, \& {Sterzik}}]{DoDuy2020}
{Do-Duy}, T., {Wright}, C.~M., {Fujiyoshi}, T., {et~al.} 2020, \mnras, 493, 4463

\bibitem[{{Dorschner} {et~al.}(1995){Dorschner}, {Begemann}, {Henning}, {Jaeger}, \& {Mutschke}}]{Dorschner1995}
{Dorschner}, J., {Begemann}, B., {Henning}, T., {Jaeger}, C., \& {Mutschke}, H. 1995, \aap, 300, 503

\bibitem[{{Ehrenfreund} {et~al.}(1997){Ehrenfreund}, {Boogert}, {Gerakines}, {Tielens}, \& {van Dishoeck}}]{Ehrenfreund1997}
{Ehrenfreund}, P., {Boogert}, A.~C.~A., {Gerakines}, P.~A., {Tielens}, A.~G.~G.~M., \& {van Dishoeck}, E.~F. 1997, \aap, 328, 649

\bibitem[{{Ehrenfreund} {et~al.}(1999){Ehrenfreund}, {Kerkhof}, {Schutte}, {Boogert}, {Gerakines}, {Dartois}, {D'Hendecourt}, {Tielens}, {van Dishoeck}, \& {Whittet}}]{Ehrenfreund1999}
{Ehrenfreund}, P., {Kerkhof}, O., {Schutte}, W.~A., {et~al.} 1999, \aap, 350, 240

\bibitem[{Erb(2024)}]{pybaselines}
Erb, D. 2024, {pybaselines}: A {Python} library of algorithms for the baseline correction of experimental data

\bibitem[{{Fedoseev} {et~al.}(2016){Fedoseev}, {Chuang}, {van Dishoeck}, {Ioppolo}, \& {Linnartz}}]{Fedoseev2016M}
{Fedoseev}, G., {Chuang}, K.~J., {van Dishoeck}, E.~F., {Ioppolo}, S., \& {Linnartz}, H. 2016, \mnras, 460, 4297

\bibitem[{{Fedoseev} {et~al.}(2012){Fedoseev}, {Ioppolo}, {Lamberts}, {Zhen}, {Cuppen}, \& {Linnartz}}]{Fedoseev2012}
{Fedoseev}, G., {Ioppolo}, S., {Lamberts}, T., {et~al.} 2012, \jcp, 137, 054714

\bibitem[{{Fogerty} {et~al.}(2016){Fogerty}, {Forrest}, {Watson}, {Sargent}, \& {Koch}}]{Fogerty2016}
{Fogerty}, S., {Forrest}, W., {Watson}, D.~M., {Sargent}, B.~A., \& {Koch}, I. 2016, \apj, 830, 71

\bibitem[{{Fuchs} {et~al.}(2009){Fuchs}, {Cuppen}, {Ioppolo}, {Romanzin}, {Bisschop}, {Andersson}, {van Dishoeck}, \& {Linnartz}}]{Fuchs2009}
{Fuchs}, G.~W., {Cuppen}, H.~M., {Ioppolo}, S., {et~al.} 2009, \aap, 505, 629

\bibitem[{{Fulvio} {et~al.}(2019){Fulvio}, {Baratta}, {Sivaraman}, {Mason}, {da Silveira}, {de Barros}, {Pandoli}, {Strazzulla}, \& {Palumbo}}]{Fulvio2019}
{Fulvio}, D., {Baratta}, G.~A., {Sivaraman}, B., {et~al.} 2019, \mnras, 483, 381

\bibitem[{{Fulvio} {et~al.}(2009){Fulvio}, {Sivaraman}, {Baratta}, {Palumbo}, \& {Mason}}]{Fulvio2009}
{Fulvio}, D., {Sivaraman}, B., {Baratta}, G.~A., {Palumbo}, M.~E., \& {Mason}, N.~J. 2009, Spectrochimica Acta Part A: Molecular Spectroscopy, 72, 1007

\bibitem[{{Furlan} {et~al.}(2008){Furlan}, {McClure}, {Calvet}, {Hartmann}, {D'Alessio}, {Forrest}, {Watson}, {Uchida}, {Sargent}, {Green}, \& {Herter}}]{Furlan2008}
{Furlan}, E., {McClure}, M., {Calvet}, N., {et~al.} 2008, \apjs, 176, 184

\bibitem[{{G{\'a}lvez} {et~al.}(2010){G{\'a}lvez}, {Mat{\'e}}, {Herrero}, \& {Escribano}}]{Galvez2010}
{G{\'a}lvez}, O., {Mat{\'e}}, B., {Herrero}, V.~J., \& {Escribano}, R. 2010, \apj, 724, 539

\bibitem[{Garland {et~al.}(2005)Garland, Wise, Beaver, DeWitt, Aiken, Jimenez, \& Tolbert}]{Garland2005}
Garland, R.~M., Wise, M.~E., Beaver, M.~R., {et~al.} 2005, Atmospheric Chemistry and Physics, 5, 1951

\bibitem[{{Gerakines} {et~al.}(1999){Gerakines}, {Whittet}, {Ehrenfreund}, {Boogert}, {Tielens}, {Schutte}, {Chiar}, {van Dishoeck}, {Prusti}, {Helmich}, \& {de Graauw}}]{Gerakines1999}
{Gerakines}, P.~A., {Whittet}, D.~C.~B., {Ehrenfreund}, P., {et~al.} 1999, \apj, 522, 357

\bibitem[{{Gibb} \& {Whittet}(2002)}]{Gibb2002}
{Gibb}, E.~L. \& {Whittet}, D.~C.~B. 2002, \apjl, 566, L113

\bibitem[{{Gibb} {et~al.}(2004){Gibb}, {Whittet}, {Boogert}, \& {Tielens}}]{Gibb2004}
{Gibb}, E.~L., {Whittet}, D.~C.~B., {Boogert}, A.~C.~A., \& {Tielens}, A.~G.~G.~M. 2004, \apjs, 151, 35

\bibitem[{{Grim} {et~al.}(1989){Grim}, {Greenberg}, {de Groot}, {Baas}, {Schutte}, \& {Schmitt}}]{Grim1989}
{Grim}, R.~J.~A., {Greenberg}, J.~M., {de Groot}, M.~S., {et~al.} 1989, \aaps, 78, 161

\bibitem[{{Grimble} {et~al.}(2024){Grimble}, {Kastner}, {Pinte}, {Sargent}, {Principe}, {Dickson-Vandervelde}, {Aguayo}, {Caceres}, {Schreiber}, \& {Stassun}}]{Grimble2024}
{Grimble}, W., {Kastner}, J., {Pinte}, C., {et~al.} 2024, arXiv e-prints, arXiv:2405.11061

\bibitem[{Guiu {et~al.}(2021)Guiu, Escatllar, \& Bromley}]{Marinoso2021}
Guiu, J.~M., Escatllar, A.~M., \& Bromley, S.~T. 2021, ACS Earth and Space Chemistry, 5, 812

\bibitem[{{Henning} {et~al.}(2024){Henning}, {Kamp}, {Samland}, {Arabhavi}, {Kanwar}, {van Dishoeck}, {G{\"u}del}, {Lagage}, {Waelkens}, {Abergel}, {Absil}, {Barrado}, {Boccaletti}, {Bouwman}, {Caratti o Garatti}, {Geers}, {Glauser}, {Lahuis}, {Mueller}, {Nehm{\'e}}, {Olofsson}, {Pantin}, {Ray}, {Scheithauer}, {Vandenbussche}, {Waters}, {Wright}, {Argyriou}, {Christiaens}, {Franceschi}, {Gasman}, {Grant}, {Guadarrama}, {Jang}, {Morales-Calder{\'o}n}, {Pawellek}, {Perotti}, {Rodgers-Lee}, {Schreiber}, {Schwarz}, {Tabone}, {Temmink}, {Vlasblom}, {Colina}, {Greve}, \& {{\"O}stlin}}]{Henning2024}
{Henning}, T., {Kamp}, I., {Samland}, M., {et~al.} 2024, \pasp, 136, 054302

\bibitem[{{Henning} \& {Semenov}(2013)}]{Henning2013}
{Henning}, T. \& {Semenov}, D. 2013, Chemical Reviews, 113, 9016

\bibitem[{{Hiramatsu} {et~al.}(2007){Hiramatsu}, {Hayakawa}, {Tatematsu}, {Kamegai}, {Onishi}, {Mizuno}, {Yamaguchi}, \& {Hasegawa}}]{Hiramatsu2007}
{Hiramatsu}, M., {Hayakawa}, T., {Tatematsu}, K., {et~al.} 2007, \apj, 664, 964

\bibitem[{Holland(1975)}]{Holland1975}
Holland, J.~H. 1975, Adaptation in natural and artificial systems : an introductory analysis with applications to biology, control, and artificial intelligence (Ann Arbor: University of Michigan Press)

\bibitem[{{Hudgins} {et~al.}(1993){Hudgins}, {Sandford}, {Allamandola}, \& {Tielens}}]{Hudgins1993}
{Hudgins}, D.~M., {Sandford}, S.~A., {Allamandola}, L.~J., \& {Tielens}, A.~G.~G.~M. 1993, \apjs, 86, 713

\bibitem[{{Hudson} {et~al.}(2014{\natexlab{a}}){Hudson}, {Ferrante}, \& {Moore}}]{Hudson2014c2h2}
{Hudson}, R.~L., {Ferrante}, R.~F., \& {Moore}, M.~H. 2014{\natexlab{a}}, \icarus, 228, 276

\bibitem[{{Hudson} \& {Gerakines}(2019)}]{Hudson2019}
{Hudson}, R.~L. \& {Gerakines}, P.~A. 2019, \mnras, 485, 861

\bibitem[{{Hudson} {et~al.}(2014{\natexlab{b}}){Hudson}, {Gerakines}, \& {Moore}}]{Hudson2014eth}
{Hudson}, R.~L., {Gerakines}, P.~A., \& {Moore}, M.~H. 2014{\natexlab{b}}, \icarus, 243, 148

\bibitem[{{Hudson} {et~al.}(2017){Hudson}, {Loeffler}, \& {Gerakines}}]{Hudson2017bs}
{Hudson}, R.~L., {Loeffler}, M.~J., \& {Gerakines}, P.~A. 2017, \jcp, 146, 024304

\bibitem[{{Hudson} {et~al.}(2005){Hudson}, {Moore}, \& {Cook}}]{Hudson2005}
{Hudson}, R.~L., {Moore}, M.~H., \& {Cook}, A.~M. 2005, Advances in Space Research, 36, 184

\bibitem[{James {et~al.}(2021)James, Ioppolo, Hoffmann, Jones, Mason, \& Dawes}]{James2021}
James, R.~L., Ioppolo, S., Hoffmann, S.~V., {et~al.} 2021, RSC Adv., 11, 33055

\bibitem[{{Janssen} {et~al.}(2023){Janssen}, {Woitke}, {Herbort}, {Min}, {Chubb}, {Helling}, \& {Carone}}]{Janssen2023}
{Janssen}, L.~J., {Woitke}, P., {Herbort}, O., {et~al.} 2023, Astronomische Nachrichten, 344, e20230075

\bibitem[{{Juh{\'a}sz} {et~al.}(2010){Juh{\'a}sz}, {Bouwman}, {Henning}, {Acke}, {van den Ancker}, {Meeus}, {Dominik}, {Min}, {Tielens}, \& {Waters}}]{Juhasz2010}
{Juh{\'a}sz}, A., {Bouwman}, J., {Henning}, T., {et~al.} 2010, \apj, 721, 431

\bibitem[{{Keane} {et~al.}(2001){Keane}, {Tielens}, {Boogert}, {Schutte}, \& {Whittet}}]{Keane2001}
{Keane}, J.~V., {Tielens}, A.~G.~G.~M., {Boogert}, A.~C.~A., {Schutte}, W.~A., \& {Whittet}, D.~C.~B. 2001, \aap, 376, 254

\bibitem[{{Kemper} {et~al.}(2004){Kemper}, {Vriend}, \& {Tielens}}]{Kemper2004}
{Kemper}, F., {Vriend}, W.~J., \& {Tielens}, A.~G.~G.~M. 2004, \apj, 609, 826

\bibitem[{{Kim} {et~al.}(2012){Kim}, {Evans}, {Dunham}, {Lee}, \& {Pontoppidan}}]{Kim2012}
{Kim}, H.~J., {Evans}, Neal~J., I., {Dunham}, M.~M., {Lee}, J.-E., \& {Pontoppidan}, K.~M. 2012, \apj, 758, 38

\bibitem[{Koza(1992)}]{Koza1992}
Koza, J.~R. 1992, Genetic programming : on the programming of computers by means of natural selection, Complex adaptive systems. 09800994X (Cambridge, MA [etc.]: The MIT Press)

\bibitem[{{Kristensen} {et~al.}(2012){Kristensen}, {van Dishoeck}, {Bergin}, {Visser}, {Y{\i}ld{\i}z}, {San Jose-Garcia}, {J{\o}rgensen}, {Herczeg}, {Johnstone}, {Wampfler}, {Benz}, {Bruderer}, {Cabrit}, {Caselli}, {Doty}, {Harsono}, {Herpin}, {Hogerheijde}, {Karska}, {van Kempen}, {Liseau}, {Nisini}, {Tafalla}, {van der Tak}, \& {Wyrowski}}]{Kristensen2012}
{Kristensen}, L.~E., {van Dishoeck}, E.~F., {Bergin}, E.~A., {et~al.} 2012, \aap, 542, A8

\bibitem[{{Lacy} {et~al.}(1998){Lacy}, {Faraji}, {Sandford}, \& {Allamandola}}]{Lacy1998}
{Lacy}, J.~H., {Faraji}, H., {Sandford}, S.~A., \& {Allamandola}, L.~J. 1998, \apjl, 501, L105

\bibitem[{{Lahuis} {et~al.}(2010){Lahuis}, {van Dishoeck}, {J{\o}rgensen}, {Blake}, \& {Evans}}]{Lahuis2010}
{Lahuis}, F., {van Dishoeck}, E.~F., {J{\o}rgensen}, J.~K., {Blake}, G.~A., \& {Evans}, N.~J. 2010, \aap, 519, A3

\bibitem[{{Leemker} {et~al.}(2023){Leemker}, {Booth}, {van Dishoeck}, {van der Marel}, {Tabone}, {Ligterink}, {Brunken}, \& {Hogerheijde}}]{Leemker2023}
{Leemker}, M., {Booth}, A.~S., {van Dishoeck}, E.~F., {et~al.} 2023, \aap, 673, A7

\bibitem[{{Lehtinen} {et~al.}(2001){Lehtinen}, {Haikala}, {Mattila}, \& {Lemke}}]{Lehtinen2001}
{Lehtinen}, K., {Haikala}, L.~K., {Mattila}, K., \& {Lemke}, D. 2001, \aap, 367, 311

\bibitem[{{Manoj} {et~al.}(2011){Manoj}, {Kim}, {Furlan}, {McClure}, {Luhman}, {Watson}, {Espaillat}, {Calvet}, {Najita}, {D'Alessio}, {Adame}, {Sargent}, {Forrest}, {Bohac}, {Green}, \& {Arnold}}]{Manoj2011}
{Manoj}, P., {Kim}, K.~H., {Furlan}, E., {et~al.} 2011, \apjs, 193, 11

\bibitem[{{Mathis} {et~al.}(1977){Mathis}, {Rumpl}, \& {Nordsieck}}]{Mathis1977}
{Mathis}, J.~S., {Rumpl}, W., \& {Nordsieck}, K.~H. 1977, \apj, 217, 425

\bibitem[{{McClure} {et~al.}(2017){McClure}, {Bailey}, {Beck}, {Boogert}, {Brown}, {Caselli}, {Chiar}, {Egami}, {Fraser}, {Garrod}, {Gordon}, {Ioppolo}, {Jimenez-Serra}, {Jorgensen}, {Kristensen}, {Linnartz}, {McCoustra}, {Murillo}, {Noble}, {Oberg}, {Palumbo}, {Pendleton}, {Pontoppidan}, {Van Dishoeck}, \& {Viti}}]{McClure2017}
{McClure}, M., {Bailey}, J., {Beck}, T., {et~al.} 2017, {IceAge: Chemical Evolution of Ices during Star Formation}, JWST Proposal ID 1309. Cycle 0 Early Release Science

\bibitem[{{McClure} {et~al.}(2023){McClure}, {Rocha}, {Pontoppidan}, {Crouzet}, {Chu}, {Dartois}, {Lamberts}, {Noble}, {Pendleton}, {Perotti}, {Qasim}, {Rachid}, {Smith}, {Sun}, {Beck}, {Boogert}, {Brown}, {Caselli}, {Charnley}, {Cuppen}, {Dickinson}, {Drozdovskaya}, {Egami}, {Erkal}, {Fraser}, {Garrod}, {Harsono}, {Ioppolo}, {Jim{\'e}nez-Serra}, {Jin}, {J{\o}rgensen}, {Kristensen}, {Lis}, {McCoustra}, {McGuire}, {Melnick}, {{\~A}-berg}, {Palumbo}, {Shimonishi}, {Sturm}, {van Dishoeck}, \& {Linnartz}}]{McClure2023}
{McClure}, M.~K., {Rocha}, W.~R.~M., {Pontoppidan}, K.~M., {et~al.} 2023, Nature Astronomy, 7, 431

\bibitem[{{Moore} {et~al.}(2007){Moore}, {Ferrante}, {Hudson}, \& {Stone}}]{Moore2007}
{Moore}, M.~H., {Ferrante}, R.~F., {Hudson}, R.~L., \& {Stone}, J.~N. 2007, \icarus, 190, 260

\bibitem[{{Nazari} {et~al.}(2024{\natexlab{a}}){Nazari}, {Rocha}, {Rubinstein}, {Slavicinska}, {Rachid}, {van Dishoeck}, {Megeath}, {Gutermuth}, {Tyagi}, {Brunken}, {Narang}, {Manoj}, {Watson}, {Evans}, {Federman}, {Muzerolle Page}, {Anglada}, {Beuther}, {Klaassen}, {Looney}, {Osorio}, {Stanke}, \& {Yang}}]{Nazari2024}
{Nazari}, P., {Rocha}, W.~R.~M., {Rubinstein}, A.~E., {et~al.} 2024{\natexlab{a}}, \aap, 686, A71

\bibitem[{{Nazari} {et~al.}(2023){Nazari}, {Tabone}, \& {Rosotti}}]{Nazari2023}
{Nazari}, P., {Tabone}, B., \& {Rosotti}, G.~P. 2023, \aap, 671, A107

\bibitem[{{Nazari} {et~al.}(2024{\natexlab{b}}){Nazari}, {Tabone}, {Rosotti}, \& {van Dishoeck}}]{Nazari2024gas}
{Nazari}, P., {Tabone}, B., {Rosotti}, G.~P., \& {van Dishoeck}, E.~F. 2024{\natexlab{b}}, arXiv e-prints, arXiv:2404.10045

\bibitem[{{Nazari} {et~al.}(2022){Nazari}, {Tabone}, {Rosotti}, {van Gelder}, {Meshaka}, \& {van Dishoeck}}]{Nazari2022}
{Nazari}, P., {Tabone}, B., {Rosotti}, G.~P., {et~al.} 2022, \aap, 663, A58

\bibitem[{{Noble} {et~al.}(2024){Noble}, {Fraser}, {Smith}, {Dartois}, {Boogert}, {Cuppen}, {Dickinson}, {Dulieu}, {Egami}, {Erkal}, {Giuliano}, {Husquinet}, {Lamberts}, {Maté}, {McClure}, {Palumbo}, {Shimonishi}, {Sun}, {Bergner}, {Brown}, {Caselli}, {Congiu}, {Drozdovskaya}, {Herrero}, {Ioppolo}, {Jimenez-Serra}, {Linnartz}, {Melnick}, {McGuire}, {{\"O}berg}, {Perotti}, {Qasim}, {Rocha}, \& {Urso}}]{Noble2024}
{Noble}, J.~A., {Fraser}, H.~J., {Smith}, Z.~L., {et~al.} 2024, Nature Astronomy

\bibitem[{{Noriega-Crespo} {et~al.}(2004){Noriega-Crespo}, {Morris}, {Marleau}, {Carey}, {Boogert}, {van Dishoeck}, {Evans}, {Keene}, {Muzerolle}, {Stapelfeldt}, {Pontoppidan}, {Lowrance}, {Allen}, \& {Bourke}}]{NoriegaCrespo2004}
{Noriega-Crespo}, A., {Morris}, P., {Marleau}, F.~R., {et~al.} 2004, \apjs, 154, 352

\bibitem[{{Novozamsky} {et~al.}(2001){Novozamsky}, {Schutte}, \& {Keane}}]{Novozamsky2001}
{Novozamsky}, J.~H., {Schutte}, W.~A., \& {Keane}, J.~V. 2001, \aap, 379, 588

\bibitem[{{{\"O}berg} {et~al.}(2008){{\"O}berg}, {Boogert}, {Pontoppidan}, {Blake}, {Evans}, {Lahuis}, \& {van Dishoeck}}]{Oberg2008}
{{\"O}berg}, K.~I., {Boogert}, A.~C.~A., {Pontoppidan}, K.~M., {et~al.} 2008, \apj, 678, 1032

\bibitem[{{{\"O}berg} {et~al.}(2011){{\"O}berg}, {Boogert}, {Pontoppidan}, {van den Broek}, {van Dishoeck}, {Bottinelli}, {Blake}, \& {Evans}}]{Oberg2011}
{{\"O}berg}, K.~I., {Boogert}, A.~C.~A., {Pontoppidan}, K.~M., {et~al.} 2011, \apj, 740, 109

\bibitem[{{{\"O}berg} {et~al.}(2007){{\"O}berg}, {Fraser}, {Boogert}, {Bisschop}, {Fuchs}, {van Dishoeck}, \& {Linnartz}}]{Oberg2007}
{{\"O}berg}, K.~I., {Fraser}, H.~J., {Boogert}, A.~C.~A., {et~al.} 2007, \aap, 462, 1187

\bibitem[{{Ohashi} {et~al.}(2023){Ohashi}, {Tobin}, {J{\o}rgensen}, {Takakuwa}, {Sheehan}, {Aikawa}, {Li}, {Looney}, {Williams}, {Aso}, {Sharma}, {Sai}, {Yamato}, {Lee}, {Tomida}, {Yen}, {Encalada}, {Flores}, {Gavino}, {Kido}, {Han}, {Lin}, {Narayanan}, {Phuong}, {Santamar{\'\i}a-Miranda}, {Thieme}, {van't Hoff}, {de Gregorio-Monsalvo}, {Koch}, {Kwon}, {Lai}, {Lee}, {Plunkett}, {Saigo}, {Hirano}, {Lam}, \& {Mori}}]{Ohashi2023}
{Ohashi}, N., {Tobin}, J.~J., {J{\o}rgensen}, J.~K., {et~al.} 2023, \apj, 951, 8

\bibitem[{{Olofsson} {et~al.}(2009){Olofsson}, {Augereau}, {van Dishoeck}, {Mer{\'\i}n}, {Lahuis}, {Kessler-Silacci}, {Dullemond}, {Oliveira}, {Blake}, {Boogert}, {Brown}, {Evans}, {Geers}, {Knez}, {Monin}, \& {Pontoppidan}}]{Olofsson2009}
{Olofsson}, J., {Augereau}, J.~C., {van Dishoeck}, E.~F., {et~al.} 2009, \aap, 507, 327

\bibitem[{{Pereira} {et~al.}(2018){Pereira}, {de Barros}, {Fulvio}, {Boduch}, {Rothard}, \& {da Silveira}}]{Pereira2018}
{Pereira}, R.~C., {de Barros}, A.~L.~F., {Fulvio}, D., {et~al.} 2018, \mnras, 478, 4939

\bibitem[{{Persi} {et~al.}(2001){Persi}, {Marenzi}, {G{\'o}mez}, \& {Olofsson}}]{Persi2001}
{Persi}, P., {Marenzi}, A.~R., {G{\'o}mez}, M., \& {Olofsson}, G. 2001, \aap, 376, 907

\bibitem[{{Pilling} {et~al.}(2010){Pilling}, {Seperuelo Duarte}, {Domaracka}, {Rothard}, {Boduch}, \& {da Silveira}}]{Pilling2010}
{Pilling}, S., {Seperuelo Duarte}, E., {Domaracka}, A., {et~al.} 2010, \aap, 523, A77

\bibitem[{{Pontoppidan} {et~al.}(2008){Pontoppidan}, {Boogert}, {Fraser}, {van Dishoeck}, {Blake}, {Lahuis}, {{\"O}berg}, {Evans}, \& {Salyk}}]{Pontoppidan2008}
{Pontoppidan}, K.~M., {Boogert}, A.~C.~A., {Fraser}, H.~J., {et~al.} 2008, \apj, 678, 1005

\bibitem[{{Pontoppidan} \& {Dullemond}(2005)}]{Pontoppidan2005circ}
{Pontoppidan}, K.~M. \& {Dullemond}, C.~P. 2005, \aap, 435, 595

\bibitem[{{Pontoppidan} {et~al.}(2024){Pontoppidan}, {Evans}, {Bergner}, \& {Yang}}]{pontoppidan2024}
{Pontoppidan}, K.~M., {Evans}, N., {Bergner}, J., \& {Yang}, Y.-L. 2024, Research Notes of the American Astronomical Society, 8, 68

\bibitem[{{Pontoppidan} {et~al.}(2003){Pontoppidan}, {Fraser}, {Dartois}, {Thi}, {van Dishoeck}, {Boogert}, {d'Hendecourt}, {Tielens}, \& {Bisschop}}]{Pontoppidan2003}
{Pontoppidan}, K.~M., {Fraser}, H.~J., {Dartois}, E., {et~al.} 2003, \aap, 408, 981

\bibitem[{{Poteet} {et~al.}(2011){Poteet}, {Megeath}, {Watson}, {Calvet}, {Remming}, {McClure}, {Sargent}, {Fischer}, {Furlan}, {Allen}, {Bjorkman}, {Hartmann}, {Muzerolle}, {Tobin}, \& {Ali}}]{Poteet2011}
{Poteet}, C.~A., {Megeath}, S.~T., {Watson}, D.~M., {et~al.} 2011, \apjl, 733, L32

\bibitem[{{Poteet} {et~al.}(2013){Poteet}, {Pontoppidan}, {Megeath}, {Watson}, {Isokoski}, {Bjorkman}, {Sheehan}, \& {Linnartz}}]{Poteet2013}
{Poteet}, C.~A., {Pontoppidan}, K.~M., {Megeath}, S.~T., {et~al.} 2013, \apj, 766, 117

\bibitem[{{Qasim} {et~al.}(2018){Qasim}, {Chuang}, {Fedoseev}, {Ioppolo}, {Boogert}, \& {Linnartz}}]{Qasim2018}
{Qasim}, D., {Chuang}, K.~J., {Fedoseev}, G., {et~al.} 2018, \aap, 612, A83

\bibitem[{{Qasim} {et~al.}(2019){Qasim}, {Lamberts}, {He}, {Chuang}, {Fedoseev}, {Ioppolo}, {Boogert}, \& {Linnartz}}]{Qasim2019}
{Qasim}, D., {Lamberts}, T., {He}, J., {et~al.} 2019, \aap, 626, A118

\bibitem[{{Rachid} {et~al.}(2021){Rachid}, {Brunken}, {de Boe}, {Fedoseev}, {Boogert}, \& {Linnartz}}]{Rachid2021}
{Rachid}, M.~G., {Brunken}, N., {de Boe}, D., {et~al.} 2021, \aap, 653, A116

\bibitem[{{Rachid} {et~al.}(2022){Rachid}, {Rocha}, \& {Linnartz}}]{Rachid2022}
{Rachid}, M.~G., {Rocha}, W., \& {Linnartz}, H. 2022, arXiv e-prints, arXiv:2207.12502

\bibitem[{{Rachid} {et~al.}(2020){Rachid}, {Terwisscha van Scheltinga}, {Koletzki}, \& {Linnartz}}]{Rachid2020}
{Rachid}, M.~G., {Terwisscha van Scheltinga}, J., {Koletzki}, D., \& {Linnartz}, H. 2020, \aap, 639, A4

\bibitem[{{Rocha} {et~al.}(2021){Rocha}, {Perotti}, {Kristensen}, \& {J{\o}rgensen}}]{Rocha2021}
{Rocha}, W.~R.~M., {Perotti}, G., {Kristensen}, L.~E., \& {J{\o}rgensen}, J.~K. 2021, \aap, 654, A158

\bibitem[{{Rocha} \& {Pilling}(2015)}]{Rocha2015}
{Rocha}, W.~R.~M. \& {Pilling}, S. 2015, \apj, 803, 18

\bibitem[{{Rocha} {et~al.}(2017){Rocha}, {Pilling}, {de Barros}, {Andrade}, {Rothard}, \& {Boduch}}]{Rocha2017}
{Rocha}, W.~R.~M., {Pilling}, S., {de Barros}, A.~L.~F., {et~al.} 2017, \mnras, 464, 754

\bibitem[{{Rocha} {et~al.}(2020){Rocha}, {Pilling}, {Domaracka}, {Rothard}, \& {Boduch}}]{Rocha2020}
{Rocha}, W.~R.~M., {Pilling}, S., {Domaracka}, A., {Rothard}, H., \& {Boduch}, P. 2020, Spectrochimica Acta Part A: Molecular Spectroscopy, 228, 117826

\bibitem[{{Rocha} {et~al.}(2024{\natexlab{a}}){Rocha}, {Rachid}, {McClure}, {He}, \& {Linnartz}}]{Rocha2024_ice}
{Rocha}, W.~R.~M., {Rachid}, M.~G., {McClure}, M.~K., {He}, J., \& {Linnartz}, H. 2024{\natexlab{a}}, \aap, 681, A9

\bibitem[{{Rocha} {et~al.}(2022){Rocha}, {Rachid}, {Olsthoorn}, {van Dishoeck}, {McClure}, \& {Linnartz}}]{Rocha2022}
{Rocha}, W.~R.~M., {Rachid}, M.~G., {Olsthoorn}, B., {et~al.} 2022, \aap, 668, A63

\bibitem[{{Rocha} {et~al.}(2024{\natexlab{b}}){Rocha}, {van Dishoeck}, {Ressler}, {van Gelder}, {Slavicinska}, {Brunken}, {Linnartz}, {Ray}, {Beuther}, {Caratti o Garatti}, {Geers}, {Kavanagh}, {Klaassen}, {Justtanont}, {Chen}, {Francis}, {Gieser}, {Perotti}, {Tychoniec}, {Barsony}, {Majumdar}, {le Gouellec}, {Chu}, {Lew}, {Henning}, \& {Wright}}]{Rocha2024}
{Rocha}, W.~R.~M., {van Dishoeck}, E.~F., {Ressler}, M.~E., {et~al.} 2024{\natexlab{b}}, \aap, 683, A124

\bibitem[{{Rubin} {et~al.}(2019){Rubin}, {Altwegg}, {Balsiger}, {Berthelier}, {Combi}, {De Keyser}, {Drozdovskaya}, {Fiethe}, {Fuselier}, {Gasc}, {Gombosi}, {H{\"a}nni}, {Hansen}, {Mall}, {R{\`e}me}, {Schroeder}, {Schuhmann}, {S{\'e}mon}, {Waite}, {Wampfler}, \& {Wurz}}]{Rubin2019}
{Rubin}, M., {Altwegg}, K., {Balsiger}, H., {et~al.} 2019, \mnras, 489, 594

\bibitem[{{Rubinstein} {et~al.}(2023){Rubinstein}, {Tyagi}, {Nazari}, {Gutermuth}, {Federman}, {Narang}, {Rocha}, {Brunken}, {Slavicinska}, {Evans}, {Green}, {Watson}, {Beuther}, {Bourke}, {Garatti}, {Hartmann}, {Klaassen}, {Linz}, {Looney}, {Manoj}, {Megeath}, {Muzerolle Page}, {Stanke}, {Tobin}, {van Dishoeck}, {Wolk}, \& {Yang}}]{Rubinstein2023}
{Rubinstein}, A.~E., {Tyagi}, H., {Nazari}, P., {et~al.} 2023, arXiv e-prints, arXiv:2312.07807

\bibitem[{{Sai} {et~al.}(2023){Sai}, {Yen}, {Ohashi}, {Tobin}, {J{\o}rgensen}, {Takakuwa}, {Saigo}, {Aso}, {Lin}, {Koch}, {Aikawa}, {Flores}, {de Gregorio-Monsalvo}, {Han}, {Kido}, {Kwon}, {Lai}, {Lee}, {Lee}, {Li}, {Looney}, {Mori}, {Phuong}, {Santamar{\'\i}a-Miranda}, {Sharma}, {Thieme}, {Tomida}, \& {Williams}}]{Sai2023}
{Sai}, J., {Yen}, H.-W., {Ohashi}, N., {et~al.} 2023, \apj, 954, 67

\bibitem[{{Sandford} {et~al.}(1988){Sandford}, {Allamandola}, {Tielens}, \& {Valero}}]{Sandford1988}
{Sandford}, S.~A., {Allamandola}, L.~J., {Tielens}, A.~G.~G.~M., \& {Valero}, G.~J. 1988, \apj, 329, 498

\bibitem[{{Sargent} {et~al.}(2009){Sargent}, {Forrest}, {Tayrien}, {McClure}, {Watson}, {Sloan}, {Li}, {Manoj}, {Bohac}, {Furlan}, {Kim}, \& {Green}}]{Sargent2009}
{Sargent}, B.~A., {Forrest}, W.~J., {Tayrien}, C., {et~al.} 2009, \apjs, 182, 477

\bibitem[{{Schutte}(1999)}]{Schutte1999}
{Schutte}, W.~A. 1999, in NATO Advanced Study Institute (ASI) Series C, Vol. 523, Formation and Evolution of Solids in Space, ed. J.~M. {Greenberg} \& A.~{Li}, 177

\bibitem[{{Schutte} {et~al.}(1993){Schutte}, {Allamandola}, \& {Sandford}}]{Schutte1993}
{Schutte}, W.~A., {Allamandola}, L.~J., \& {Sandford}, S.~A. 1993, \icarus, 104, 118

\bibitem[{{Schutte} {et~al.}(1999){Schutte}, {Boogert}, {Tielens}, {Whittet}, {Gerakines}, {Chiar}, {Ehrenfreund}, {Greenberg}, {van Dishoeck}, \& {de Graauw}}]{Schutte1999_weak}
{Schutte}, W.~A., {Boogert}, A.~C.~A., {Tielens}, A.~G.~G.~M., {et~al.} 1999, \aap, 343, 966

\bibitem[{{Schutte} \& {Khanna}(2003)}]{Schutte2003}
{Schutte}, W.~A. \& {Khanna}, R.~K. 2003, \aap, 398, 1049

\bibitem[{{Skrutskie} {et~al.}(2006){Skrutskie}, {Cutri}, {Stiening}, {Weinberg}, {Schneider}, {Carpenter}, {Beichman}, {Capps}, {Chester}, {Elias}, {Huchra}, {Liebert}, {Lonsdale}, {Monet}, {Price}, {Seitzer}, {Jarrett}, {Kirkpatrick}, {Gizis}, {Howard}, {Evans}, {Fowler}, {Fullmer}, {Hurt}, {Light}, {Kopan}, {Marsh}, {McCallon}, {Tam}, {Van Dyk}, \& {Wheelock}}]{Skrutskie2006}
{Skrutskie}, M.~F., {Cutri}, R.~M., {Stiening}, R., {et~al.} 2006, \aj, 131, 1163

\bibitem[{{Slavicinska} {et~al.}(2024{\natexlab{a}}){Slavicinska}, {Boogert}, {Tychoniec}, {van Dishoeck}, {van Gelder}, {Santos}, {Klaassen}, {Kavanagh}, \& {Chuang}}]{Slavicinska2024nh4}
{Slavicinska}, K., {Boogert}, A., {Tychoniec}, {\L}., {et~al.} 2024{\natexlab{a}}, arXiv e-prints, arXiv:2410.02860

\bibitem[{{Slavicinska} {et~al.}(2023){Slavicinska}, {Rachid}, {Rocha}, {Chuang}, {van Dishoeck}, \& {Linnartz}}]{Slavicinska2023}
{Slavicinska}, K., {Rachid}, M.~G., {Rocha}, W.~R.~M., {et~al.} 2023, \aap, 677, A13

\bibitem[{{Slavicinska} {et~al.}(2024{\natexlab{b}}){Slavicinska}, {van Dishoeck}, {Tychoniec}, {Nazari}, {Rubinstein}, {Gutermuth}, {Tyagi}, {Chen}, {Brunken}, {Rocha}, {Manoj}, {Narang}, {Thomas Megeath}, {Yang}, {Looney}, {Tobin}, {Beuther}, {Bourke}, {Linnartz}, {Federman}, {Watson}, \& {Linz}}]{Slavicinska2024hdo}
{Slavicinska}, K., {van Dishoeck}, E.~F., {Tychoniec}, {\L}., {et~al.} 2024{\natexlab{b}}, \aap, 688, A29

\bibitem[{{Smith} {et~al.}(2024){Smith}, {Dickinson}, {Fraser}, \& {McClure}}]{Smith2024sub}
{Smith}, Z.~L., {Dickinson}, H.~J., {Fraser}, H.~J., \& {McClure}, M.~K. 2024, Nat. Astr., submitted

\bibitem[{{Sturm} {et~al.}(2023){Sturm}, {McClure}, {Beck}, {Harsono}, {Bergner}, {Dartois}, {Boogert}, {Chiar}, {Cordiner}, {Drozdovskaya}, {Ioppolo}, {Law}, {Linnartz}, {Lis}, {Melnick}, {McGuire}, {Noble}, {{\"O}berg}, {Palumbo}, {Pendleton}, {Perotti}, {Pontoppidan}, {Qasim}, {Rocha}, {Terada}, {Urso}, \& {van Dishoeck}}]{Sturm2023}
{Sturm}, J.~A., {McClure}, M.~K., {Beck}, T.~L., {et~al.} 2023, \aap, 679, A138

\bibitem[{{Sun} {et~al.}(2023){Sun}, {Egami}, {Pirzkal}, {Rieke}, {Baum}, {Boyer}, {Boyett}, {Bunker}, {Cameron}, {Curti}, {Eisenstein}, {Gennaro}, {Greene}, {Jaffe}, {Kelly}, {Koekemoer}, {Kumari}, {Maiolino}, {Maseda}, {Perna}, {Rest}, {Robertson}, {Schlawin}, {Smit}, {Stansberry}, {Sunnquist}, {Tacchella}, {Williams}, \& {Willmer}}]{Sun2023}
{Sun}, F., {Egami}, E., {Pirzkal}, N., {et~al.} 2023, \apj, 953, 53

\bibitem[{{Sun} {et~al.}(2022){Sun}, {Egami}, {Pirzkal}, {Rieke}, {Boyer}, {Correnti}, {Gennaro}, {Girard}, {Greene}, {Kelly}, {Koekemoer}, {Leisenring}, {Misselt}, {Nikolov}, {Roellig}, {Stansberry}, {Williams}, {Willmer}, \& {Members of the JWST/NIRCam Commissioning Team}}]{Sun2022}
{Sun}, F., {Egami}, E., {Pirzkal}, N., {et~al.} 2022, \apjl, 936, L8

\bibitem[{{Terwisscha van Scheltinga} {et~al.}(2018){Terwisscha van Scheltinga}, {Ligterink}, {Boogert}, {van Dishoeck}, \& {Linnartz}}]{Scheltinga2018}
{Terwisscha van Scheltinga}, J., {Ligterink}, N.~F.~W., {Boogert}, A.~C.~A., {van Dishoeck}, E.~F., \& {Linnartz}, H. 2018, \aap, 611, A35

\bibitem[{{Terwisscha van Scheltinga} {et~al.}(2021){Terwisscha van Scheltinga}, {Marcandalli}, {McClure}, {Hogerheijde}, \& {Linnartz}}]{Scheltinga2021}
{Terwisscha van Scheltinga}, J., {Marcandalli}, G., {McClure}, M.~K., {Hogerheijde}, M.~R., \& {Linnartz}, H. 2021, arXiv e-prints, arXiv:2105.02226

\bibitem[{{Tielens} {et~al.}(1991){Tielens}, {Tokunaga}, {Geballe}, \& {Baas}}]{Tielens1991}
{Tielens}, A.~G.~G.~M., {Tokunaga}, A.~T., {Geballe}, T.~R., \& {Baas}, F. 1991, \apj, 381, 181

\bibitem[{{Tyagi} {et~al.}(2024){Tyagi}, {Manoj}, {Narang}, {Megeath}, {Rocha}, {Brunken}, {Rubinstein}, {Gutermuth}, {Evans}, {van Dishoeck}, {Federman}, {Watson}, {Neufeld}, {Anglada}, {Beuther}, {Garatti}, {Looney}, {Nazari}, {Osorio}, {Stanke}, {Yang}, {Bourke}, {Fischer}, {Furlan}, {Green}, {Habel}, {Klaassen}, {Karnath}, {Linz}, {Muzzerolle}, {Tobin}, {Atnagulov}, {Rahatgaonkar}, {Sheehan}, {Slavicinska}, {Stutz}, {Tychoniec}, \& {Wolk}}]{Tyagi2024}
{Tyagi}, H., {Manoj}, P., {Narang}, M., {et~al.} 2024, arXiv e-prints, arXiv:2410.06697

\bibitem[{{van Broekhuizen} {et~al.}(2005){van Broekhuizen}, {Pontoppidan}, {Fraser}, \& {van Dishoeck}}]{vanBroekhuizen2005}
{van Broekhuizen}, F.~A., {Pontoppidan}, K.~M., {Fraser}, H.~J., \& {van Dishoeck}, E.~F. 2005, \aap, 441, 249

\bibitem[{{van Gelder}(2022)}]{vgelder2022PhD}
{van Gelder}, M. 2022, PhD thesis, Leiden Observatory

\bibitem[{{van Gelder} {et~al.}(2024){van Gelder}, {Ressler, M. E.}, {van Dishoeck, E. F.}, {Nazari, P.}, {Tabone, B.}, {Black, J. H.}, {Tychoniec, Ł.}, {Francis, L.}, {Barsony, M.}, {Beuther, H.}, {Caratti o Garatti, A.}, {Chen, Y.}, {Gieser, C.}, {le Gouellec, V. J. M.}, {Kavanagh, P. J.}, {Klaassen, P. D.}, {Lew, B. W. P.}, {Linnartz, H.}, {Majumdar, L.}, {Perotti, G.}, \& {Rocha, W. R. M.}}]{vangelder2024}
{van Gelder}, M.~L., {Ressler, M. E.}, {van Dishoeck, E. F.}, {et~al.} 2024, A\&A, 682, A78

\bibitem[{{van Kempen} {et~al.}(2009){van Kempen}, {van Dishoeck}, {Hogerheijde}, \& {G{\"u}sten}}]{vankempen2009}
{van Kempen}, T.~A., {van Dishoeck}, E.~F., {Hogerheijde}, M.~R., \& {G{\"u}sten}, R. 2009, \aap, 508, 259

\bibitem[{{Woitke} {et~al.}(2016){Woitke}, {Min}, {Pinte}, {Thi}, {Kamp}, {Rab}, {Anthonioz}, {Antonellini}, {Baldovin-Saavedra}, {Carmona}, {Dominik}, {Dionatos}, {Greaves}, {G{\"u}del}, {Ilee}, {Liebhart}, {M{\'e}nard}, {Rigon}, {Waters}, {Aresu}, {Meijerink}, \& {Spaans}}]{Woitke2016}
{Woitke}, P., {Min}, M., {Pinte}, C., {et~al.} 2016, \aap, 586, A103

\bibitem[{{Wright} {et~al.}(2016){Wright}, {Do Duy}, \& {Lawson}}]{Wright2016}
{Wright}, C.~M., {Do Duy}, T., \& {Lawson}, W. 2016, \mnras, 457, 1593

\bibitem[{{Yang} {et~al.}(2022){Yang}, {Green}, {Pontoppidan}, {Bergner}, {Cleeves}, {Evans}, {Garrod}, {Jin}, {Kim}, {Kim}, {Lee}, {Sakai}, {Shingledecker}, {Shope}, {Tobin}, \& {van Dishoeck}}]{Yang2022}
{Yang}, Y.-L., {Green}, J.~D., {Pontoppidan}, K.~M., {et~al.} 2022, \apjl, 941, L13

\bibitem[{{Yarnall} {et~al.}(2020){Yarnall}, {Gerakines}, \& {Hudson}}]{Yarnall2020}
{Yarnall}, Y.~Y., {Gerakines}, P.~A., \& {Hudson}, R.~L. 2020, \mnras, 494, 4606

\bibitem[{{Zasowski} {et~al.}(2009){Zasowski}, {Kemper}, {Watson}, {Furlan}, {Bohac}, {Hull}, \& {Green}}]{Zasowski2009}
{Zasowski}, G., {Kemper}, F., {Watson}, D.~M., {et~al.} 2009, \apj, 694, 459

\bibitem[{{Zinnecker} {et~al.}(1999){Zinnecker}, {Krabbe}, {McCaughrean}, {Stanke}, {Stecklum}, {Brandner}, {Padgett}, {Stapelfeldt}, \& {Yorke}}]{Zinnecker1999}
{Zinnecker}, H., {Krabbe}, A., {McCaughrean}, M.~J., {et~al.} 1999, \aap, 352, L73

\bibitem[{{Zubko} {et~al.}(1996){Zubko}, {Mennella}, {Colangeli}, \& {Bussoletti}}]{Zubko1996}
{Zubko}, V.~G., {Mennella}, V., {Colangeli}, L., \& {Bussoletti}, E. 1996, \mnras, 282, 1321

\end{thebibliography}


\clearpage
\onecolumn

\appendix

\section{NIRCam spectrum}
\label{psfnirc}

\begin{figure}
   \centering
   \includegraphics[width=12cm]{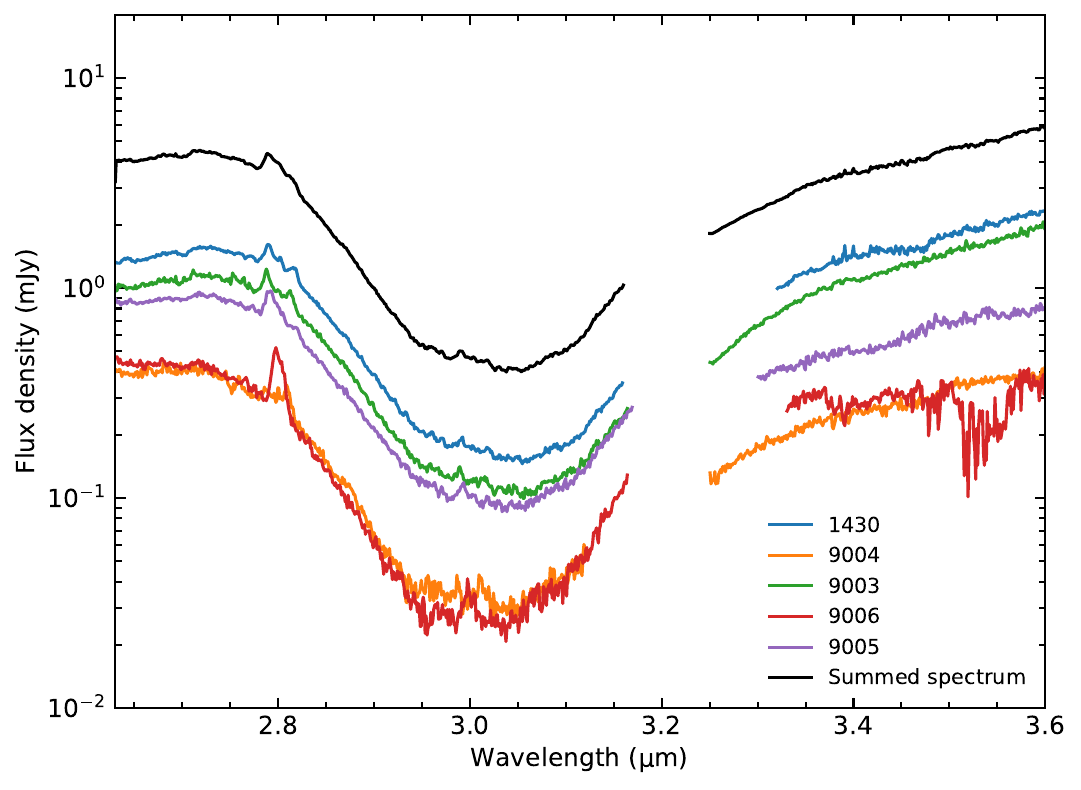}
      \caption{Extracted spectrum using the PSF method of different regions around Ced~110~IRS4A as indicated in Figure~\ref{FigNIRCAM}. Individual spectra are shown in colour and the sum of these individual data is presented in black. While the emission line at 2.8~$\mu$m is due to H$_2$ O(3) line, the strong absorption features at 3.55~$\mu$m in the region 9006 (red) is due to extraction artifact.}
         \label{figpsfnc}
   \end{figure}

The protostars Ced~110~IRS4A and IRS4B were observed using the NIRCam Wide Field Slitless Spectroscopy. Spectral extraction was performed at different positions of the two sources as indicated in Figure~\ref{FigNIRCAM}. The individual and summed spectra are shown in Figure~\ref{figpsfnc}. The spectral gaps are due to the physical separation between detectors in the field of view. The summed NIRCam spectrum is added together with the MIRI spectrum, and the combined data is used in the ice analysis demonstrated in this paper.

\section{Comparison between MIRI/MRS and Spitzer/IRS}
\label{JWST_Spitzer_sec}
\begin{figure}
   \centering
   \includegraphics[width=12cm]{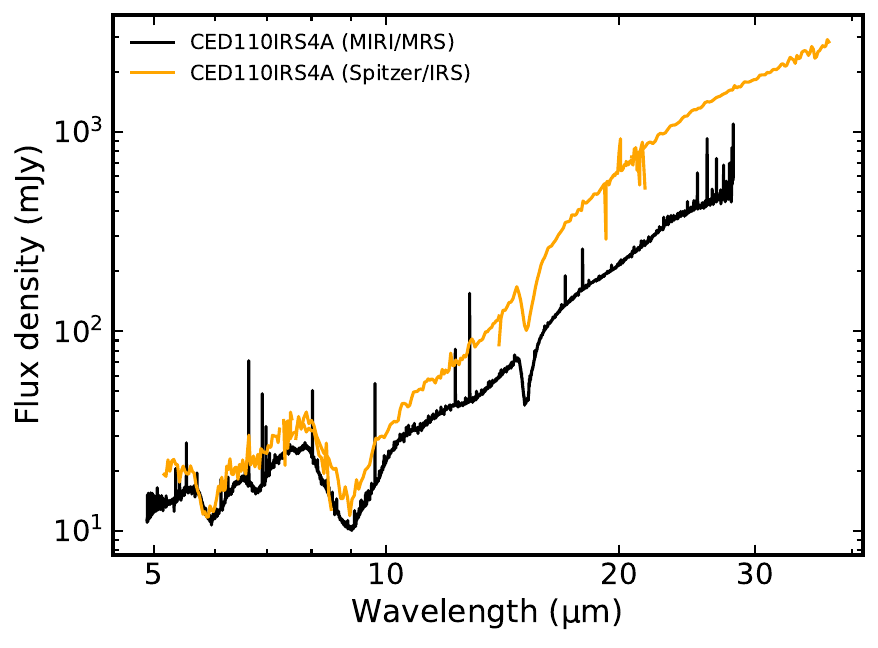}
      \caption{Comparison between the MIRI/MRS and \textit{Spitzer}/IRS spectrum of Ced~110~IRS4A. The nominal spectral resolving power of both data at 7~$\mu$m are $\sim$3000 and 600, respectively.}
         \label{spitzer_jwst}
   \end{figure}

The primary source, Ced~110~IRS4A was observed with \textit{Spitzer} in 2005, as part of the program ``Spectroscopy of protostellar, protoplanetary and debris disks'' (P.I. J. R. Houck). This spectrum was originally published by \citet{Manoj2011}, where the data reduction steps are detailed. The spectral data was downloaded from the Spitzer Heritage Archive catalogue\footnote{\url{https://irsa.ipac.caltech.edu/data/SPITZER/docs/spitzerdataarchives/}} based on the astronomical observation request (AOR) 12692224. A comparison between the \textit{Spitzer}/InfraRed Spectrometer (IRS) and the MIRI spectrum is shown in Figure~\ref{spitzer_jwst}. When roughly spectrally resolved, both data show similar absorption features. The \textit{Spitzer} spectrum has an enhanced flux longwards of 10~$\mu$m due to the larger slit width compared to MIRI (e.g., Spitzer 14$-$36~$\mu$m: 10.2"; MIRI/MRS: 4"). The spectral range between 6.5 and 10~$\mu$m is fully resolved in the MIRI range, which allowed the analysis performed in Section~\ref{sec7_10}. In addition, a clear feature at 6.8~$\mu$m is visible in the MIRI/MRS spectrum, as well as multiple emission lines which are not labelled in this paper. 

\section{List of laboratory data}
\label{list_lab}
A comprehensive list of molecules was used in this paper to search for the best fit of the region between 6.8 and 10~$\mu$m for Ced~110~IRS4A and between 2.5$-$5.0~$\mu$m for Ced~110IRS4B. This list includes COMs in different mixtures as well as simple molecules. The full list is shown in Table~\ref{tab_list}.

\begin{longtable}{lcccc}
\caption{\label{tab_list} Laboratory data tested in the global fit performed with \texttt{ENIIGMA}.}\\
\hline
\hline
Label & Temperature (K) & Resolution (cm$^{-1}$) & Database$^a$ & Reference\\
\hline
\endfirsthead
\caption{Continued.}\\
\hline
Label & Temperature$^a$ (K) & Resolution (cm$^{-1}$) & Database$^b$ & Reference\\
\endhead
\hline
\endfoot
\hline
\endlastfoot
\multicolumn{5}{c}{\bf{Simple molecules (less than 6 atoms) and hydrocarbons}}\\
\hline
H$_2$O & 15$-$160 & 2.0 & LIDA & [1]\\
CH$_4$ & 10$-$30 & 1.0  & OCdb  & [2]\\
CO  & 10 &  1.0 & LIDA  & [3]\\
HCOOH  & 15$-$165 &  1.0 & LIDA  & [4]\\
NH$_3$:CO$_2$ (3:1)  & 20 &  1.0 & ...  & [5]\\
H$_2$O:CO$_2$ (10:1) & 10 & 1.0 & LIDA & [6]\\
H$_2$O:CO$_2$ (1:10) & 10 & 1.0 & LIDA & [6]\\
H$_2$O:CO$_2$ (1:6) & 10 & 1.0 & LIDA & [6]\\
H$_2$O:CO$_2$ (1:1) & 10 & 1.0 & LIDA &  [6]\\
H$_2$O:CO$_2$:CH$_4$ (10:1:1) & 10 & 2.0 & UNIVAP & [7]\\
H$_2$O:CO (1:0.14) & 10 & 1.0 & LIDA & [3]\\
SO$_2$:CH$_3$OH (1:1)  & 15 &  1.0 & LIDA  & [8]\\
H$_2$O:CH$_4$ (10:1)  & 15 &  1.0 & UNIVAP  & [7]\\
H$_2$O:NH$_3$ (10:1.6)  & 10 &  2.0 & LIDA  & [9]\\
CO:CO$_2$ (2:1) & 40$-$80 & 0.5 & LIDA & [10]\\
CO:CO$_2$ (1:0.7) & 10 & 0.5 & LIDA & [10]\\
CO$_2$:CH$_3$OH (1:1) & 10 & 1.0 & LIDA & [6]\\
CO$_2$:CH$_3$OH (3:1) & 10 & 1.0 & LIDA & [6]\\
CO$_2$:CH$_3$OH (1:3) & 10 & 1.0 & LIDA & [6]\\
C$_2$H$_2$  & 15 &  1.0 & NASA  & [11]\\
C$_2$H$_4$  & 15 &  1.0 & NASA  & [12]\\
C$_2$H$_6$  & 15 &  1.0 & NASA  & [12]\\
HCOOH:H$_2$O:CO$_2$ (0.1:1.0:0.4) & 15 & 1.0 & LIDA & [4]\\
N$_2$O (irr.) & 14 & 1.0 & ... & [13]\\
N$_2$O & 10 & 1.0 & ... & [13]\\
N$_2$O:H$_2$O (1:30) & 14 & 1.0 & ... & [13]\\
NO$_2$:N$_2$O$_4$ (1:1) & 16 & 1.0 & ... & [14]\\
NO$_2$:N$_2$O$_4$ (1:1) & 60 & 1.0 & ... & [14]\\
\hline
\multicolumn{5}{c}{\bf{Ions$^c$}}\\
\hline
OCN$^-$: HNCO:NH$_3$ (1:1) & 15 & 1.0 & LIDA & [15]\\
HCOO$^-$: H$_2$O:NH$_3$:HCOOH (100:2.6:2) & 14$-$210 & 1.0 & LIDA & [16]\\
HCOO$^-$: NH$_3$:HCOOH (1.3:1) & 14$-$210 & 1.0 & LIDA & [16]\\
\hline
\multicolumn{5}{c}{\bf{COMs (more than 6 atoms)}}\\
\hline
CH$_3$OH  & 10$-$120 & 1.0 & OCdb    & [2]\\
CH$_3$CHO  & 15$-$120 &  1.0  &  LIDA    & [17]\\
CH$_3$CN  & 15$-$150 &  1.0  &  LIDA    & [18]\\
CH$_3$OCH$_3$  & 15$-$100 &  1.0 & LIDA      & [17]\\
CH$_3$COCH$_3$  & 15$-$100 &  1.0 & LIDA      & [19]\\
CH$_3$CH$_2$OH  &  15$-$150 &  1.0 & LIDA     & [17]\\
CH$_3$OCHO  &  15$-$120 &  1.0 & LIDA     & [20]\\
CH$_3$COOH  &  10 &  1.0 & NASA     & [21]\\
CH$_3$NH$_2$  &  10 &  1.0 & LIDA     & [22]\\
CH$_3$CH$_2$CH$_2$OH  &  13 &  1.0 & NASA   & [23]\\
HC(O)CH$_2$CH$_3$  &  10 &  1.0 & NASA  & [24]\\
CH$_3$CH$_2$OH:H$_2$O (1:20) & 15$-$160 & 1.0 & LIDA &[17] \\
CH$_3$CH$_2$OH:CO (1:20) & 15, 30 & 1.0 & LIDA  &[17] \\
CH$_3$CH$_2$OH:CH$_3$OH (1:20) & 15$-$150 & 1.0 & LIDA  &[17] \\
CH$_3$CH$_2$OH:CO:CH$_3$OH (1:20:20) & 15$-$150 & 1.0 & LIDA  &[17] \\
CH$_3$CHO:H$_2$O (1:20) & 15$-$120 & 1.0 & LIDA &[17] \\
CH$_3$CHO:CO (1:20) & 15, 30 & 1.0 & LIDA  &[17] \\
CH$_3$CHO:CH$_3$OH (1:20) & 15$-$140 & 1.0 & LIDA  &[17] \\
CH$_3$CHO:CO:CH$_3$OH (1:20:20) & 15$-$120 & 1.0 & LIDA  &[17]\\
CH$_3$OCH$_3$:H$_2$O (1:20) & 15$-$160 & 1.0 & LIDA &[17] \\
CH$_3$OCH$_3$:CO (1:20) & 15, 30 & 1.0 & LIDA  &[17] \\
CH$_3$OCH$_3$:CH$_3$OH (1:20) & 15$-$120 & 1.0 & LIDA  &[17] \\
CH$_3$OCH$_3$:CO:CH$_3$OH (1:20:20) & 15$-$100 & 1.0 & LIDA &[17]\\
CH$_3$COCH$_3$:H$_2$O (1:20) & 15$-$160 & 1.0 & LIDA &[19] \\
CH$_3$COCH$_3$:CO (1:20) & 15, 30 & 1.0 & LIDA &[19] \\
CH$_3$COCH$_3$:CO$_2$ (1:20) & 15$-$100 & 1.0 & LIDA &[19] \\
CH$_3$COCH$_3$:CH$_3$OH (1:20) & 15$-$140 & 1.0 & LIDA &[19]\\
CH$_3$COCH$_3$:H$_2$O:CO$_2$ (1:2.5:2.5) & 15$-$160 & 1.0 & LIDA &[19]\\
CH$_3$COCH$_3$:CO:CH$_3$OH (1:2.5:2.5) & 15$-$140 & 1.0 & LIDA &[19]\\
CH$_3$OCHO:H$_2$O (1:20) & 15$-$120 & 1.0 & LIDA &[20]\\
CH$_3$OCHO:CO (1:20) & 15$-$120 & 1.0 & LIDA &[20]\\
CH$_3$OCHO:H$_2$CO (1:20) & 15$-$120 & 1.0 & LIDA &[20]\\
CH$_3$OCHO:CO:H$_2$CO:CH$_3$OH (1:20:20:20) & 15$-$120 & 1.0 & LIDA &[20]\\
CH$_3$COOH:H$_2$O (1:20) & 10 & 1.0 & NASA &[21]\\
CH$_3$NH$_2$:H$_2$O (1:20) & 15$-$150 & 1.0 & LIDA &[22]\\
CH$_3$NH$_2$:NH$_3$ (1:20) & 15$-$150 & 1.0 & LIDA &[22]\\
CH$_3$NH$_2$:CH$_4$ (1:20) & 15$-$150 & 1.0 & LIDA &[22]\\
CH$_3$CN:H$_2$O (1:20) & 15$-$150 & 1.0 & LIDA &[22]\\
CH$_3$CN:CO (1:20) & 15$-$100 & 1.0 & LIDA &[18]\\
NH$_2$CHO:H$_2$O (7:100) & 15$-$160 & 1.0 & LIDA &[25]\\
NH$_2$CHO:CO (4:100) & 15$-$34 & 1.0 & LIDA &[25]\\
HCOCH$_2$OH:H$_2$O (1:18) & 10 & 1.0 & NASA &[26]\\
CH$_3$OH:CO (1:4) & 15 & 1.0 & LIDA &[27]\\
CH$_3$OH:NH$_3$ (1:1) & 15 & 1.0 & UNIVAP &[28]\\
CH$_3$OH:H$_2$O (10:0.8) & X & X & ... & [29]\\
HCOOH:H$_2$O:CH$_3$OH (0.1:1.0:0.4) & 15 & 1.0 & LIDA &[4]\\
\hline
\hline
\multicolumn{5}{c}{\tablefoot{[1] \citet{Oberg2007}; [2] \citet{Hudgins1993}; [3] \citet{Ehrenfreund1997}; [4] \citet{Bisschop2007}; [5] \citet{James2021}; [6] \citet{Ehrenfreund1999}; [7] \citet{Rocha2017}; [8] \citet{Boogert1997}; [9] \url{http://strw.leidenuniv.nl/~schutte/database/H2O_NH3_20.10K}; [10] \citet{Schutte1999}; [11] \citet{Hudson2014c2h2}; [12] \citet{Hudson2014eth}; [13] \citet{Bergantini2022}; [14] \citet{Fulvio2009}; [15] \citet{Novozamsky2001}; [16] \citet{Galvez2010}; [17] \citet{Scheltinga2018}; [18] \citet{Rachid2022}; [19] \citet{Rachid2020}; [20] \citet{Scheltinga2021}; [21] No reference found - taken from the NASA Ice Database (Pure: \url{https://science.gsfc.nasa.gov/691/cosmicice/spectra/refspec/Acids/CH3COOH/ACETIC-W.txt}, Mixture: \url{https://science.gsfc.nasa.gov/691/cosmicice/spectra/8_compounds/Combined_spectra_2018-12-20.xlsx}); [22] \citet{Rachid2021}; [23] \citet{Hudson2019}; [24] \citet{Yarnall2020}; [25] \citet{Slavicinska2023}; [26] \citet{Hudson2005}; [27] \citet{Cuppen2011}; [28] \citet{Rocha2020}; [29] \citet{Dawes2016}, $^a$ Each spectral data was recorded for the temperature indicated in this table. $^b$LIDA: The Leiden Ice Database for Astrochemistry (\url{https://icedb.strw.leidenuniv.nl/}); OCdb: The Optical Constant Database (\url{https://ocdb.smce.nasa.gov/}); UNIVAP: \url{https://www1.univap.br/gaa/nkabs-database/data.htm}; NASA Cosmic Ice Laboratory: \url{https://science.gsfc.nasa.gov/691/cosmicice/spectra.html}. $^c$ Only in the case of the ions, the ratios of the mixtures correspond to the deposition values. The ratios of the products are not available.}}
\end{longtable}

\section{CO:CO$_2$ and H$_2$O:CO$_2$ ice mixtures}
\label{bin_mix}

\begin{figure*}
   \centering
   \includegraphics[width=\hsize]{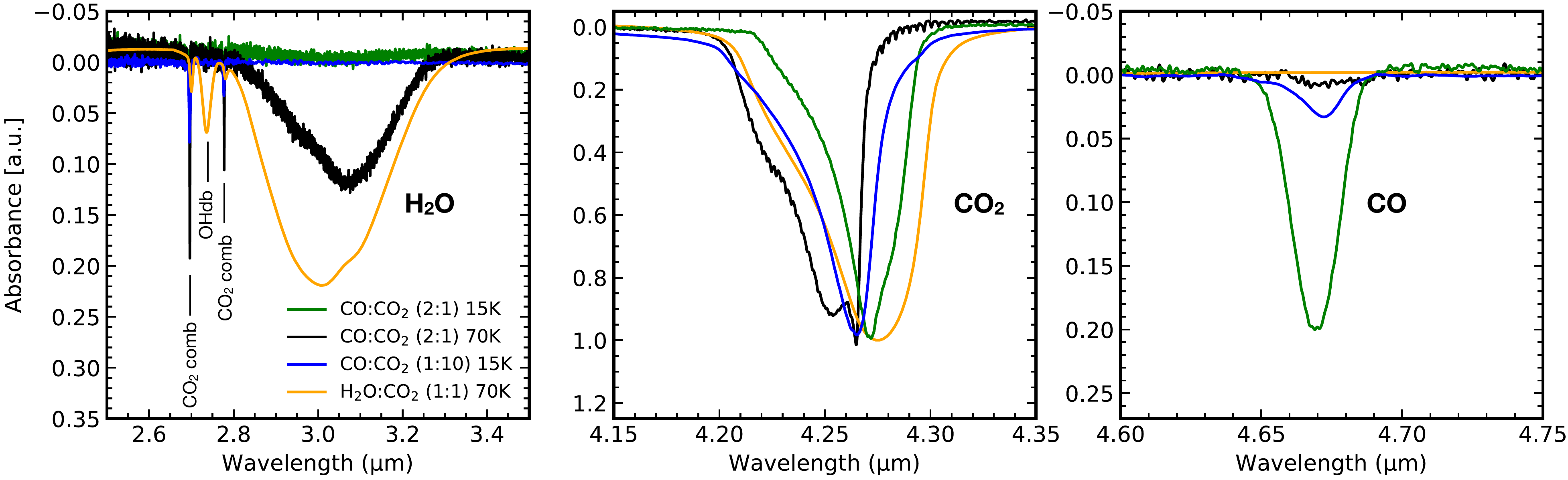}
      \caption{Comparison between laboratory spectra of CO:CO$_2$ and H$_2$O:CO$_2$. The individual features of these molecules are presented.}
         \label{coco2comp}
   \end{figure*}

Section~\ref{sec_nirB} shows that CO:CO$_2$ at 70~K, combined with other ice mixtures, fits well the CO$_2$ ice band at 4.27~$\mu$m for Ced~110~IRS4B. In Figure~\ref{coco2comp} we show a brief comparison of this IR spectrum with other laboratory data with similar conditions. Left, middle and right panels highlight the specific absorption features of H$_2$O, CO$_2$ and CO ice, respectively. At 70~K, CO is no longer in the ice as indicated in the right panel. Nevertheless, the CO$_2$ band of this CO:CO$_2$ mixture shows a prominent enhancement at the blue wing of the CO$_2$ ice feature, which can be due to ice distillation due to the CO sublimation. We highlight, however, that this data has H$_2$O ice contamination, as presented in the left panel, which corresponds to 70\% of the CO$_2$ ice column density. By visual inspection of the spectra, we did not found evidence of contamination from other species.  

To check whether the presence of H$_2$O in the ice can induce the blue wing excess of the CO$_2$ band in the CO:CO$_2$ data, we compare this profile with that of H$_2$O:CO$_2$ (1:1) at 70~K. Middle panel shows no enhancement of the CO$_2$ blue wing profile in that case. In this case, it is possible to conclude that the ice distillation can affect the CO$_2$ IR feature, and therefore it is a tracer of thermal evolution of the ice.

\begin{figure*}
   \centering
   \includegraphics[width=\hsize]{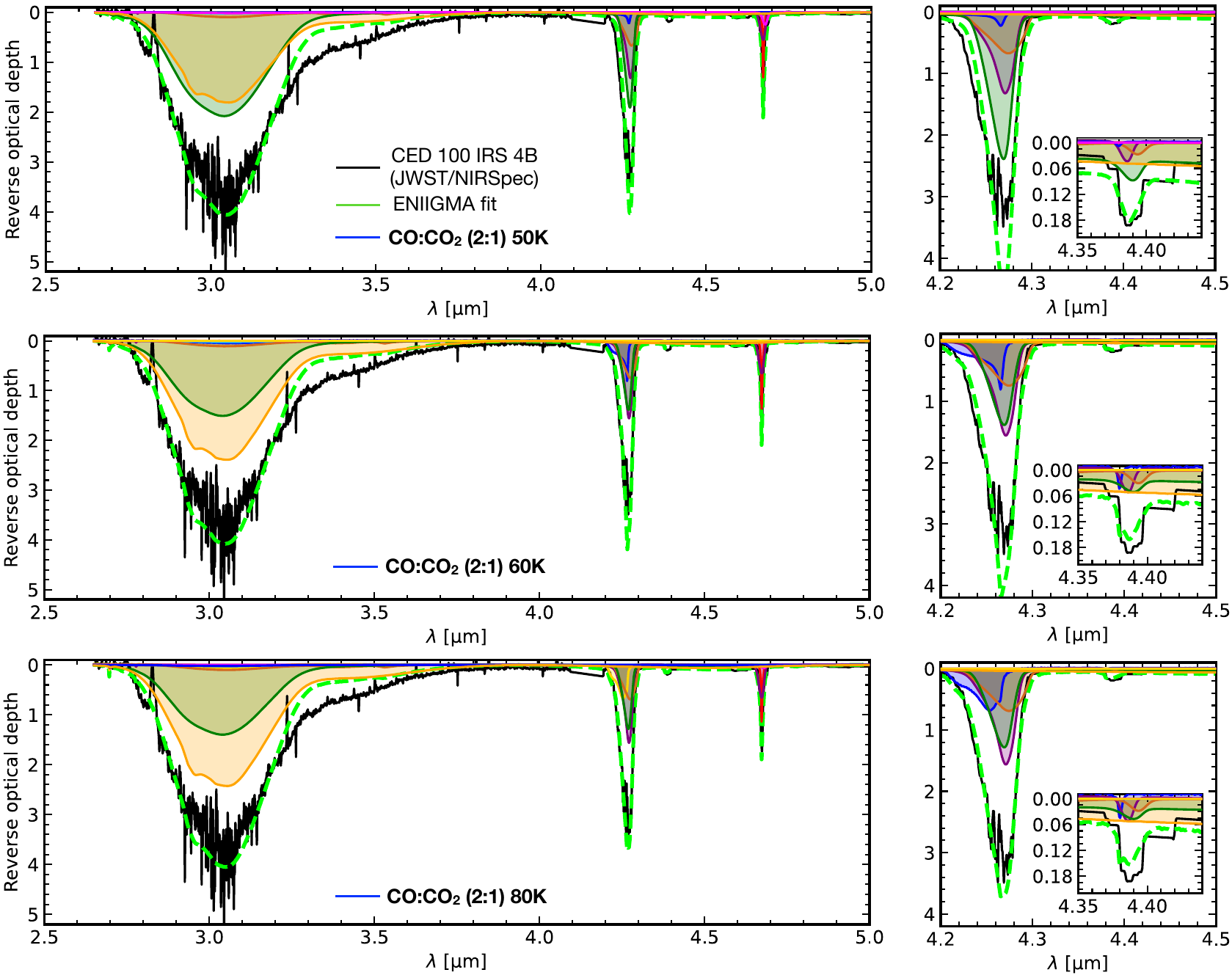}
      \caption{\texttt{ENIIGMA} fits of the NIRSpec range of Ced~110~IRS4B by assuming IR spectra of the CO:CO$_2$ ice mixture at 50 (top), 60 (middle) and 80~K (bottom). Left panels shows the full spectral range, and the right plots show a zoom-in of the ${^12}$CO$_2$ and $^{13}$CO$_2$ ice absorption features.}
         \label{coco2oT}
   \end{figure*}

In Figure~\ref{coco2oT} we replace the CO:CO$_2$ IR spectra at 70~K (best fit) by other spectra of the same ice mixture at 50, 60 and 80~K to have a sense of the variability of the fit. No calculation was performed using data recorded at 90~K because the $^{12}$CO$_2$ ice feature is no longer visible in the laboratory IR spectrum. As presented in the right panels, the fits of the $^{12}$CO$_2$ ice band changes according to the temperature of the CO:CO$_2$ ice spectra. The blue shoulder of $^{12}$CO$_2$ is prominent at 60 and 80~K, but not at 50K. This causes the fit of $^{12}$CO$_2$ at 50~K to be dominated by the H$_2$O:CO$_2$ ice component, which is not enough to explain the blue sholder profile in the data of Ced~110~IRS4B. Better fits are obtained in the cases data from CO:CO$_2$ at 60 and 80~K. The fitness score of both fits compared to the best fit is 15 and 8\%, respectively. Given the saturation of the $^{12}$CO$_2$ ice band, these differences are not significant. Therefore, the solutions with CO$_2$ ice at 60, 70, and 80~K are acceptable since a better constrain requires the detection of the peak of $^{12}$CO$_2$ ice band.  

\section{CO rotational–vibrational line subtraction for Ced~110~IRS4B}
\label{corovib}
\begin{figure}
   \centering
   \includegraphics[width=15cm]{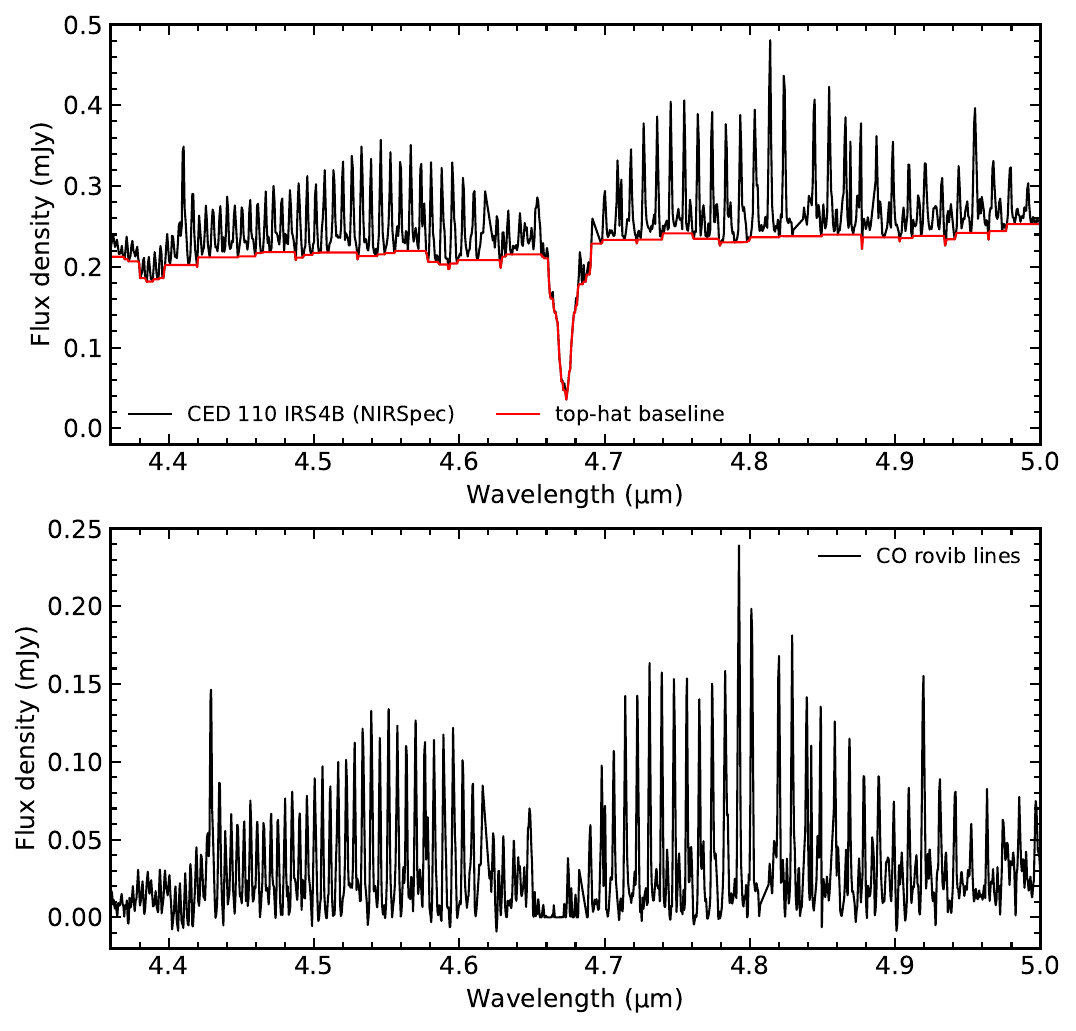}
      \caption{Subtraction of the CO gas rotational-vibrational lines in the NIRSpec data of Ced~110IRS4B. Both P (4.3 $\leq$ $\lambda$ ($\mu$m) $\leq$ 4.56) and Q (4.65 $\leq$ $\lambda$ ($\mu$m) $\leq$ 5.0) branches are visible in this spectrum. Top panel shows the observational data with the baseline, and bottom panel shows the continuum-subtracted spectrum.}
         \label{figcorovib}
   \end{figure}

The CO rovibrational lines are removed before tracing the continuum SED. This step is necessary to disentangle the ice features that are blended with emission lines. The continuum is traced by using a function called ``top-hat'' \citep{pybaselines, Rubinstein2023} to perform a baseline under the bottom of the emission features, but avoiding the regions where multiple emission features create a quasi-continuum profile. The baseline profile and the continuum subtracted CO lines are shown in Figure~\ref{figcorovib}. No further analysis of these CO forest lines that includes extinction correction and rotational diagrams are presented in this paper, but they will be discussed in a future work. 

\section{Emission profile at 7.18~$\mu$m}
\label{CH3ion}
\begin{figure}
   \centering
   \includegraphics[width=15cm]{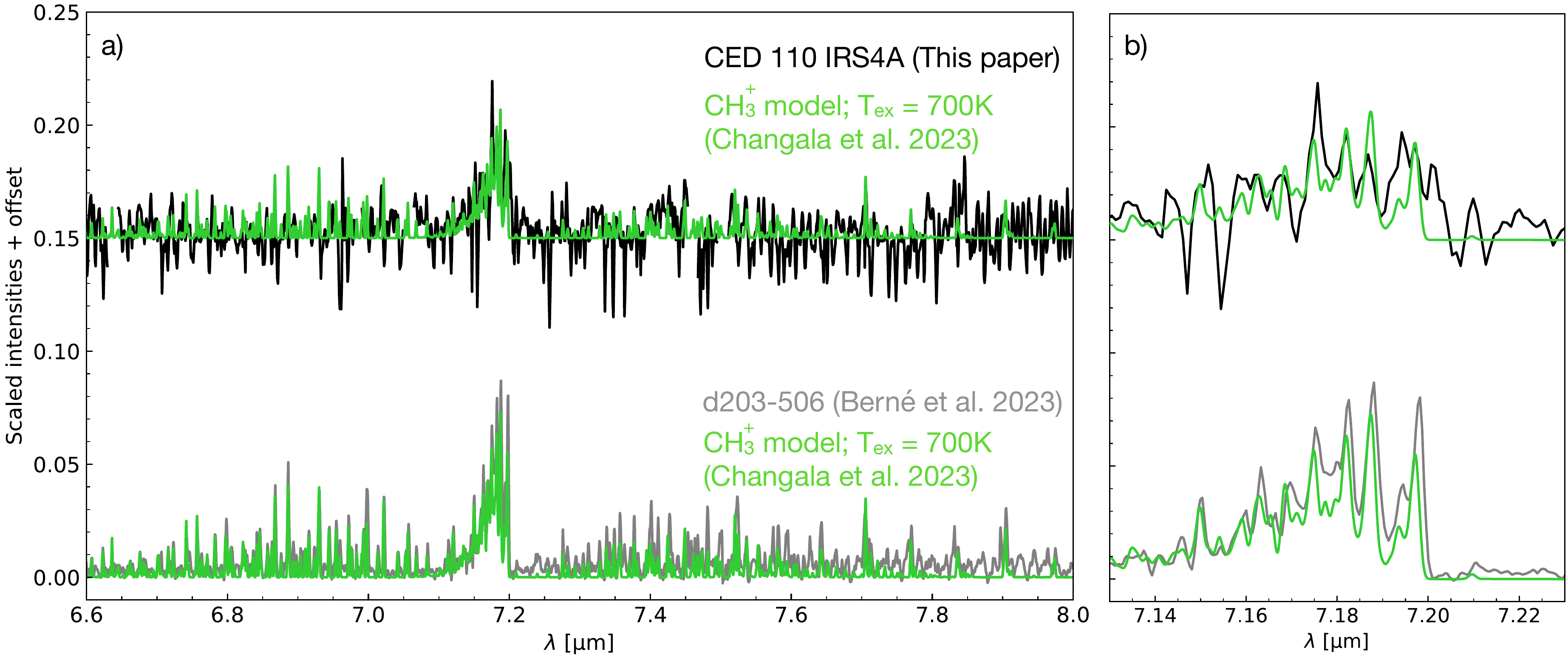}
      \caption{Comparison between the methyl cation feature from the Orion proplyd \citep[][bottom, R = 3800]{Berne2023} and from Ced~110~IRS4A (top) overlaid with the gas-phase model from \citet{Changala2023}. Panel b shows a zoom-in of panel a around 7.18~$\mu$m.}
         \label{CH3fig}
   \end{figure}

In recent works, methyl cation (CH$_3^+$) was detected in a protoplanetary disk, called d203-506, located in the Orion star-forming region \citep{Berne2023} and in the protoplanetary disk TW Hya \citep{Henning2024}. Accurate molecular data was calculated by \citet{Changala2023} that allowed further study of the CH$_3^+$ features in both d203-506 and TW Hya. In Figure~\ref{CH3fig} (bottom), we presents the comparison between the observation and model from \citet{Changala2023} of the 7.18~$\mu$m band in the case of d203-506. The model assumes an excitation temperature of $T_{\rm{ex}} = 700$~K. A similar emission profile was observed in the MIRI/MRS spectrum of Ced~110~IRS4A, which is different from the profile from d203-506 when compared by eye. In Figure~\ref{CH3fig}, we overlaid the same gas-phase model of CH$_3^+$ from \citet{Changala2023}. By visual examination, this model seems to also match the protostar observation. The continuum subtracted spectrum also display absorption lines, which will require further study to interpret them. 

\citet{Berne2023} shows that this molecular ion is formed via a two-step process. Molecular H$_2$ is excited by a strong FUV field, and reacts with C$^+$. Next, a sequence of hydrogen abstraction reactions leads to the formation of CH$_3^+$. Addressing the best model to fit the data in the case of the CH$_3^+$ is beyond the scope of this work. For this reason, we consider that CH$_3^+$ is only tentatively detected in the protostar Ced~110~IRS4A. Further works will address this possibility, as well as discuss the likely chemical and physical origin of this ion.    

\section{Local continuum around 7~$\mu$m: Ced~110~IRS4A}
\label{localfit78}
A local continuum method is used to isolate the ice features between 7 and 10~$\mu$m in the MIRI/MRS spectrum of Ced~110~IRS4A. This methodology was first used by \citet{Schutte1999_weak} to isolate the same ice features, and reproduced again in \citet{Yang2022}. In \citet{Rocha2024} we used this method to analyse the ice features in the same spectral region as shown in this paper. The difference between this paper and \citet{Rocha2024} with previous works is that the local continuum is subtracted in optical depth scale after silicate removal and not with the spectrum in flux scale. This enhances the visibility of shallow ice features longwards of 8~$\mu$m. This method consists of tracing a fourth-order polynomial to guiding points anchored in the observational spectrum. The physical interpretation of this polynomial is that broad and strong solid-phase features of other components (e.g., NH$_4^+$, H$_2$O ice bending mode, organic refractory material) can create a similar continuum that is similar to the polynomial profile.   

\begin{figure}
   \centering
   \includegraphics[width=10cm]{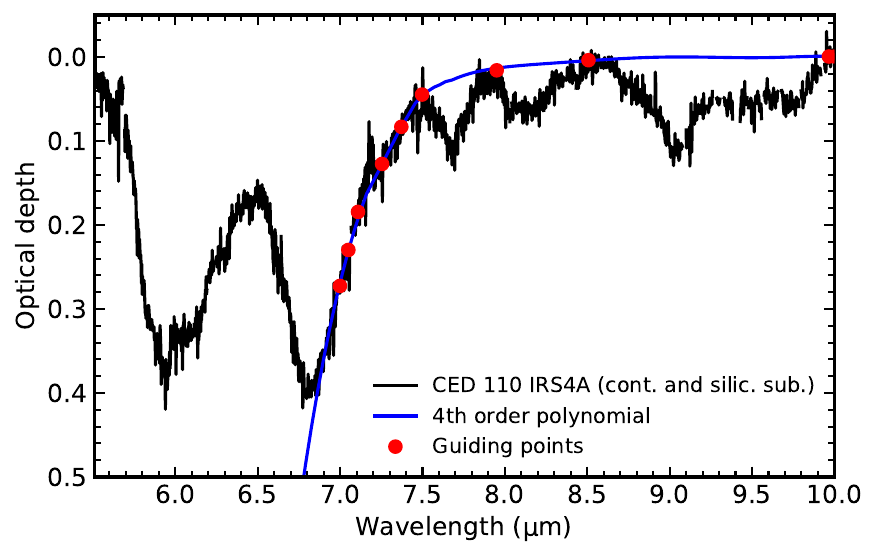}
      \caption{Local continuum subtraction between 7 and 10~$\mu$m using a 4th order polynomial. Guiding points are anchored in the observational spectrum.}
         \label{local_710mic}
   \end{figure}

\section{Statistical analysis and consistency checks}
\label{stats_analysis}
We derive the confidence intervals for the fits using the \texttt{ENIIGMA} fitting tool \citep{Rocha2021}. Briefly, the code explores how coefficient values in the linear combination changes the goodness of the fit. From this analysis, the code calculates $\chi^2$ values for each new solutions and the confidence intervals are derived based on a $\Delta\chi^2$ map \citep{Avni1980}, which is given by: 
\begin{subequations}
\begin{eqnarray}
\chi^2 &=& \frac{1}{dof}\sum_{i=0}^{n-1} \left(\frac{\tau_{\nu,i}^{\rm{obs}}  - \sum_{j=0}^{m-1} w_j \tau_{\nu,j}^{\rm{lab}}}{\gamma_{\nu,i}^{\rm{obs}}} \right)^2 ,\\
\Delta \chi^2(\alpha, \epsilon) &=& \chi^2 - \chi_{min}^2
,\end{eqnarray}
\label{deltachi}
\end{subequations}
where $dof$ is the number of degrees of freedom, $\gamma$ is the error in the observational optical depth spectrum propagated from the flux error assumed to be 5\%, $\alpha$ and $\epsilon$ are the statistical significance and the number of free parameters, respectively. $\chi_{min}^2$ corresponds to the goodness-of-fit in the global minimum solution.

\begin{figure}
   \centering
   \includegraphics[width=14cm]{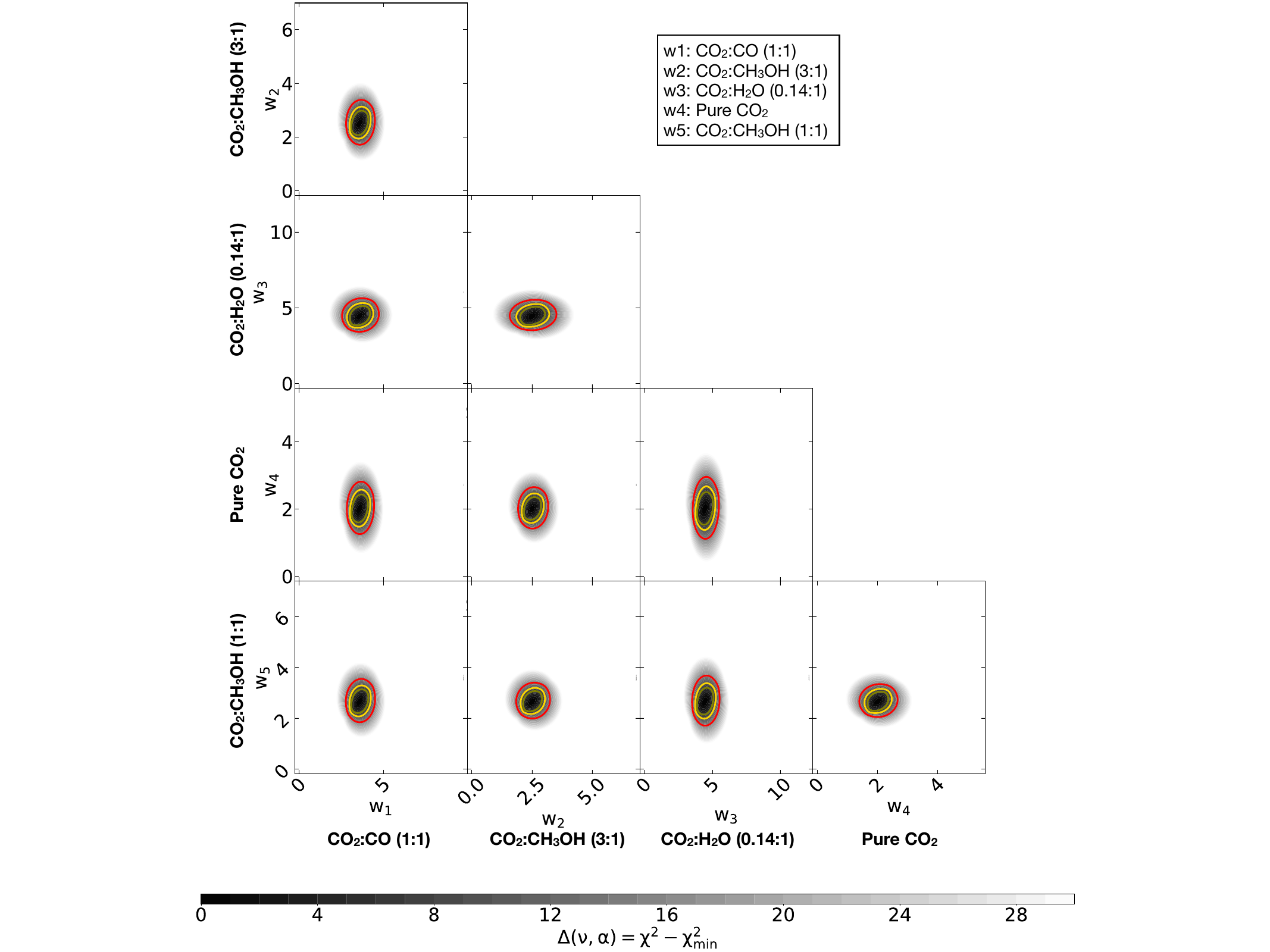}
      \caption{Confidence interval analysis for the 15.2~$\mu$m band. Green, yellow and red contours indicate 1, 2, and 3~$\sigma$ confidence interval for a fit using four components.}
         \label{confco2}
   \end{figure}

\begin{figure}
   \centering
   \includegraphics[width=8cm]{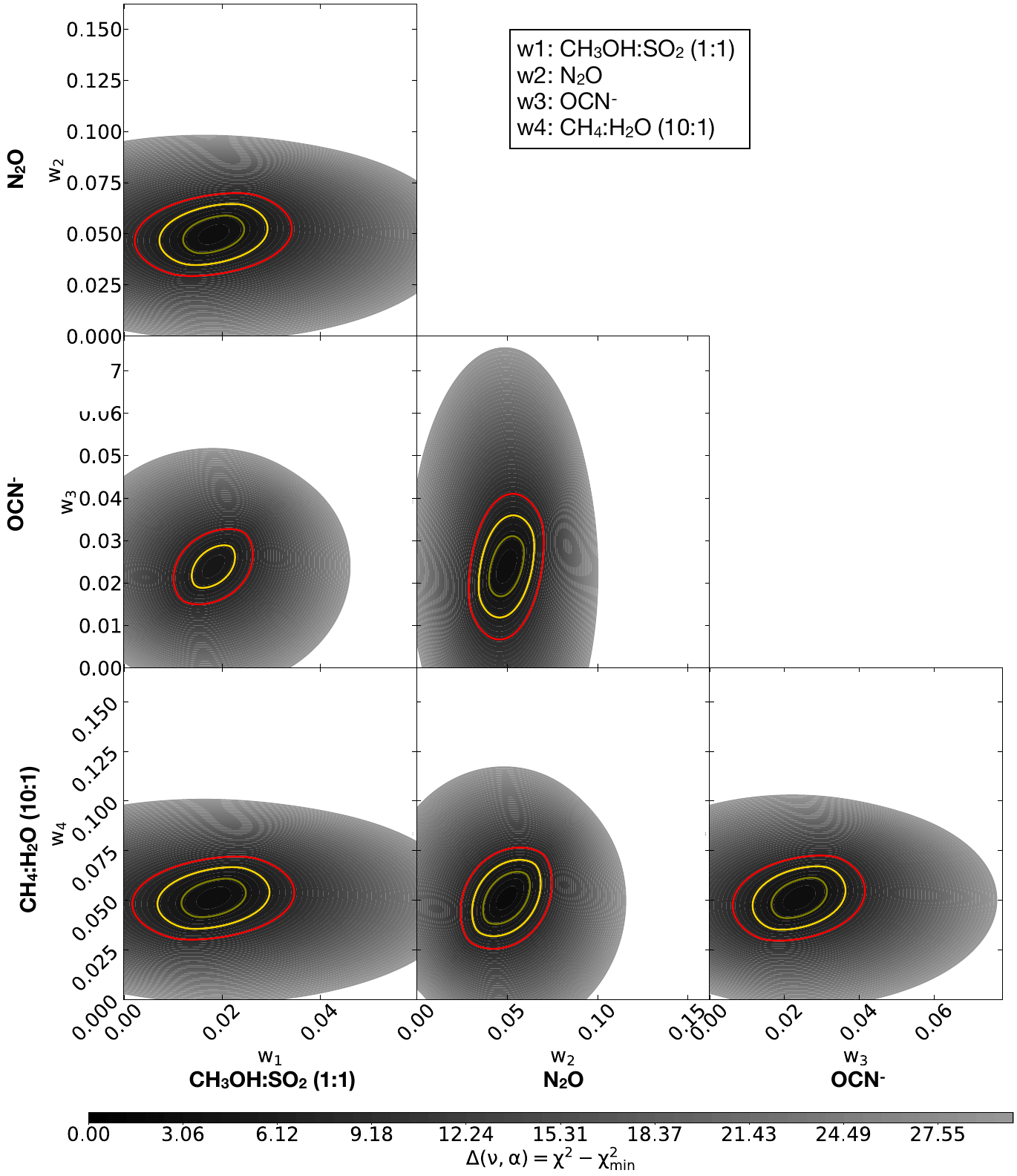}
      \caption{Same as Figure~\ref{confco2}, but for the 7~$\mu$m band of Ced~110~IRS4A.}
         \label{7micanalysis}
   \end{figure}

\begin{figure}
   \centering
   \includegraphics[width=10cm]{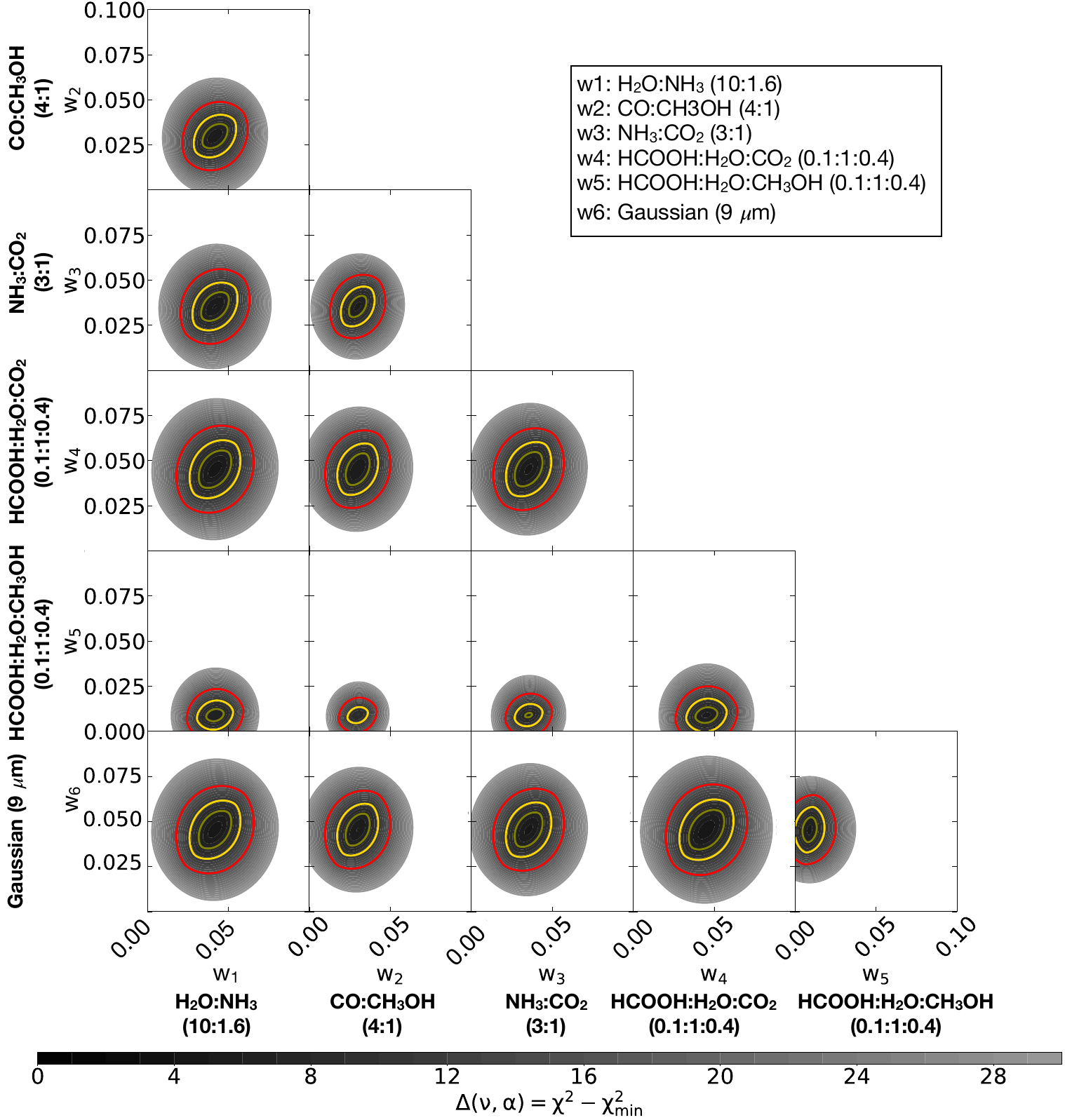}
      \caption{Same as Figure~\ref{confco2}, but for the 8$-$10~$\mu$m range of Ced~110~IRS4A.}
         \label{710mic}
   \end{figure}

\begin{figure}
   \centering
   \includegraphics[width=10cm]{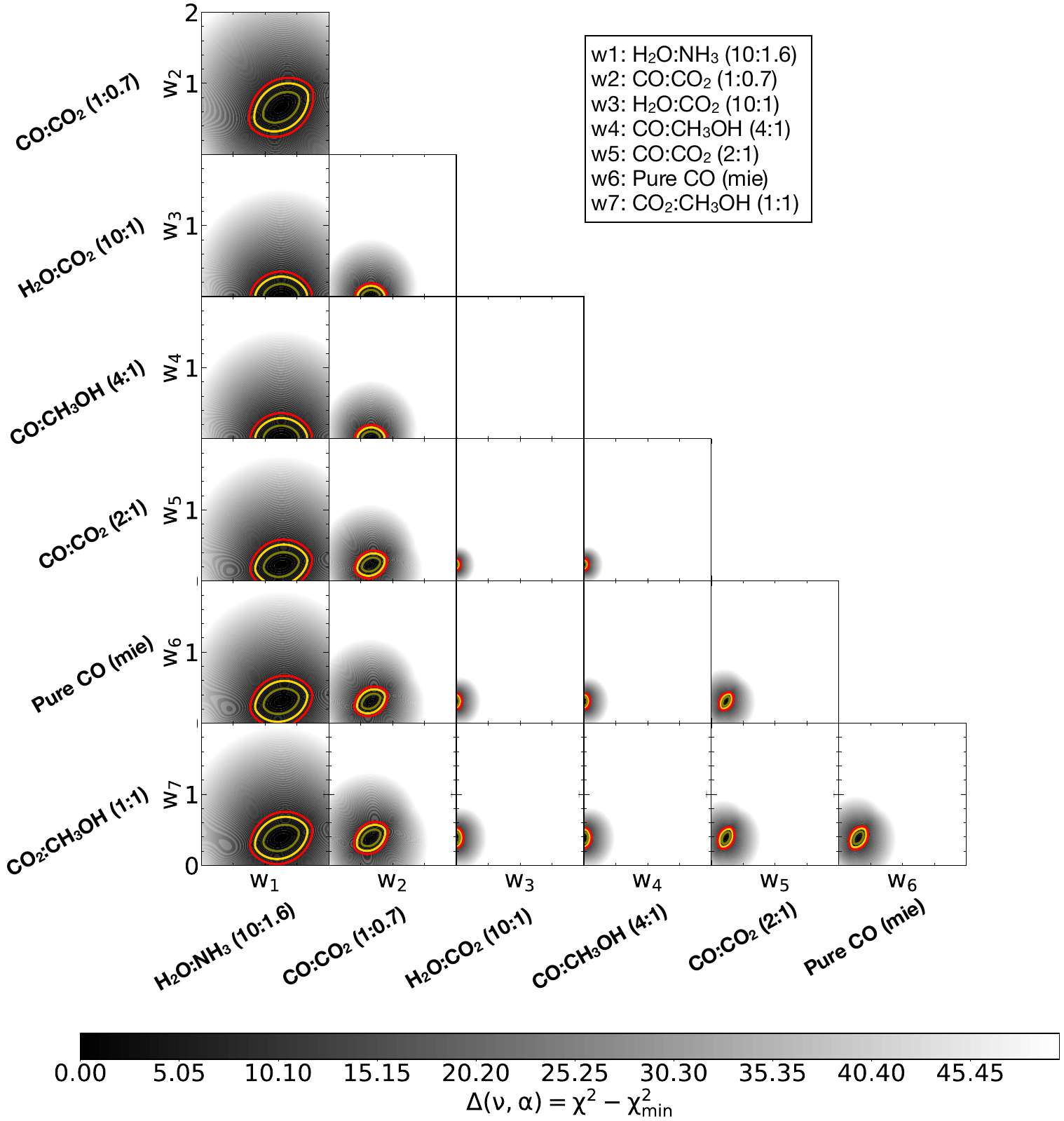}
      \caption{Same as Figure~\ref{confco2}, but for the 2.5$-$5~$\mu$m range of Ced~110~IRS4B.}
         \label{irs4b_conf}
   \end{figure}

Figures~\ref{confco2}, \ref{7micanalysis}, \ref{710mic} and \ref{irs4b_conf} show the confidence intervals analysis of Ced~110~IRS4A at 15.2~$\mu$m, 7~$\mu$m, 8$-$10~$\mu$m and for Ced~110~IRS4B between 2.5 and 5~$\mu$m, respectively. The lower and upper boundaries from these contours are used to derive the statistical error bars on the ice column densities presented in Table~\ref{ice_cd}.

\begin{figure*}
   \centering
   \includegraphics[width=\hsize]{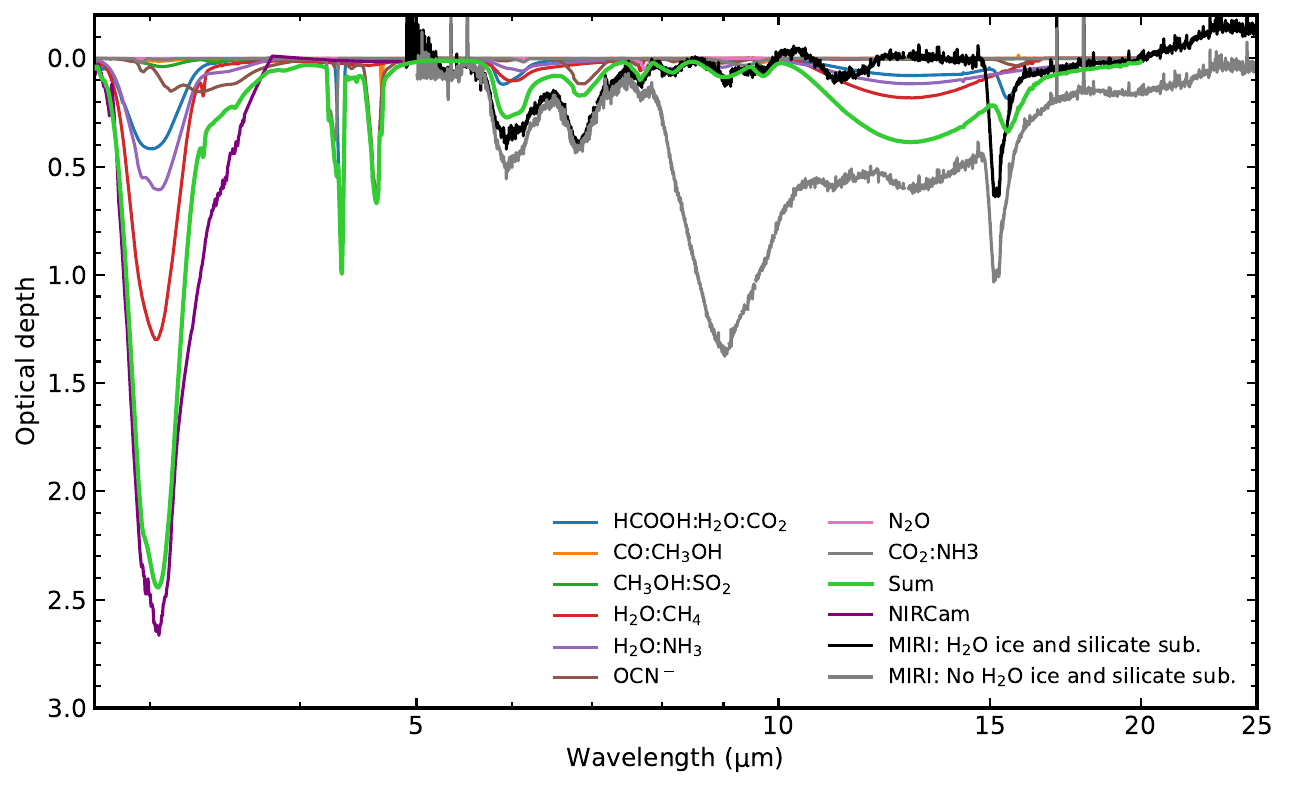}
      \caption{Consistency check for IRS4A: Sum of ice components fitting the range between 7-10~$\mu$m (Fig.~\ref{eniigmafitSA}) compared to the full MIRI spectrum before and after continuum, silicate and H$_2$O ice subtraction.}
         \label{con_check}
   \end{figure*}

\begin{figure*}
   \centering
   \includegraphics[width=\hsize]{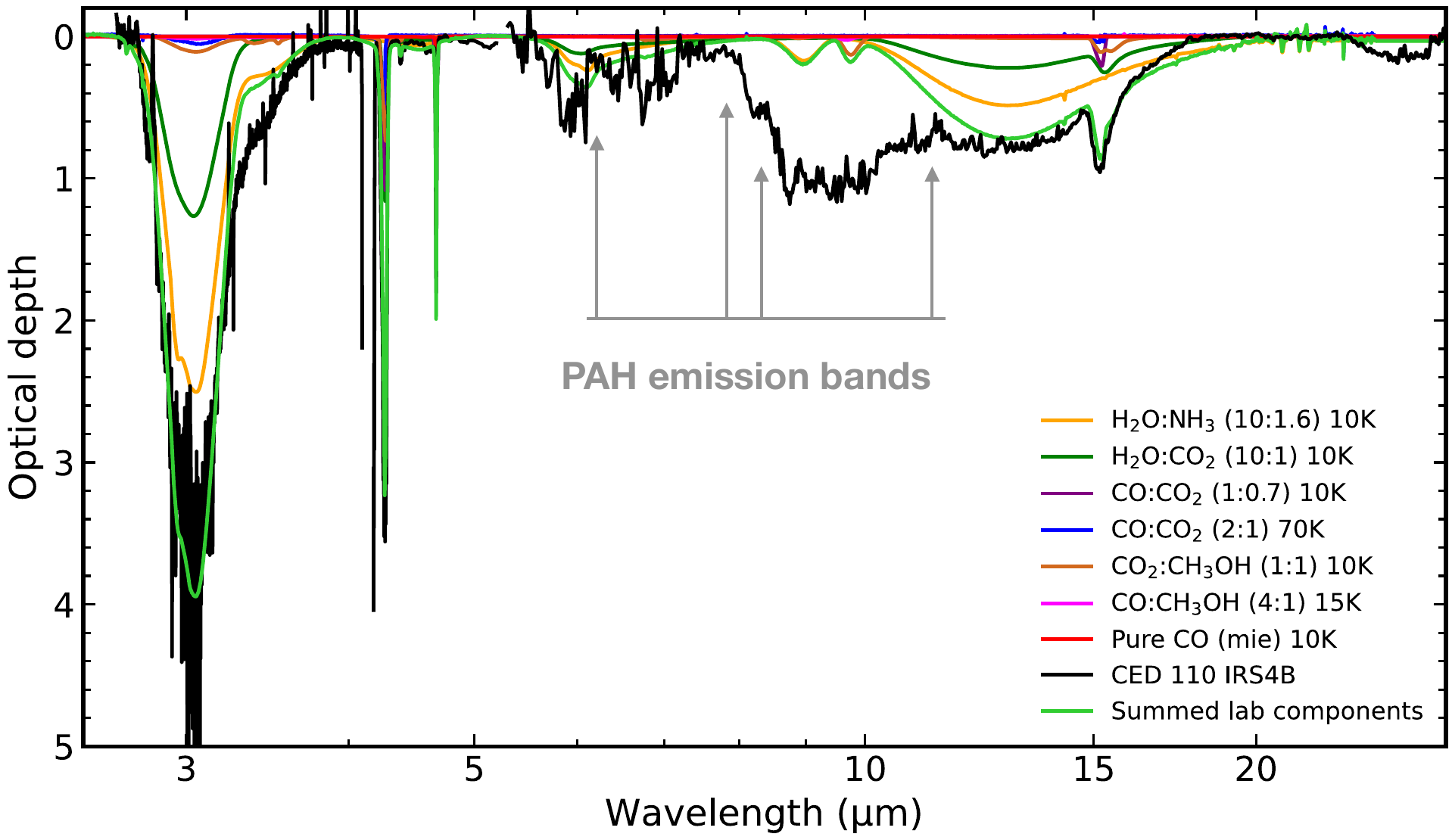}
      \caption{Consistency check for IRS4B: Sum of ice components fitting the range between 2.5-5~$\mu$m (Fig.~\ref{eniigmafitSB}) compared to the full MIRI/MRS spectrum. Some PAH emission bands are indicated in this figure.}
         \label{nir_check}
   \end{figure*}

In addition to these analyses, we also perform consistency checks of the small spectral range fits. For example, in Figure~\ref{con_check} we overplot the same components used in the fit of the 7$-$10~$\mu$m range of Ced~110~IRS4A, but now covering the full spectral range between 2.5 and 25~$\mu$m. The goal of this approach is to assess whether the small range fit would over-predict the ice absorption in another spectral ranges. Similarly, in Figure~\ref{nir_check}, the components fitting the 2.5$-$5~$\mu$m data of Ced~110~IRS4B are compared to the observational data between 5 and 25~$\mu$m. For both sources, we conclude that the \texttt{ENIIGMA} fits are consistent with the full spectral range. We note that between 4 and 5~$\mu$m, where the CO$_2$ (CO stretch, 4.27~$\mu$m), CO (CO stretch, 4.67~$\mu$m) and OCN$^-$ (CN stretch, 4.6~$\mu$m) ice bands are, no observational data is available. 

Finally, we derive Akaike Information Criterion (AIC) values for the fit of the spectral range between 7 and 10~$\mu$m. AIC is a powerful statistical parameter to assess the quality of the fit because it adds a penalty to the $\chi^2$ value when more parameters are included in the model. It can defined as:
\begin{equation}
   AIC = \chi^2 + 2p + \frac{2p(p+a)}{N - p - 1}, 
\end{equation}
where $p$ is the number of parameters and $N$ is the number of bins. The $\Delta$AIC is defined as $AIC_i$ - $min(AIC)$, where $AIC_i$ represent values for different models, and $min(AIC)$ corresponds to the minimum AIC value. The $\Delta$AIC scores are interpreted as follows \citep{burnham2002model}:
\begin{itemize}
    \item 0 < $\Delta$AIC < 2: substantial support of the model;
    \item 4 < $\Delta$AIC < 7: considerably less support of the model;
    \item $\Delta$AIC > 10: essentially no support of the model.
\end{itemize}

\begin{figure}
   \centering
   \includegraphics[width=15cm]{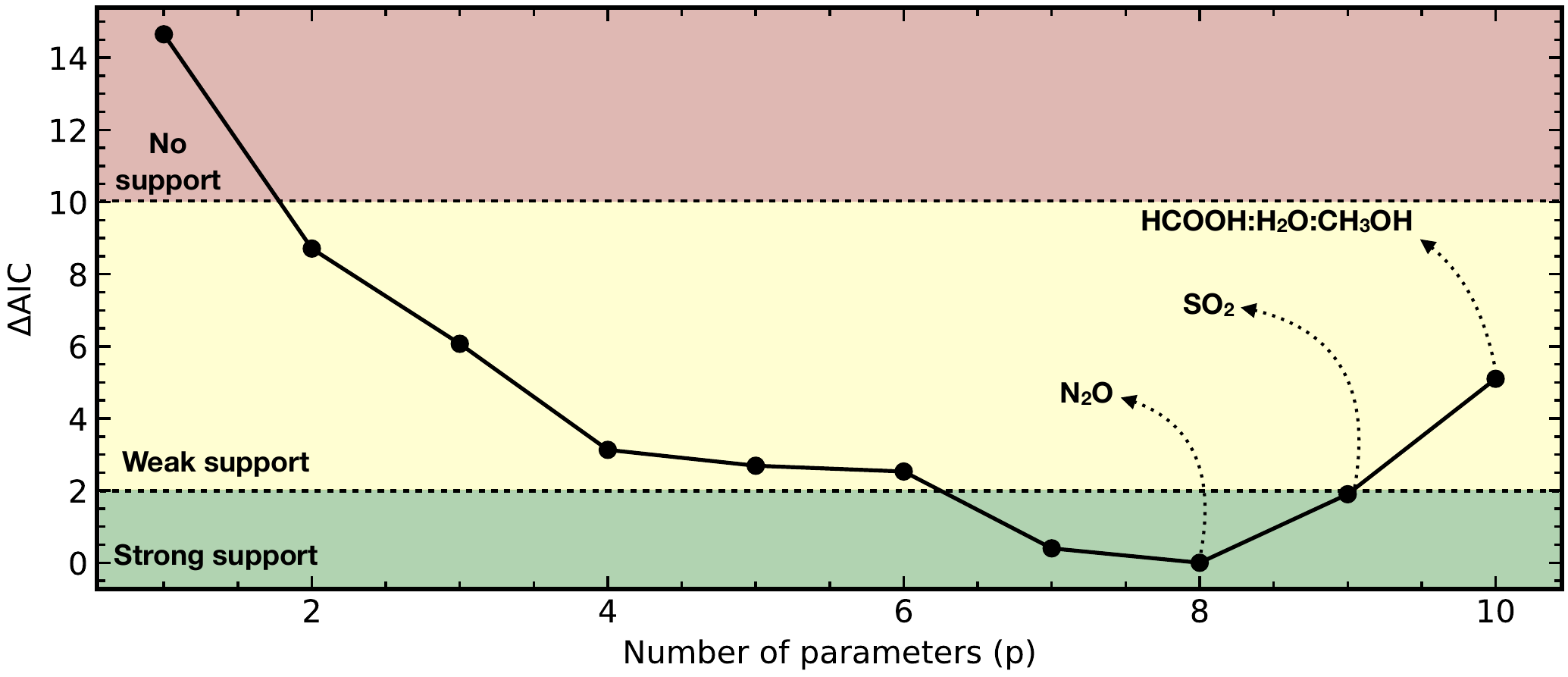}
      \caption{Relative Akaike Information Criterion for the fit between 7 and 10~$\mu$m. Coloured regions indicate the support to the model.}
         \label{aic_model}
   \end{figure}

Figure~\ref{aic_model} shows the $\Delta$AIC value when more parameters are added to fit the 7-10~$\mu$m region. The lowest $\Delta$AIC is reached when the eighth component is added to the mode, namely, N$_2$O. By adding SO$_2$ (CH$_3$OH:SO$_2$) and HCOOH:H$_2$O:CH$_3$OH the $\Delta$AIC increases, which would lead to a weak support of the model. Nevertheless, adding SO$_2$ keeps the $\Delta$AIC value in range of substantial support of the model. This is well aligned with the confidence interval analysis shown in Figure~\ref{7micanalysis}, where SO$_2$ is not excluded in the 3$\sigma$ confidence interval, whereas HCOOH:H$_2$O:CH$_3$OH can be ruled out (Fig.~\ref{710mic}). 

\begin{figure}
   \centering
   \includegraphics[width=15cm]{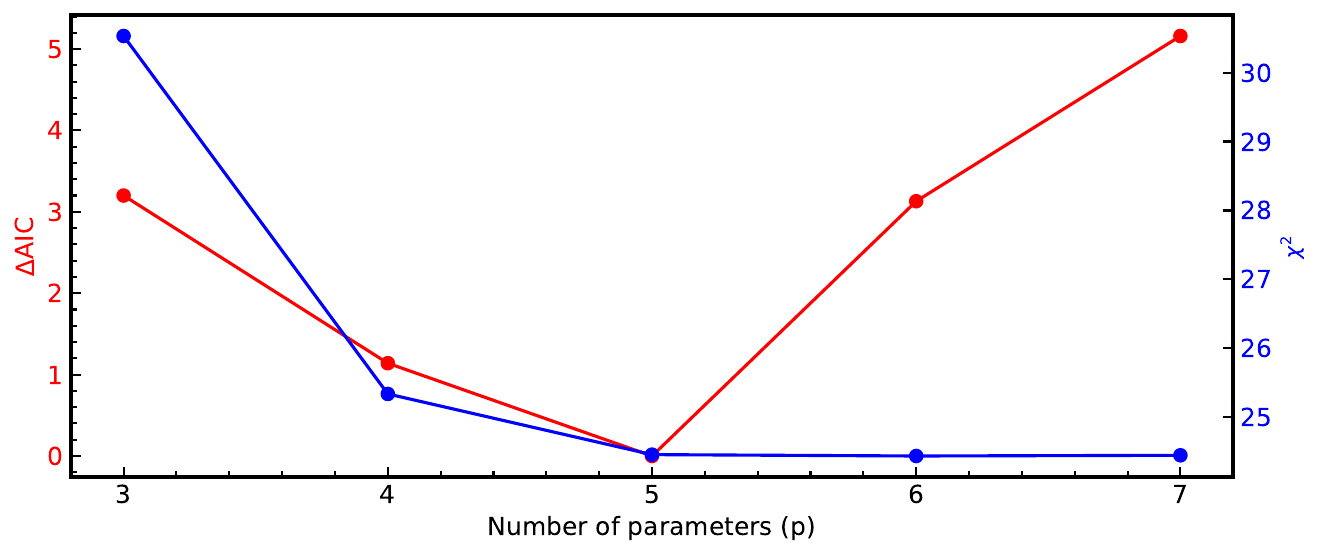}
      \caption{Relative Akaike Information Criterion and $\chi^2$ for the fit of the CO$_2$ ice bending mode at 15.2~$\mu$m.}
         \label{CO2aic}
   \end{figure}

Figure~\ref{CO2aic} displays both $\chi^2$ and relative AIC values for different number of components in the fit. The $\chi^2$ values remain the same for the fits with 5$-$7 components, which indicates that adding more IR spectra does not improve the fit. The AIC values allows the same conclusion. In addition, it shows that extra components to the fit increases the AIC values. Therefore, a fit with five components is ideal for Ced~110~IRS4A.  

\end{document}